\documentclass[10pt,english,aps,notitlepage,twocolumn,superscriptaddress,longbibliography,footinbib,prb,floatfix]{revtex4-2}

\usepackage{mathtools,amsthm,amsmath,amsfonts,amssymb}
\usepackage{graphicx}
\usepackage{subfigure}
\usepackage[colorlinks = true,
            linkcolor = blue,
            urlcolor  = blue,
            citecolor = blue,
            anchorcolor = blue]{hyperref}
\usepackage{physics}
\usepackage{comment}
\usepackage{pgfplots}
    \usetikzlibrary{positioning}
\usepackage[normalem]{ulem}
\usepackage[noend]{algpseudocode}
\usepackage[dvipsnames]{xcolor}
\usepackage{orcidlink}
\usepackage{tikz}
    \usetikzlibrary{math}

\newtheorem*{theorem}{Theorem}

\DeclareMathOperator{\mpsi}{\hat{\chi}}
\DeclareMathOperator{\E}{\mathbf{E}}

\newcommand{\tmh}{t_{\mathrm{H}}}
\newcommand{\tmt}{\tau_{\mathrm{Th}}}

\newcommand{\bb}[1]{\textcolor{red}{#1}}

\newcommand{\wes}[1]{\textcolor{violet}{#1}}

\newcommand{\pcsadd}{Center for Theoretical Physics of Complex Systems, Institute for Basic Science(IBS), Daejeon 34126, Republic of Korea}
\newcommand{\ustadd}{Basic Science Program, Korea University of Science and Technology (UST), Daejeon 34113, Republic of Korea}
\newcommand{\uvladd}{Universidad del Valle,Ciudad Universitaria Melendez, Calle 13 No. 100-00 Cali 25360 Valle del Cauca Colombia}
\newcommand{\ictpsaifradd}{ICTP South American Institute for Fundamental Research \\
                           Instituto de F\'{i}sica Te\'{o}rica, UNESP - Univ. Estadual Paulista \\
                           Rua Dr. Bento Teobaldo Ferraz 271, 01140-070, S\~{a}o Paulo, SP, Brazil}

\begin{document}

\title{{L\'evy Sachdev-Ye-Kitaev Model}}

\author{Budhaditya Bhattacharjee\,\orcidlink{0000-0003-1982-1346}}
    \altaffiliation{Current address: Department of Physics and Materials Science, University of Luxembourg, Ave. de la F\"aiencerie, Luxembourg 1511}
    \email{budhaditya.bhattacharjee@uni.lu}
    \affiliation{\pcsadd}

\author{William E. Salazar\,\orcidlink{xx}}
    \email{salazar.william@correounivalle.edu.co}
    \affiliation{\uvladd}
    \affiliation{\ictpsaifradd}

\author{Dario Rosa\,\orcidlink{0000-0001-9747-1033}}
    \email{dario\_rosa@ictp-saifr.org}
    \affiliation{\ictpsaifradd}

\author{Alexei Andreanov\,\orcidlink{0000-0002-3033-0452}}
    \altaffiliation{Current address: Center for Trapped Ions Quantum Science, Institute for Basic Science, Daejeon 34126, Republic of Korea}
    \email{aalexei@ibs.re.kr}
    \affiliation{\pcsadd}
    \affiliation{\ustadd}

\begin{abstract}
    We explore the spectral properties of the \(4\)-fermion Sachdev-Ye-Kitaev model with interaction sourced from a L\'evy Stable (fat-tailed) distribution.
    We study the eigenvalue distribution, focusing on long- and short-range correlations and extreme statistics.
    Our analysis uncovers a crossover from chaotic to integrable behavior (in the spectral correlations) as the distribution becomes increasingly fat-tailed.
    We investigate this crossover through a hierarchical analysis of the eigenvalue spectrum, based on the multi-fractal hierarchy of the L\'evy Stable distribution.
    The crossover is explained in terms of a genuine many-body effect, which is distinct from the single-particle transition (controlled by a mobility edge) of the underlying L\'evy random matrices.
    We conclude with comments on the model's solvability and discussion of possible models with exact transitions.
\end{abstract}

\maketitle

\section{Introduction}

The study of interacting many-body quantum systems has been an area of active research over the last few decades.
Closely related to the study of quantum chaotic systems and Random Matrix Theory (RMT), there has been extensive investigation of interacting many-body systems with disordered interactions.
One such system is the celebrated Sachdev-Ye-Kitaev (SYK) model~\cite{sachdev1993spin,kitaevLectures}.
The SYK model, in its most common form, is a system of \emph{all-to-all} interacting components (i.e. fermions/bosons/spins etc.) in \((0 + 1)\)-dimensions with \emph{random and uncorrelated} interaction.
In this paper, we consider all-to-all interacting Majorana fermions with random interaction described by the following Hamiltonian
\begin{align}
    \hat{H} = i^{\lfloor q/2\rfloor}\sum_{1 = i_1 < i_2 < \dots < i_q}^{N}J_{i_1 i_2 \dots i_q}\mpsi_{i_1}\mpsi_{i_2}\dots \mpsi_{i_q}\,.
    \label{eq:sykmodel}
\end{align}
where $\mpsi_i$ represents a Majorana fermion, \textit{i.e.} a quantum mechanical operator satisfying the anti-commutation algebra $\{\mpsi_{i},\mpsi_{j}\} = 2\delta_{i j}$ and $\mpsi^\dagger = \mpsi$.
The coupling constants $J_{i_1 i_2 \dots i_q}$ are chosen randomly from some probability distribution.
Each term in the Hamiltonian represents the interaction between $q$ of the system's total $N$ Majorana fermions.
In the most extensively studied version of the model, the interactions are chosen from a \emph{Gaussian} distribution with $\E(J_{i_1 i_2 \dots i_q}) = 0$ and $\E(\vert J_{i_1 i_2 \dots i_q} \vert^2) = \mathcal{K}(N)$ for some real $N$-dependent constant $\mathcal{K}$ (the dependence on \(N\) being fixed by the request of extensivity of eigenenerges).

One of the central features of this model is its \emph{solvability} in the limit of large $N$, arising from the interplay between the all-to-all interaction, the Majorana algebra and the Gaussian distribution of $J_{i_1 i_2 \dots i_q}$~\cite{maldacena2016remarks}.
Thanks to this property, the SYK model has found wide applications related to various physical phenomena~\cite{rossini2020quantum}.
Several extensions of the SYK model have been proposed and studied~\cite{fu2017susy,witten2019an,sunderhauf2019quantum,krajewski2019non,tezuka2023binary,hanada2024a,andreanov2025from}.
A general observation, which eventually can be traced to the central limit theorem (CLT), is that modifications of the model with non-Gaussian disorder (with finite variance) leave the underlying physics largely invariant.
In contrast, genuinely nonequivalent modifications of the original model are often obtained by imposing additional structure on the random interactions, thus escaping from the universality properties of CLT.
Some well-known examples are the ones with sparse~\cite{garcia2021sparse,xu2020a,caceras2022spectral} and binary-sparse~\cite{tezuka2023binary} disorder, to name just two.
The disadvantage of this approach is that it renders the model largely unsolvable, due to the imposed additional structure.
It is then natural to wonder if one can construct a generalization of SYK, that formally retains its solvability while violating CLT for the couplings \(J_{i_1 i_2 \dots i_q}\).

A natural choice in this direction is to consider a distribution of the random couplings escaping from the basin of attraction of CLT - while keeping the all-to-all feature.
The L\'evy distribution is one such choice for the underlying probability distribution from which the interactions are sampled.
This distribution is characterized by divergent moments, thus making it fundamentally different from the Gaussian distribution.
Under the \emph{Generalized Central Limit Theorem} (GCLT, also known as \emph{Stable Law})~\cite{kolmogorov1954limit}, the L\'evy distribution serves as the limiting distribution (analogous to the Gaussian distribution in CLT) for all the probability distributions with diverging second (or even the first) moment.

In this manuscript, we introduce the corresponding SYK model, which is dubbed \emph{L\'evy SYK} model (LSYK).
As we will show via semi-analytical arguments, supported by extensive numerical simulations, the model displays a \emph{genuinely many-body} chaos/integrable crossover as a function of the Levy parameter $\mu$, thus promoting the LSYK to an excellent playground to study non-ergodic behaviors in quantum many-body systems.

\section{The model and distribution}

The L\'evy stable distribution is described by its characteristic function
\begin{align}
    \mathrm{d}P_{\mu}[X] =\frac{\mathrm{d}X}{2\pi} 
    \int \mathrm{d}k \, \exp\left(ikX-|\sigma k|^{\mu}\right)\,.
    \label{eq:def_stable_dist_1}
\end{align}
The stability index $\mu \in [0,2]$ controls the asymptotic behavior (i.e. the fatness of the tail) of $P_{\mu}[X]$. 
For $\mu = 2$, the distribution becomes Gaussian, with variance given by the stability index $\sigma$.
For $\mu < 2$, the generalization of the variance is $\sigma$, which is also related to the ``typical'' value of $P_{\mu}[X]$.
The full L\'evy distribution~\cite{nolan2020univariate} is more general, with $2$ additional parameters (skewness and shift) which we set to $0$ as they are not relevant to our discussion. 
Fat tails of Levy distribution lead us to expect that, at least at the level of spectral correlations, the LSYK could mimic some aspects of the SYK models with sparse interactions (SYK models defined on hypergraphs)~\cite{janzen2010thermodynamics} -- without sacrificing the solvability properties.
Random matrices based on L\'evy distributions have been extensively studied~\cite{cizeau1994theory,biroli2021levy,burda2002free,burda2007free}.
In the following, without losing generality on its main physical features, we focus on the Hamiltonian in Eq.~\eqref{eq:sykmodel} with \(q = 4\) and study numerically various system sizes \(N\).
The interaction tensor\footnote{Which has $\mathcal{N} = \binom{N}{q}\Big\vert_{q = 4}$ independent elements.} \(J_{i_1 i_2 i_3 i_4}\) is sampled from a L\'evy Distribution (LD)~\eqref{eq:def_stable_dist_1}, where we fix the scale as
\begin{align}
    \sigma = J\left(\frac{(q-1)!}{N^{q-1}}\right)^{1/\mu}\Big\vert_{q = 4}
\end{align}
and the range of the parameter \(\mu\) is \(0 < \mu \leq 2\).
For \(\mu = 2\), the corresponding LSYK reduces to the usual Gaussian SYK~\cite{maldacena2016remarks}, with the variance ensuring extensivity of the ground state energy.

Any set of random variables which are sampled from a LD, follow a hierarchy~\cite{cizeau1994theory}.
Following~\cite{monthus2016localization}, it is convenient to represent this hierarchy using the following 
\begin{theorem}[L\'evy Hierarchy]
    \label{levyhierarchy}
    Given a set of $\mathcal{N} = \binom{N}{q}$ L\'evy random variables (for $\mu < 2$)\, it contains $O(\mathcal{N}^{1-z})$ number of random variables with magnitude $O(N^{1/\mu}\mathcal{N}^{\frac{z-1}{\mu}})$, where $0 \leq z \leq 1$. 
\end{theorem}
The proof of this theorem is straightforward (and follows from the \emph{multifractal} analysis of Ref.~\onlinecite{monthus2016localization}).
For completeness, we sketch the proof in the Appendix.~\ref{app:AppendixA}.
This implies 
(1) there are almost $\mathcal{N}$ random variables of magnitude $O(N^{1/\mu}\mathcal{N}^{-1/\mu})$, forming a very weak but dense background,
(2) there are $O(N)$ random variables of magnitude $O(1)$, forming a relatively stable mid layer and
(3) there are $O(1)$ random variables of magnitude $O(N^{1/\mu})$, comprising the set of \emph{parametrically large} outliers.
For a L\'evy random matrix, a mobility edge exists for $\mu \leq 1^{-}$~\cite{cizeau1994theory,tarquini2016level}.

For the spectral analysis of the LSYK model, we employ an approach that uses the so-called \emph{spectrum of fractal dimension} of random variables~\cite{kutlin2024anatomy}. 
We present a self-contained description of this approach and the application to the eigenvalue spectrum of LSYK in Appendixes~\ref{app:AppendixB1}-\ref{app:AppendixB2}.
Here we mention the main results. 
The central idea is to arrange a sample of $N$ random numbers $\{x_1,x_2,\dots,x_N\}$ (sourced from some distribution $P_X(x)$) in decreasing order of magnitude $\vert x _1 \vert \geq \vert x_2 \vert \geq \cdots \geq \vert x_N \vert$. 
To quantify their magnitudes, the numbers are scaled so that the largest number is $\mathcal{O}(1)$. 
Then the magnitudes can be parameterized\footnote{The scaling and parameterization is not unique and not crucial to the approach. 
Different scalings and orderings can also be used as long as appropriate modifications are made to the rest of the analysis.} in powers of $N$ as $N^{-\alpha}$ where $\alpha \geq 0$. 
The next step is to estimate the fraction of random variables of size $N^{-\alpha}$ in a typical set of $N$ random variables. 
This fraction is denoted by $N^{f_{X}(\alpha) - 1}$, where we introduce the \emph{spectrum of fractal dimensions} (SFD) $f_X(\alpha) \leq 1$. 
This function is determined by the probability distribution $P_X(x)$. This parameterization of a probability distribution glosses over the finer details of a random sample set from the distribution. 
In particular, logarithmic corrections to the ordering of magnitude are lost in the process. 
However, as we shall see, this coarse-grained approximation will be sufficient to describe the correlations in LSYK. 


{
The functional form of then SFD \(f_L(\alpha)\), corresponding to the L\'evy and Gaussian distributed interactions $j_I$, is represented graphically below.
}
\begin{center}
\centering
\begin{tikzpicture}[scale=0.75]

\def\a{0}

\node at (2+\a, -1.4) {\textbf{(a)} L\'evy};

\draw[->] (0+\a,0) -- (4.2+\a,0);
\draw[->] (0+\a,0) -- (0+\a,3.2);

\draw[-,dashed,gray,line width=0.5mm] (0+\a,2) -- (4.2+\a,2);

\draw[-,dotted,brown,line width=0.5mm] (2+\a,2) -- (2+\a,0);

\draw[-,blue,line width=0.8mm] (0+\a,0) -- (2+\a,2);
\draw[-,blue,line width=0.8mm] (2+\a,2) -- (3+\a,0);

\node[rotate=90] at (-0.5+\a, 1.1) {$f_L(\alpha)$};
\node at (4.2+\a, -0.4) {$\alpha$};

\node at (-0.25+\a, 2.0) {$1$};
\node at (2+\a, -0.35) {$1/\mu$};

\node[rotate=45]  at (0.8+\a, 1.2) {$\mu\alpha$};
\node[rotate=-63.43, scale = 0.6] at (2.8+\a, 1.0) {$1-\alpha+(1/\mu)$};

\node at (2+\a, 3.0) {$X \sim |j_I|$};

\def\b{5.5}

\node at (1.75+\b, -1.4) {\textbf{(b)} Gaussian};

\draw[->] (0+\b,0) -- (3.7+\b,0);
\draw[->] (0+\b,0) -- (0+\b,3.2);

\draw[-,dashed,gray,line width=0.5mm] (0+\b,2) -- (3.7+\b,2);

\draw[-,brown,line width=0.8mm] (0+\b,2) -- (3+\b,0);

\node[rotate=90] at (-0.5+\b, 1.1) {$f_G(\alpha)$};
\node at (3.7+\b, -0.4) {$\alpha$};

\node at (-0.25+\b, 2.0) {$1$};

\node[rotate=-33.43] at (1.8+\b, 1.2) {$1-\alpha$};

\node at (1.75+\b, 3.0) {$X \sim |x_G|$};

\end{tikzpicture}
\end{center}
{The above panel (a)} is the graphical representation of the L\'evy Hierarchy Theorem.
The SFD has the following functional form
\begin{align}
    f_L(\alpha) = \mu\alpha\theta(1-\mu\alpha) + \left(\frac{1}{\mu} + 1 - \alpha\right) \theta(\mu\alpha-1)\,,
    \label{eq:sfd-lvd}
\end{align}
where $\mu$ is the scale parameter as defined previously. 
This holds strictly for $\mu < 2$.
For $\mu = 2$, the probability distribution is Gaussian with the SFD $f_G(\alpha) = 1 - \alpha$ for a Gaussian random variable $x_G$.
{
The non-monotonicity of the SFD in Eqn.~\eqref{eq:sfd-lvd}, also shown in panel (a) captures the difference between the \emph{typical} and \emph{maximal} (largest) random numbers that one may find in random sample set.
Complete details of the derivation of Eqn.~\eqref{eq:sfd-lvd} are presented in Appendices.~\ref{app:AppendixB1} and~\ref{app:AppendixB2}
}. 

The SFD is to be interpreted as follows: the value of $\alpha = \alpha_0$ at which $f(\alpha_0) = 1$ represents the \emph{typical} size of random variable from the corresponding distribution.
The fraction of random numbers with the largest size is in the set (corresponding to $\alpha = 0$) is found at $f(0)$. 
Given the SFD $f_L(\alpha)$ of L\'evy random variables, the fusion rules of random variables~\cite{kutlin2024anatomy} are used to compute the SFD of functions of L\'evy random variables. 
In the following section, we shall use the L\'evy distribution SFD and fusion rules to study the scaling of level correlations in LSYK.

\section{Effective integrability and the SFD of level correlations}

We first discuss the physics expected to emerge due to the LD/fat tails.
For LSYK (with general \(q\)), one samples \(\mathcal{N} = \binom{N}{q}\) random variables. {
The numerical details, and the extra difficulties associated with L\'evy distribution of the couplings, are discussed in detail in the Appendices~\ref{app:AppendixA} and~\ref{app:AppendixB}.
In Appendix.~\ref{app:AppendixA}, we review the derivation of the hierarchy of L\'evy random numbers. 
In Appendix.~\ref{app:AppendixB}, first, the exact many-body picture is described. 
The Hamiltonian is considered in the basis where the largest $J_{i j k l}$ terms are on the diagonal. 
Then then rest of the spectrum is calculated as perturbations to these very large terms. 
Next (Appendix.~\ref{app:AppendixB1} and~\ref{app:AppendixB2}) we review the multifractal spectrum (and present the associated diagrammatic analysis, following~\cite{kutlin2024anatomy}), and apply it to calculate the scaling of the level spacing distribution.
The derivation concludes in the result presented in Eqn.~\eqref{eq:sfd-lvd} and the figures above and below.
The paragraph before that outlines the exact meaning of the spectrum of fractal dimensions.
Finally at the end of Appendix.~\ref{app:AppendixB2} we outline the main argument, which is also mentioned in the main text. 
The following Appendices.~\ref{app:AppendixB3} and~\ref{app:AppendixB4} are devoted to demonstrating the robustness of this approach.
}

For large $N$, random matrices composed of these variables demonstrate an energy dependent transition from an ergodic to localized phase for \emph{any} $\mu < 1$.
The key difference between a L\'evy random matrix and the Hamiltonian~\eqref{eq:sykmodel} is that the latter is \emph{many-body}, \textit{i.e.} it encodes a tensor product structure.
In what follows, we estimate the nature of the \emph{many-body} spectrum.

\begin{figure}
    \centering
    \subfigure[]{
    \includegraphics[width=0.45\linewidth]{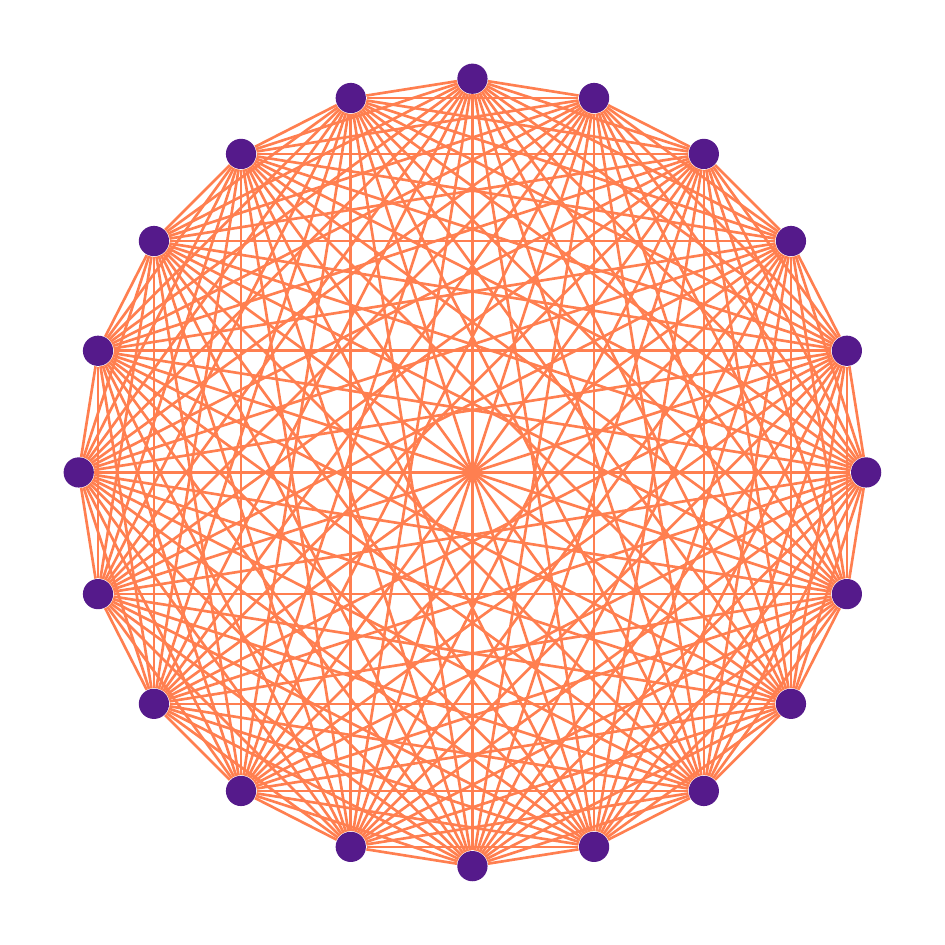}}
    \subfigure[]{
    \includegraphics[width=0.45\linewidth]{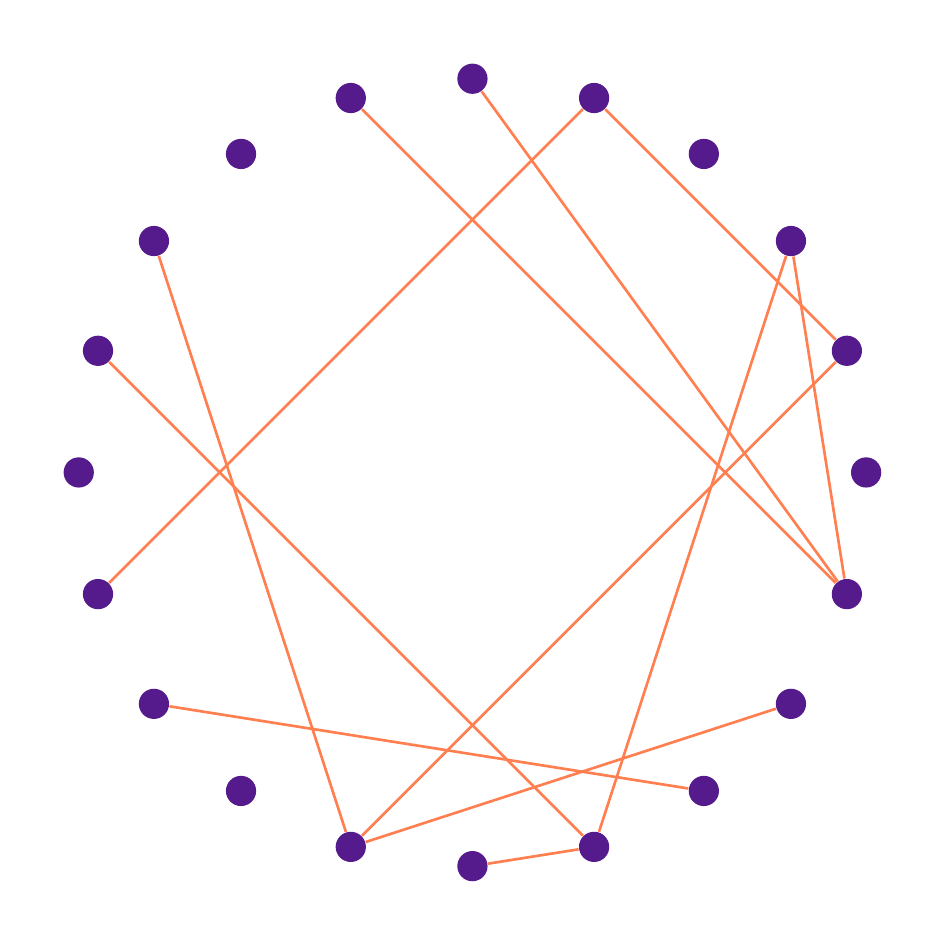}}
    \caption{
        Interaction graph for \(20\) Majorana fermions (corresponding to \(q=2\) for representation purpose), for
        (a) \(\mu = 2.0\) corresponding to the usual Gaussian SYK and 
        (b) \(\mu = 1.0\) corresponding to strong L\'evy disorder.
        The cutoff for removed interactions is set to \(10^{-5}\). 
    }
    \label{fig:levygraph}
\end{figure}

As we will show, the spectral correlations of the LSYK model are significantly different from those of a L\'evy random matrix (for strong L\'evy-ness, \textit{i.e.} small $\mu$). Let us denote the Hamiltonian as $H = \sum_{I}J_{I}\Psi_{I}$, where $I \equiv i_1,i_2,\dots ,i_q$ and $\Psi_I \equiv \mpsi_{i_1}\mpsi_{i_2}\dots\mpsi_{i_q}$.
Since there is a hierarchy in $J_{I}$, let us represent the parametrically large outliers by $\mathbf{J}_{I}$, the remaining terms by $\mathbf{j}_I$. According to the hierarchy of L\'evy random variables, there are $O(1)$ of parametrically large random numbers $\mathbf{J}_I$.
The Hamiltonian then decomposes as 
\begin{align}
    H = i^{\lfloor q/2 \rfloor}\sum_{I}\mathbf{J}_{I}\Psi_{I} + i^{\lfloor q/2 \rfloor}\sum_{I'}\mathbf{j}_{I'}\Psi_{I'}\,,
    \label{eq:split_ham_lv_1}
\end{align}
where $I, I'$ run over the indices corresponding to the parametrically large and the remaining random variables (in the hierarchy), respectively.
A representative example for $N = 20$ fermions (with $q = 2$ body interactions) is presented in Fig.~\ref{fig:levygraph}, where the parametrically smaller interactions $\mathbf{j}_i$ are deleted.
Recall that given two strings of Majorana fermions of length $q$ (even) each, having $k$ common fermions, they commute with each other iff $k$ is even. 

Since the number of large $J_I$ is $O(1)$, their terms in Eq.~\eqref{eq:split_ham_lv_1} commute with each other almost surely.
The Hamiltonian can then be cast in the common eigenbasis of these terms.
For simplicity, let us assume that there is $1$ such term, say $J$.
In the eigenbasis of this term (which we denote by $\ket{\psi^0}$), the Hamiltonian can be viewed as being comprised of $2$ blocks of size $\frac{\mathcal{D}}{2} \times \frac{\mathcal{D}}{2}$, where $\mathcal{D}$ is the Hilbert space dimension.
The upper-left and lower-right diagonal blocks correspond to the eigenvectors $\ket{\psi^0}$ with eigenvalues $+J$ and $-J$ respectively.
The diagonal of $H$ in this basis is given by $H_{ii} = \pm J + \sum_{k = 1}^{M}w_{i k}j_{k}$, where $j_{k}$ are some of the smaller terms (usually $M \sim O(N)$ in number) which are also simultaneously diagonal in the eigenbasis $\ket{\psi^0}$.
Further, $w_{i k}$ are $O(1)$ weights with arbitrary signs. 
According to the Ger\v{s}gorin Circle Theorem~\cite{varga2011geršgorin}, we know that the eigenvalues of $H$ will lie in intervals centered at the diagonal $H_{ii}$ with a width given by $R = \sum_{j\neq i} \vert H_{i j}\vert$.
Since the parametrically large terms dominate sums of L\'evy random numbers, the density of states is peaked at $H_{ii}$ with a spread of $O(R)$.

In units of the largest term $J$, the \emph{typical} interaction (which forms the underlying weak but densely connected hypergraph) terms are then of $O(N^{-a/\mu})$, for some constant $a > 1$.
This is a consequence of the L\'evy hierarchy.
The eigenvalues of this model can be expanded perturbatively in orders of the largest interaction (which is set to $\mathcal{O}(1)$ by normalization).
\begin{align}
    E_{i} = 1 + \sum_{k}^{N^r}w_{i k}j_{k} + \sum_{m}^{N^a}\frac{\vert j_{m} \vert^2}{e_{m}} + \sum_{n}^{N^b}\frac{\vert j_n \vert^2}{2}
\end{align}
where $w_{i k}$ are $O(1)$ phases, and $e_{m} = \sum_{k'}^{N^r}w_{m_1 k'}j_{k'} - \sum_{k}^{N^r}w_{m_2 k''}j_{k''}$ for some pair of indices $m_1, m_2$.
From this expression, we can compute the mean level spacing of the eigenvalues $\Delta = \langle E_i - E_j\rangle/\mathcal{D}$.

The mean level spacing,\footnote{Which can be visualized as weighted sums of L\'evy random variables, with CLT or $O(1)$ weights.} on the other hand, is bounded by $\frac{N^{b/\mu}}{\mathcal{D}^2}$ (again, in units of $J$), see the Supplementary for a detailed analysis.
The SFD of the mean level spacing $\Delta$ can be similarly computed.
\begin{center}
\begin{tikzpicture}[scale=0.95]
    \def\a{5}; 
    \def\b{0.5}; 
    \draw[->] (0+\a,0) -- (4+\a,0);
    \draw[->] (0+\a,0) -- (0+\a,3);
    \draw[-,dashed,gray,line width=0.5mm] (0+\a,2) -- (4+\a,2);
    \draw node[rotate=90] at (-0.5+\a,3.0) {$f_{\Delta}(\alpha)$}; 
    \draw node at (4.0+\a,-0.5) {$\alpha$};
    \draw node at (-0.25+\a,2.0) {$1$};
    \draw node at (-0.8+\a,1.0) {$\max\{b,\frac{1}{2}\}$};
    \draw node at (2+\a,-0.3) {\scriptsize $\alpha_0 = \min\{\frac{1}{\mu},\frac{2(1-b)}{\mu}\}$};
    \draw[-,dotted,brown,line width=0.5mm] (2+\a,2) -- (2+\a,0);
    \draw[-,RubineRed,line width=0.8mm] (2+\a,2) -- (2.8+\a,0);
    \draw[-,RubineRed,line width=0.8mm] (0+\a,1.0) -- (2+\a,2.0);
    \draw node[rotate = 26.57] at (0.95+\a,1.2) {\scriptsize $\frac{\mu\alpha}{2} + \max\{b,\frac{1}{2}\}$};
    \draw node[rotate = -68.2] at (2.8+\a,1.0) {${1-\alpha + \alpha_0}$};
\end{tikzpicture}
\end{center}
Here $b = 2(1-\beta)$ and $N^\beta$ is the fraction of L\'evy random variables that appear in the sum. 
Let us emphasize that this treatment essentially considers the level spacings as independent random variables. 
This is not true, strictly speaking, but is an accurate enough assumption given that we are only trying to estimate the typical size of this quantity.
This is helped by the fact that for small $\mu$, the Generalized Central Limit Theorem ensures that any sums are strongly dominated by the parametrically large values of the underlying distribution.

The emergence of integrability can be estimated by the (largest value of) mean level spacing being \emph{much greater} than the typical interaction strength, very much in the spirit of a ``resonance counting'' approach.
This means that we compare the typical interaction strength $j_{I,\mathrm{typ}}$ (the magnitude that appears most often) with the $\max(\Delta)$.
The value $\max(\Delta)$ tells us the largest order of magnitude that the mean level spacing of LSYK can be, on an average over realizations.
If the typical interaction $j_{I,\mathrm{typ}}$ is of the same order of magnitude as the $\max(\Delta)$, then each term in the Hamiltonian can cause a large number of hoppings between eigenstates of the model.
This is the feature that we identify as the signature of quantum chaotic behavior.
On the other hand, if $j_{I,\mathrm{typ}} \ll \max(\Delta)$, then there will be a much smaller fraction of such hoppings that will be possible.
This is presented in the following schematic.
\begin{center}
\begin{tikzpicture}[scale=0.95]
    \draw[-,black,line width=1mm] (0,0) -- (1,0);
    \draw[-,black,line width=1mm] (1,2) -- (2,2);
    \def\a{1.0};
    \draw[-,black,line width=0.3mm] (-0.5,1) -- (-0.5+\a,1);
    \draw[-,black,line width=0.3mm] (-0.2,0.5) -- (-0.2+\a,0.5);
    \draw[-,black,line width=0.3mm] (0.2,0.8) -- (0.2+\a,0.8);
    
    \draw[-,black,line width=0.3mm] (0.5,1.5) -- (0.5+\a,1.5);
    \draw[-,black,line width=0.3mm] (0.2,1.2) -- (0.2+\a,1.2);
    \draw[-,black,line width=0.3mm] (-0.2,1.8) -- (-0.2+\a,1.8);

    \draw[-,black,line width=0.3mm] (-0.2,2.2) -- (-0.2+\a,2.2); 
    \draw[-,black,line width=0.3mm] (0.2,-0.2) -- (0.2+\a,-0.2); 
    \draw[->,black] (-1,0) -- (-1,2);
    \draw node[rotate=90] at (-1.2,1) {$E$};  

    \draw[<->,red] (2.5,0) -- (2.5,2); 
    \draw node[rotate=-90] at (2.7,1) {$\color{red} \max(\Delta)$};  

    \draw[<->,blue,thick] (0.8,0.2) to[bend right] (1.8,1.8);
    \draw node[] at (1.8,0.4) {$\color{blue} j_{I,\mathrm{typ}}$};
\end{tikzpicture}
\end{center}
Enforcing the condition leads to the following bound
\begin{align}
    j_{I,\text{typ}} \sim \label{eq:chaos_int_transition_prediction}
    N^{-a/\mu} \ll \frac{N^{b/\mu}}{\mathcal{D}^2}
    \implies \mu \ll \frac{a' \ln N}{N\ln 2}
\end{align}
where we have set $a' = a + b$, with $a$ and $b$ being $O(1)$ functions of $q$.
Note that the conclusion remains the same even if we consider the typical mean level spacing values, which corresponds to setting $b = 0$.
The key requirement here is $b > -a$, which is almost surely true since the sum of all level spacing should be greater than the typical interaction~\footnote{This fails if the typical interaction is also the largest one, as is the case with the Gaussian (or any CLT) distribution}.
Therefore, the critical $\mu$, at which the condition for integrability is satisfied, is expected to scale inversely~\footnote{The logarithmic term in the numerator is an artifact of the multi-fractal analysis of L\'evy numbers.} with $N$.
This argument unveils the origin of the genuine \emph{many-body} effect in the spectra.
Using this approach, it is straightforward to demonstrate that if $\mathcal{D} \sim \mathcal{N}$, then $\mu$ is $O(1)$ and independent of $N$.
This is the effective single-particle L\'evy random matrix result.
In turn, this same argument suggests that an SYK model whose $q$ scales with $N$ might give rise to a spectrum replicating the L\'evy matrix mobility edge at the many-body level.

In the following sections, we will test the predictions of Eq.~\eqref{eq:chaos_int_transition_prediction} against the results on the spectral statistics of LSYK at finite values of $N$.

The specific Gaussian random matrix class that the model belongs to depends on the total number of fermions $N$~\cite{garcia2016spectral}.
For system size $N = 2, 6 \,(\text{mod} \, 8)$, the model belongs to the Gaussian Unitary Ensemble (GUE).
For system sizes $N = 4 \, (\text{mod}\, 8)$, the RMT class is the Gaussian Symplectic Ensemble (GSE).
Finally, system sizes $N = 8 k, \, k \,\in\, \mathrm{Z}_{+} $ belong to the Gaussian Orthogonal Ensemble (GOE).
The spectrum is computed via exact diagonalization.


\section{Numerical Results: Short-range statistics}

\begin{figure}[ht]
    \centering
    \includegraphics[width=\linewidth]{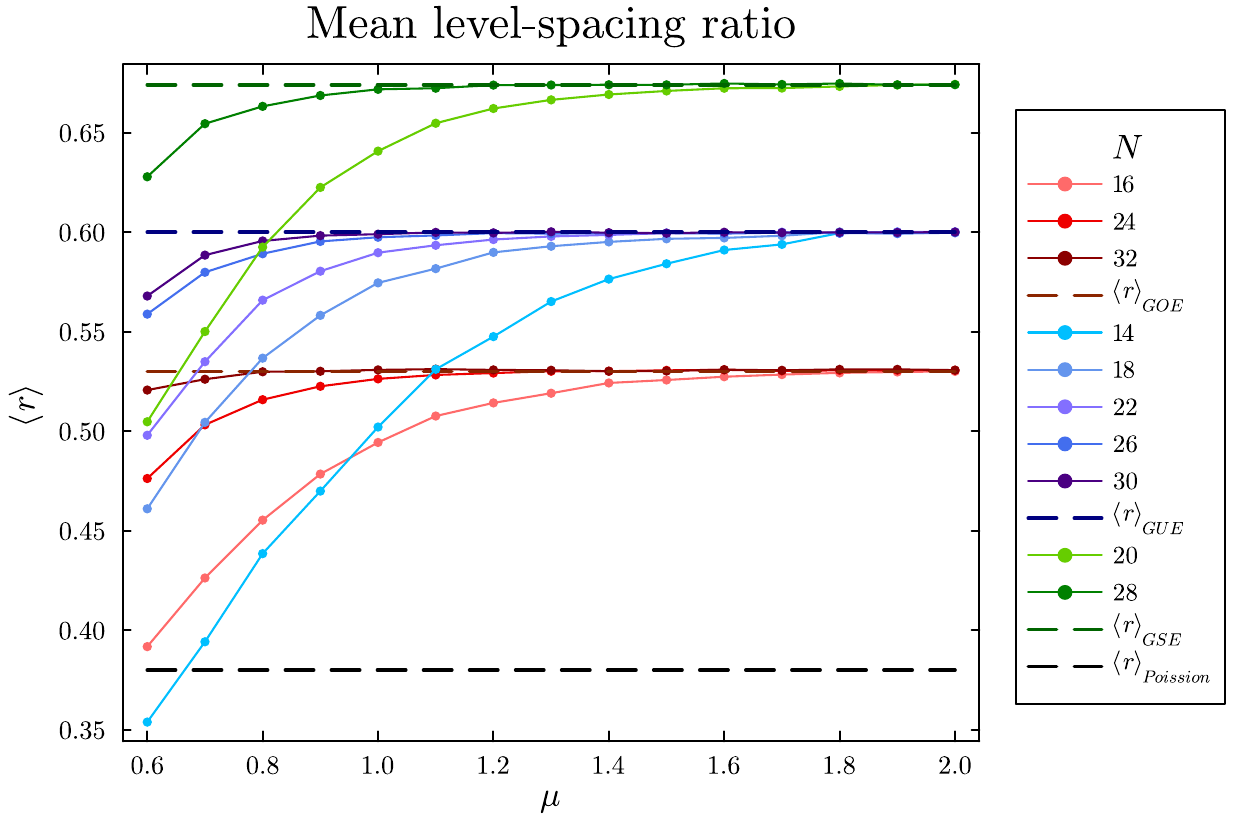}
    \caption{
        (Mean) Nearest neighbor level spacing ratio for LSYK as a function of the L\'evy scale parameter \(\mu\) for \(N = 14\) to \(30\) Majorana fermions.
        The middle \(1/3\) of the spectrum is used to compute \(\langle r \rangle\).
        The results are robust against different choices of the part of the eigenspectrum used to compute it.
    }
    \label{fig:r-ratio-usual}
\end{figure}

The short-range spectral properties are studied through the spacing $s_{k}$ of nearest neighbor eigenvalues $s_{k} = \epsilon_{k + 1} - \epsilon_{k}$.
The spacing distribution probabilities are known for Gaussian ensembles~\cite{atas2013distribution}.
We compute the mean nearest-neighbor spacing ratio \(\langle r \rangle\)~\cite{oganesyan2007localization}.
The results (averaged over many disorder realizations) are presented in Fig.~\ref{fig:r-ratio-usual}.
The behavior of $\langle r \rangle$ as a function of \(\mu\) shows a clear deviation from Gaussian RMT results (\(\mu = 2.0\), where LSYK becomes the usual Gaussian SYK).
It is interesting to note that the deviation from Gaussian RMT (upon decreasing $\mu$) is \emph{not immediate} at the level of short-range eigenvalue correlations, and a consistent, finite, deviation from $\mu = 2$ is necessary before making the deviation noticeable.
More in details, as the system size $N$ increases smaller and smaller values of $\mu$ are necessary to deviate from the RMT statistics.

\begin{figure}[ht]
    \centering
    \includegraphics[width=\linewidth]{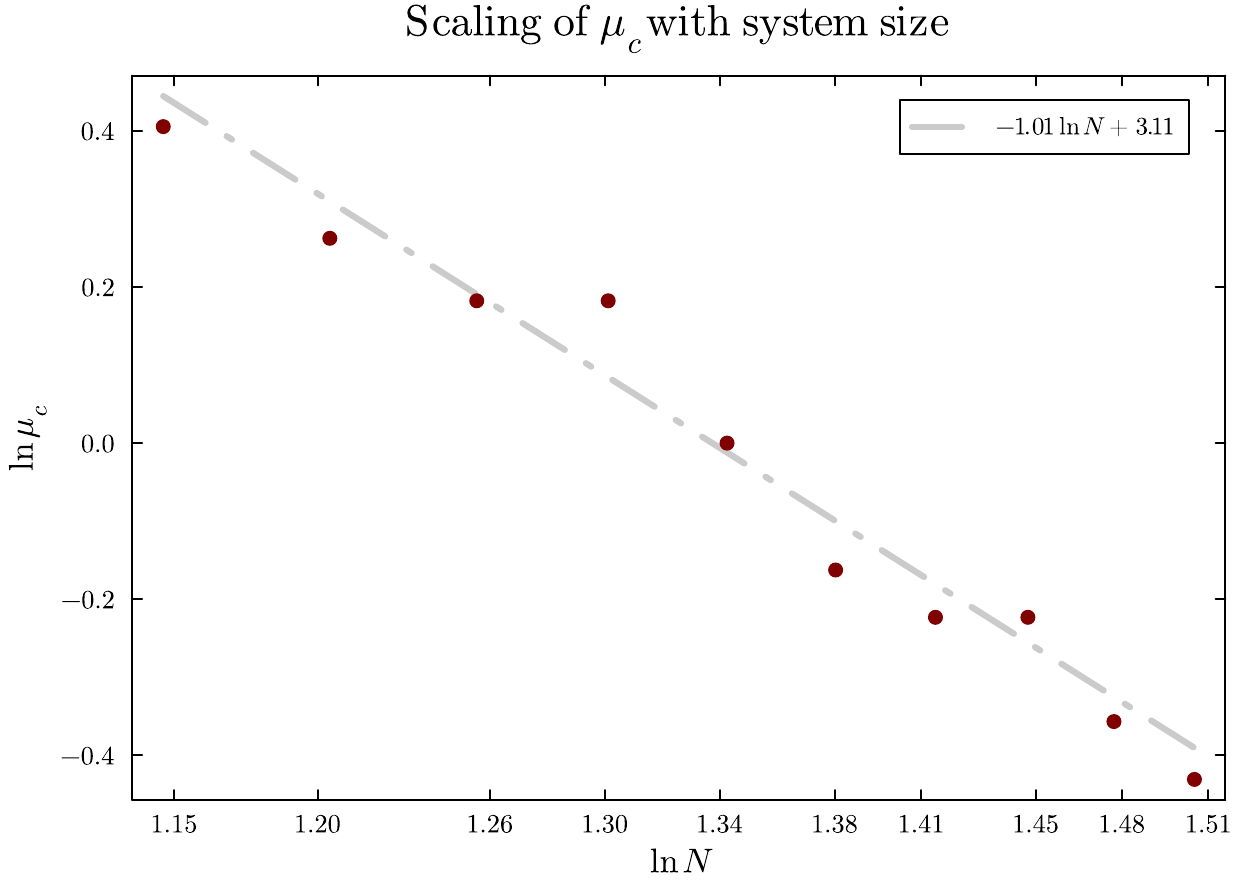}
    \caption{
        Crossover scale parameter \(\mu_{c} (N)\) at which the mean \(\langle r \rangle\) shows significant deviation away from the Gaussian RMT value.
        The observed behavior is compatible with \(\mu_{c} \sim N^{-1}\).
    }
    \label{fig:critical-mu-usual}
\end{figure}

We define a crossover value \(\mu_c\) at which the mean $\langle r \rangle$ deviates significantly from the Gaussian RMT value.
Namely, we define \(\mu_c\) as a value of \(\mu\) for which \(\langle r \rangle\) deviates away from the Gaussian RMT values beyond a threshold of \(\sim 10\%\).
We verified that this result is quite robust against different threshold values: 
we find quantitatively similar results for thresholds between \(6 - 12 \%\).
Using this definition, we determine the scaling of \(\mu_c\) as a function of \(N\), as shown in Fig.~\ref{fig:critical-mu-usual}.

We find that \(\log \mu_c\) decreases roughly linearly with $\log N$, \(\ln \mu_c = -\eta_1\ln N\), with a coefficient $\eta_1 \sim 1$.
Based on the system sizes accessible to us, we conclude that, the short-range correlations start to deviate from the Gaussian RMT at a crossover value of the scale parameter \(\mu\) which decreases with system size:
\begin{align}
    \mu_{c} \sim \# N^{-\eta_1},\;\; \eta_{1} \sim 1\,.
    \label{short-range-muc}
\end{align}
The exponent \(\eta_1\) is only known approximately, but is gapped away from \(0\) and expected to be close to \(1\).
This indicates that in the thermodynamic limit \(N \rightarrow \infty\), we expect \(\mu_c \rightarrow 0\).
The numerically obtained relation, Eq.~\eqref{short-range-muc} is in qualitative agreement with the theoretical prediction of Eq.~\eqref{eq:chaos_int_transition_prediction} and it provides confirmation of the \emph{many-body} origin of the chaos/integrable crossover in LSYK.
{
In contrast, the bulk spectral correlations of Gaussian SYK, as captured by probes such mean nearest-neighbor level statistics or (universal part of) spectral form factor/number variance, display Gaussian RMT behavior.
This is a consequence of the generic nature of the SYK model.
In particular, the many-body character of the SYK model is not reflected in these level correlations.
Concretely, Gaussian SYK spectral correlations are qualitatively very similar to the spectral correlations of Gaussian \(J_{ijkl}\) random matrices. The sole observation here is that for LSYK the spectral correlations are not the same as the L\'evy \(J_{i j k l}\) random matrices. We do not comment on the exact underlying mechanism that leads to the presence/absence of this similarity. 
Many-body effects are reflected more clearly in \emph{dynamical complexity} such as Out-of-time-ordered-correlators or Krylov complexity, or in extremal spectral statistics.
}

It is also instructive to explore the behavior of the eigenvalues at the edges of the spectrum, and study the impact of L\'evy disorder on their properties.
Edge (extreme) statistics provide valuable insight into the system's properties and have been extensively studied in random matrix theory~\cite{dean2006large,majumdar2010extreme,auffinger2016extreme,winn2023extreme}, while also having been studied quite extensively in connection with its holographic interpretation~\cite{garcia2016spectral,maldacena2016remarks,jevicki2016bilocal}.
We first study the scaling of the smallest eigenvalue \(E_{\mathrm{min}}\) (\textit{i.e.} the ground state energy) with system size.
This quantity has been studied for Gaussian SYK~\cite{garcia2016spectral}.
Since the (Gaussian SYK) Hamiltonian is constructed out of fermions, the lowest eigenstate is expected to be close to the configuration with all fermions in the lowest available single-particle levels following Pauli exclusion principle.
Therefore, the average eigenvalue $\langle E_{\mathrm{min}} \rangle$ is expected to scale linearly with $N$.
Broadly, the same expectation should hold for LSYK as well.
However, the L\'evy distribution and its fat-tailed property should introduce strong fluctuations as $\mu$ decreases.  

\begin{figure}[ht]
    \centering
    \includegraphics[width=\linewidth]{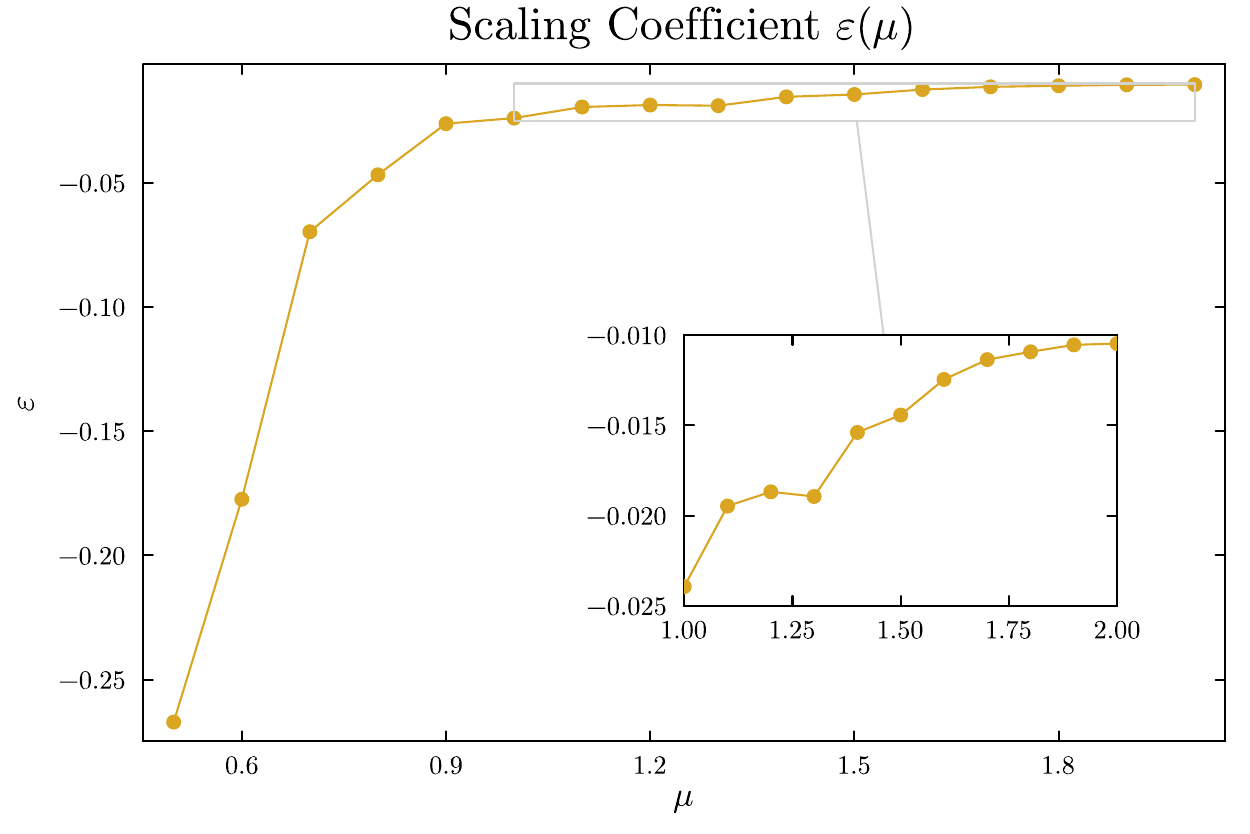}
    \caption{
        Scaling of the coefficient \(\varepsilon\), which controls the scaling of \(\langle E_{\min}\rangle\) with \(N\), as a function of \(\mu\). 
    }
    \label{fig:epsilon-scaling}
\end{figure}

To go further, we can assume a linear scaling, especially for small $\mu$, and use a log averaging to reduce the fluctuations and to get a faithful mean.
We can denote the linear coefficient by \(\varepsilon\) (defined as \(\langle E_{\min} \rangle \sim \varepsilon N\), with \(\varepsilon = -0.01\) for \(\mu = 2.0\), as a function of \(\mu\).
Note that $\varepsilon$ can also be interpreted as the ground state energy density for $N \rightarrow \infty$.
The result is presented in Fig.~\ref{fig:epsilon-scaling}.
We find that \(\varepsilon\) decreases with decreasing $\mu$.
However, until $\mu \geq 1.0$ its value decreases very weakly, while a sharp drop is evident for \(\mu < 1.0\).

The transition in the edge statistics is \emph{not} in agreement with the bulk $r$-statistics:
the edge sees a weak system size scaling (unlike the robust $N$ scaling in bulk statistics), in addition to a sharp transition at $\mu \approx 1$ (which is missing in the bulk).
This observation supports the argument that the bulk transition is a true \emph{many-body effect}.
The edge statistics are effectively a single-particle effect.
Therefore, its behavior should be dictated by the mentioned \emph{L\'evy random matrices} transition~\cite{burda2002free,burda2007free,biroli2021levy} at $\mu = 1$~\footnote{See Appendix for a multifractal analysis of the L\'evy spectrum, where we derive the $\mu = 1$ transition}.

\section{Numerical Results: Long-range statistics}

In general, long-range spectral properties of a Hamiltonian may behave differently as compared to the short-range statistics~\cite{a1995spectral,bujisman2020sensitivity}.
In the sparse SYK model, the onset of integrability breaking is revealed for different degrees of sparsity when considering short- or long-range spectral statistics~\cite{garcia2021sparse,Orman:2024mpw}.
We use the \emph{spectral form factor} (SFF) for our analysis.
Other alternatives include number variance, spectral rigidity~\cite{haake1991quantum,mehta2004random}, power spectrum~\cite{A2002quantum}. 
\begin{figure}[ht]
    \centering
    \includegraphics[width=\linewidth]{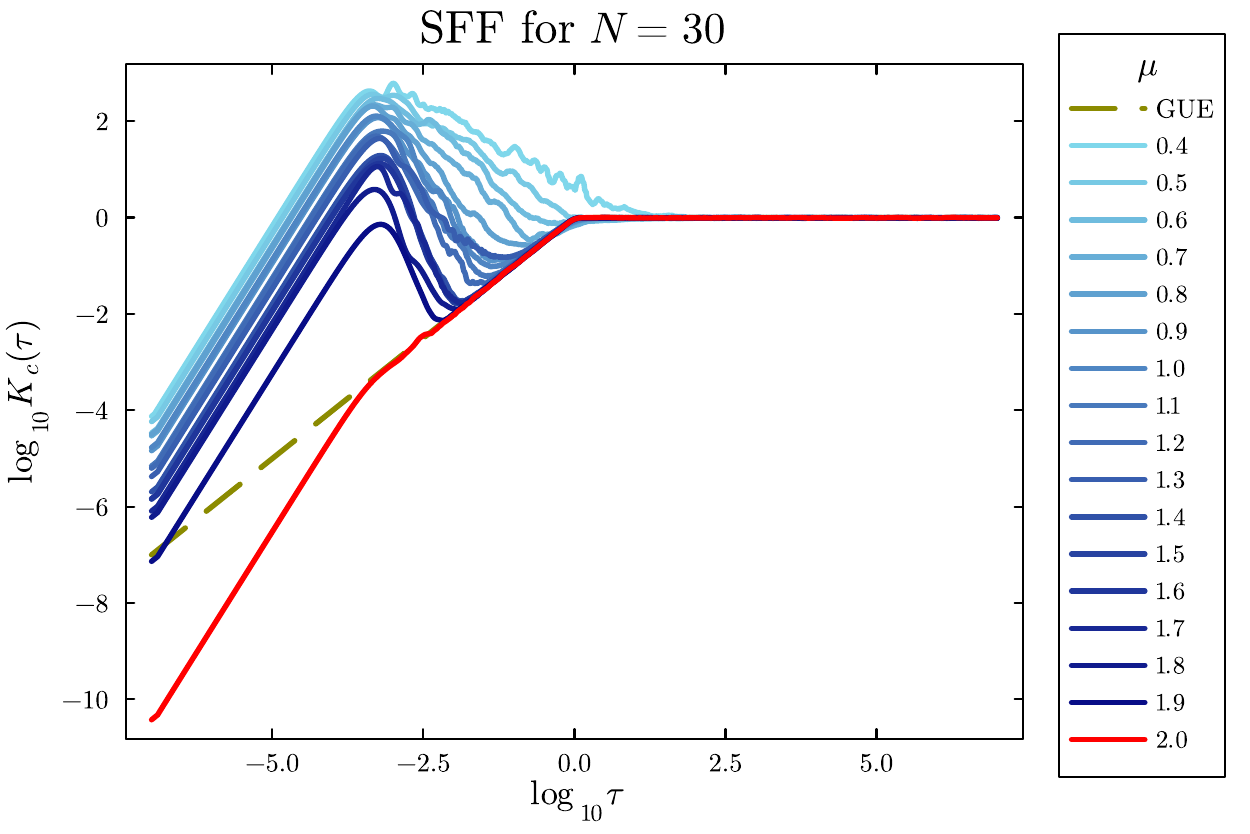}
    \caption{
        Connected SFF computed with with unfolded spectrum for $N = 30$ fermions and a range of Levy parameters \(\mu\).
        The dashed line corresponds to the Gaussian RMT theory (GUE for \(N=30\) fermions).
        The red curve corresponds to the Gaussian SYK, \(\mu=2.0\) and is always bounded from above by the RMT curve.
        There is a clear bump present for any \(\mu < 2.0\).
    }
    \label{fig:unf-sff-conn-30}
\end{figure}

The result for the connected SFF for the same \(N=30\) is given in Fig.~\ref{fig:unf-sff-conn-30}.
At the level of the disconnected SFF, a distinct feature appears for \(\mu < 2.0\): while for \(\mu = 2.0\) (Gaussian SYK), $K_{c}(\tau)$ is always bounded from above by the GUE curve~\footnote{For other system size, it will be bounded above by the corresponding RMT curve}, this is no longer true for \(\mu < 2.0\).
$K_{c}(\tau)$ approaches the GUE curve from \emph{above} for $\mu < 2.0$, rather than from below as it does for $\mu = 2.0$, as seen in Fig.~\ref{fig:unf-sff-conn-30}.
This behavior is accompanied by a few salient features shown in Fig.~\ref{fig:unf-sff-conn-30}:
(i) $K_{c}(\tau)$ develops a maxima prior to merging with the GUE curve for $\mu < 2.0$;
(ii) the time $\tau^{*}$ at which this maxima occurs is nearly independent of the value of $\mu$;
(iii) the height of this maxima $K_{c}(\tau^{*})$ increases as $\mu$ decreases.
By examining other system sizes, we find that both $\tau^{*}$ and $K_{c}(\tau^{*})$ depend on system size.
Specifically, $\tau^{*}$ decreases and $K_{c}(\tau^{*})$ increases with increasing $N$.

\begin{figure}[ht]
    \centering
    \includegraphics[width=\linewidth]{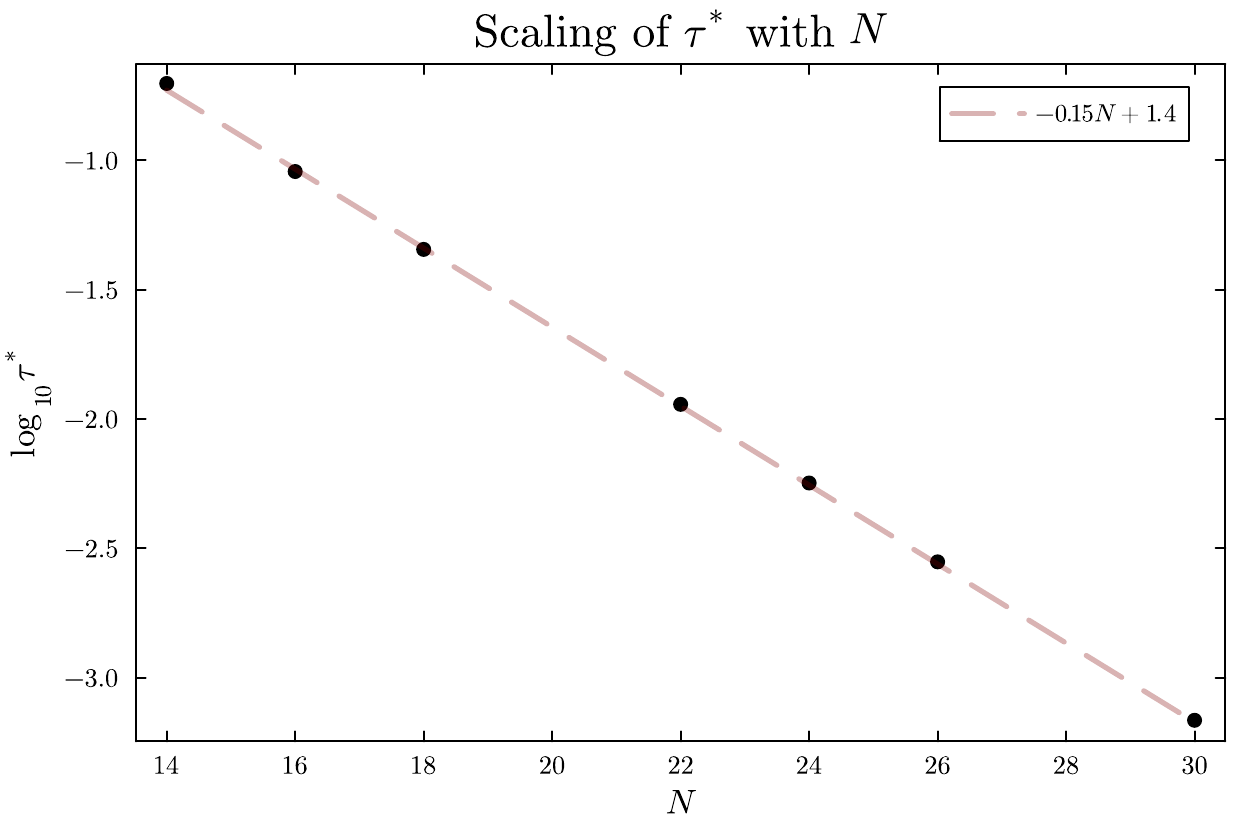}
    \caption{
        The time $\tau^{*}$, at which the connected SFF peaks before merging with the RMT curve, as a function of system size $N$.
        The decay follows a weak power-law behavior.
    }
    \label{fig:tsar-sff}
\end{figure}

The scaling of $\tau^{*}$ with system size is shown in Fig.~\ref{fig:tsar-sff}.
It obeys a weak power law given by the exponent 
\begin{align}
    \log_{10} \tau^{*} \sim -\alpha^{*}N \;\;\;,\;\; \alpha^{*} = 0.15
\end{align}
This indicates that $\tau^{*} \rightarrow 0$ in the thermodynamic limit.

Another important measure of the onset of RMT behavior in terms of the SFF is the \emph{Thouless time} \(\tmt\)~\cite{winer2022hydrodynamic} defined as the time time after which the (unfolded) SFF matches the corresponding RMT curve (i.e. the \emph{ramp}, see Fig.~\ref{fig:unf-sff-conn-30}).
We investigate the scaling of the $\tmt$ as a function of system size $N$, for a fixed L\'evy disorder strength $\mu$. 
We observe a power law decay of \(\tmt\) for different values of \(\mu\):
\begin{align}
    \log_{10} \tmt \sim -\alpha_\mu N.
\end{align}
Naturally, the condition $\alpha^{*} > \alpha_{\mu}$ must hold $\forall\;\mu$, since it corresponds to $\tau^{*} < \tmt$.
The power law scaling of \(\tmt\) with $N$ holds for all \(\mu\), with the exponent \(\alpha_\mu\) depending on $\mu$.

From the result shown in Fig.~\ref{fig:unf-sff-conn-30}, we expect the Thouless time to decrease with decreasing \(\mu\), and thus the coefficient \(\alpha_{\mu}\) should also decrease with \(\mu\).
We present these results in the Supplementary, as they are not central to the current discussion on deviation from chaoticity.
The final quantity that we can compute here is the crossover value of $\mu$ at which the Thouless time goes to $0$, or equivalently \(\alpha_\mu \to 0\).
We denote this value by $\mu_{c,2}$, to distinguish it from $\mu_{c}$ at which the $r$-ratio deviates from RMT results~\eqref{short-range-muc}.
Given the limits of the system sizes we can access and the limited reliability of the unfolding procedure for small $\mu$, our numerical result is best interpreted as an order of magnitude estimate of the actual $\mu_{c,2}$.
Further numerical details are covered in the Supplementary.
\begin{figure}[ht]
    \centering
    \includegraphics[width=\linewidth]{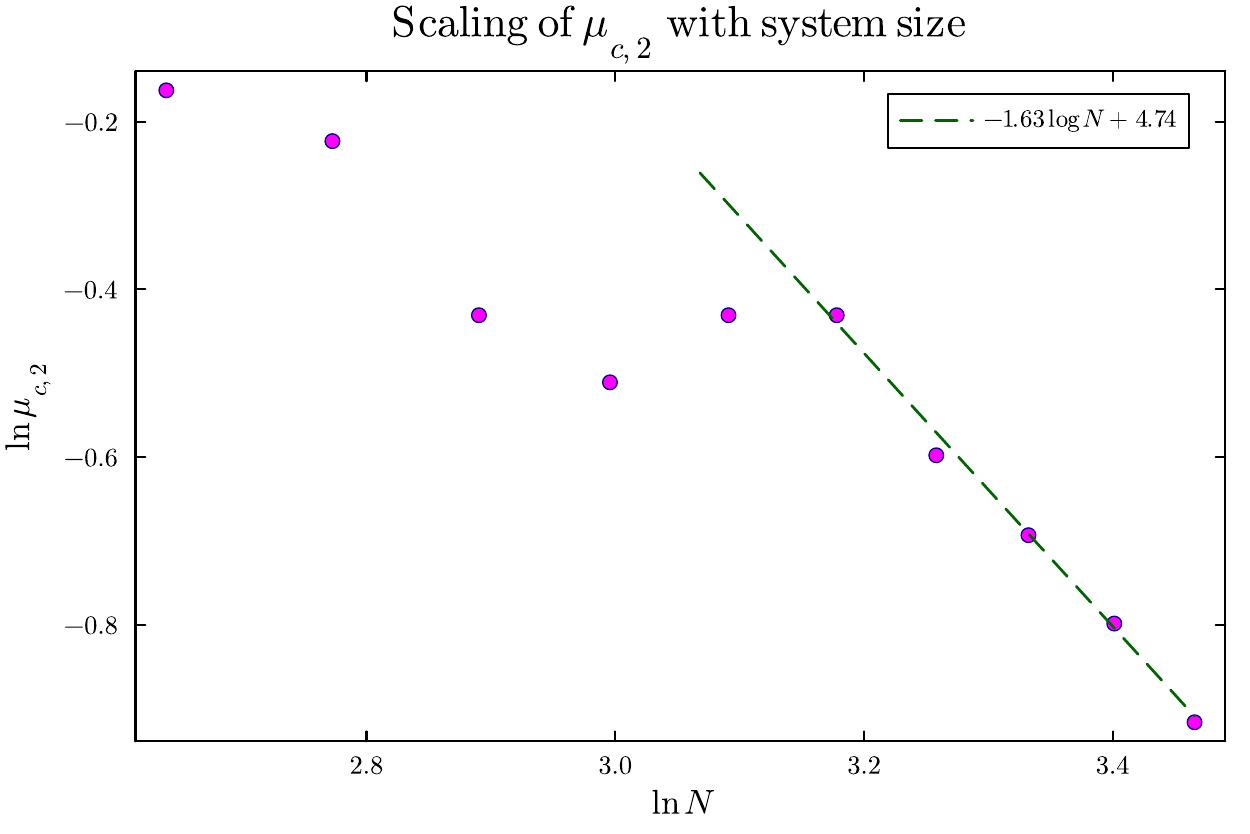}
    \caption{
        Scaling of the critical $\mu$ value (denoted by $\mu_{c,2}$) at which the Thouless time goes to $0$.
        In other words, this indicates the disappearance of the linear ramp.
    }
    \label{fig:mucrit-SFF-usual}
\end{figure}
From Fig.~\ref{fig:mucrit-SFF-usual}, we note that for large $N$ the critical value $\mu_{c,2}$ scales as a power-law in $N$ as $N^{\eta_{2}}$ where $\eta_{2} \approx -1.63$. 
\begin{align}
   \mu_{c,2} \sim \#N^{-\eta_2},\;\; \eta_{2} \sim 1.63\,.
   \label{long-range-muc}
\end{align}

From Eqs.~\eqref{short-range-muc} and~\eqref{long-range-muc} we observe that there are $2$ crossover strengths of $\mu$ at which deviations from the RMT behavior emerge, corresponding to short- and long-range spectral statistics.
These strengths are characterized by two exponents $\eta_1$~\eqref{short-range-muc} and $\eta_2$~\eqref{long-range-muc}, which according the numerical observations satisfy the relation:~$\eta_2 \gtrsim \eta_1$.
The physical implication of this relation is that the value of $\mu$ at which RMT behaviour is destroyed is smaller for long-range statistics than for short-range statistics.
In other words, the long-range RMT behavior persists for stronger (smaller \(\mu\)) L\'evy disorder as compared to short-range statistics.
This is in qualitative agreement with the sparse-SYK behavior~\cite{garcia2021sparse,Orman:2024mpw}.

\section{Conclusions}

In this work, we considered the properties of the $4$-body SYK model with $N$ Majorana fermions, where couplings are sampled from the fat-tailed L\'evy stable distribution.
We semi-analytically argue that the many-body spectral properties of the model are qualitatively different from the single-particle level statistics of the underlying L\'evy random matrix.
We employ a multifractral characterization of L\'evy random numbers together with a perturbative analysis of the SYK eigenvalues.
By this analysis, the spectrum is expected to cross over from chaotic to integrable upon decreasing the stability index $\mu$.
This is controlled by two effects:
(1) emergence of an $O(1)$ number interaction terms that are parametrically larger than the rest and
(2) a competition between the number of random interaction $\mathcal{N}$ and the Hilbert space dimension $\mathcal{D}$.
The first effect leads to a splitting of the spectrum into $O(1)$ number of sectors (called ``bubbles''), each with an $O(\mathcal{D})$ number of eigenvalues.
The second effect leads to the level-spacing within a sector to decrease exponentially with $N$, rather than polynomially (i.e. how the interaction scales).
This competition predicts a crossover instead of a transition (as compared to L\'evy random matrices Ref~\onlinecite{tarquini2016level}).
This is numerically confirmed via both short- and long-range spectral correlations, where the model is shown to deviate from RMT at a crossover value of the L\'evy stability index $\mu_c$, which scales inversely with system size $N$\footnote{Numerical codes used in this article are available in~\href{https://github.com/Budhadityab/Levy-SYK}{https://github.com/Budhadityab/Levy-SYK}}.

The LSYK model is also expected to be solvable (in the usual SYK sense), since thermodynamic quantities have been computed for L\'evy spin glasses~\cite{janzen2010thermodynamics} and similar techniques may be used for solving L\'evy SYK.
It would be interesting to construct a similar model with a \emph{many-body transition} instead of a \emph{crossover}. 
This is expected if the number of interaction terms scales similarly to the Hilbert space dimension (see Supplementary).
This might be possible if $q$ is chosen to scale with $N$ (i.e. a \emph{double-scaled L\'evy SYK}).

\section{Acknowledgement}

BB and AA acknowledge extensive discussions with Ivan Khaymovich on this and related projects, particularly regarding the multi-fractal spectrum of L\'evy random variables.
BB is grateful to Adolfo del Campo, Barbara Dietz, Sergej Flach, Dung Ngyuen, Elizabeta Safonova, Alessandro Romito, Marco Tarzia, and Masaki Tezuka for discussions and comments.
BB and AA are supported by the Institute for Basic Science under the grant IBS-R024-D1.
DR thanks FAPESP for the ICTP-SAIFR grant 2021/14335-0 and for the Young Investigator grant 2023/11832-9.
DR also acknowledges the Simons Foundation for the Targeted Grant to ICTP-SAIFR.

\onecolumngrid
\appendix

\section{Fractal Spectrum Distribution of L\'evy Disorder}\label{app:AppendixA}

Probability distributions such as the L\'evy stable distribution pose various technical challenges, particularly when computing moments of complicated combinations of corresponding random variables.
In such cases, it is often useful to focus only on the hierarchy of a random variable sampled from the distribution.
The advantage is that one can work solely with the fraction of random variables having a given magnitude.
This fraction can be interpreted as the probability of finding random variables of the corresponding magnitude when sampling from the underlying distribution.
In other words, say we sample $N$ random variables and are interested in the probability of finding a random variable of magnitude $N^{-\alpha}$ (normalized such that $\alpha = 0$ corresponds to the largest variable).
Let us denote the probability distribution as $p_X (x)$.
Then we can write
\begin{align}
    p_X (x) \mathrm{d}x \xrightarrow[x=N^{-\alpha}]{} p_{X}(N^{-\alpha})\ln (N) N^{-\alpha}\mathrm{d}\alpha \equiv \ln (N) N^{f_X (\alpha,N) - 1}\mathrm{d}\alpha\,.
\end{align}
In the thermodynamic limit $N\rightarrow \infty$, we expect the function $f_X$ to depend only on $\alpha$.
This allows us to define $f_X (\alpha) = \lim_{N\rightarrow\infty}f_X(\alpha,N)$.
If one can determine the function $f_X(\alpha)$, called the \emph{spectrum of fractal dimensions} (SFD), then it becomes possible to compute the corresponding SFDs of sums, products and other combinations of the random variable $X$.
These methods were developed extensively in Ref.~\onlinecite{kutlin2024anatomy}.
To use these for the analysis of the eigenvalues of LSYK, it is first necessary to discuss the SFD of the L\'evy distribution.

In this section, we shall derive the hierarchy of the magnitude of the random interactions $J_{i j k l}$, following the approach of~\cite{monthus2016localization}.
Let us denote the indices in the random variable $\vert J_{i_1 i_2 \dots i_q} \vert$ with the collective index $\vert J_{I} \vert$.
We sample a total of $\mathcal{N} = \binom{N}{q}$ such random variables for each realization of the Hamiltonian. Since L\'evy distributions are not known in analytical closed forms, we have to consider only the asymptotic form of the distribution.
This is given by~\cite{burda2007free}
\begin{align}
    \mathtt{P}_{\mu,\sigma}(x) \equiv \lim_{x\rightarrow \infty}\mathtt{L}_{\mu,\sigma}(x) = \frac{\gamma(\mu)\sigma}{\vert x \vert^{\mu + 1}} \;,\;\;\; \forall\;0 < \mu < 2\,.
    \label{eq:pmusigma}
\end{align}
where $\gamma(\mu) = \Gamma(1 + \mu)\sin(\pi\mu/2)$. In this notation $\sigma$ gives the ``typical value'' which we set to $\sigma = J^{\mu}\frac{(q-1)!}{N^{q-1}}$.
Arranging the $\mathcal{N}$ random variables $\vert J_n \vert$ in a descending order $\vert J_1 \vert > \vert J_2 \vert >\cdots>\vert J_n\vert > \cdots >\vert J_{\mathcal{N}}\vert$, we can argue that 
\begin{align}
\label{eq:J_n_def}
    \int_{\vert J_n \vert}^{\infty}\mathtt{P}_{\mu,\sigma}(x)\mathrm{d}x \propto \frac{n}{\mathcal{N}}\,,
\end{align}
for some $n \in \{1,\dots,\mathcal{N}\}$. Inserting Eq.~\eqref{eq:pmusigma} in Eq.~\eqref{eq:J_n_def}, we obtain the following relation
\begin{align}
    \frac{\gamma(\mu)\sigma}{\mu \vert J_n \vert^\mu} \propto \frac{n}{\mathcal{N}} \, .
\end{align}
Thus, the typical size of $\vert J_n \vert$ will be given by
\begin{align}
    \vert J_n \vert \propto \left( \frac{\gamma(\mu)\sigma \mathcal{N}}{n} \right)^{1/\mu} \, .
\end{align}
Using $\sigma \sim\frac{1}{\mathcal{N}_1}$, where $\mathcal{N}_1 = \binom{N}{q-1}$, we find that $\vert J_n \vert \propto \left( \frac{\gamma(\mu)\mathcal{N}}{n \mathcal{N}_1} \right)^{1/\mu}$.
Finally, performing the replacement $n = \mathcal{N}^{x}\,\; 0 \leq x \leq 1$, we find that the size of a random variable may be represented as 
\begin{align}
    \vert J_n \vert \propto \gamma(\mu)^{1/\mu}\left( \frac{\mathcal{N}^{1-x}}{\mathcal{N}_1} \right)^{1/\mu} \, .
\end{align}
It is convenient to further parameterize the magnitude of these terms as $\mathcal{N}^{-1/\mu}_1 \mathcal{N}^{\alpha}$ where $0 \leq \alpha \leq 1/\mu$.
We count the total number of such terms. 
\begin{align}
    \#(\vert J_n \vert \propto \mathcal{N}^{-1/\mu}_1 \mathcal{N}^\alpha) \propto \int_{0}^{1}\mathrm{d}x \mathcal{N}^{x}\delta\left(\alpha - \frac{1-x}{\mu}\right) = \mathcal{N}^{1-\alpha\mu} \, .
\end{align}
\noindent Thus, we arrive at the following fact:
\begin{center}
    There are $O(\mathcal{N}^{1-\alpha\mu})$ number of terms of strength $O(\mathcal{N}^{-1/\mu}_1 \mathcal{N}^{\alpha})$ for $0 \leq  \alpha \leq \frac{1}{\mu}$. 
\end{center}
In the main text, this result is presented as the following theorem (provided the identifications $z = \mu\alpha$ and $\mathcal{N} \sim N\mathcal{N}_1$):
\begin{theorem}[L\'evy Hierarchy]
    \label{levyhierarchy-spp}
    Given a set of $\mathcal{N} = \binom{N}{q}$ L\'evy random variables (for $\mu < 2$)\, it contains $O(\mathcal{N}^{1-z})$ number of random variables with magnitude $O(N^{1/\mu}\mathcal{N}^{\frac{z-1}{\mu}})$, where $0 \leq z \leq 1$. 
\end{theorem}

Let us now verify the validity of the theorem by counting the number of terms corresponding to each scale ($X_{typ}, X_{0}, X_{large}$ as discussed previously).
\begin{itemize}
    \item The terms of magnitude $O(\mathcal{N}_1^{-1/\mu})$ correspond to $\alpha = 0$.
    There are $O(\mathcal{N})$ such numbers.
    This is the weak (but abundant) background, which is chaotic but parametrically suppressed.
    \item The terms of  order $O(1)$ correspond to $\alpha = \frac{q-1}{q\mu}$ (using the fact that $\mathcal{N}\sim N \mathcal{N}_1$).
    There are $O(N)$ of such numbers in each ensemble realization, and they form a sparse hypergraph in the space of couplings, very much similar to the sparse hypergraphs entering the sparse versions of the SYK model. 
    \item The terms of order $O(N^{1/\mu})$ correspond to setting $\alpha = 1/\mu$.
    There are $O(1)$ such numbers, and we denote their strength as $\vert J_L \vert$.
    These are the parametrically large outliers. 
    \item The total number of terms making up the background $+$ hypergraph can be computed by the integral $\int_{0}^{\frac{q-1}{q\mu}}\mathcal{N}^{1-\alpha\mu}\mathrm{d}\alpha = \frac{\mathcal{N} - \mathcal{N}^{1/q}}{\mu\ln\mathcal{N}}$.
    In terms of $N$, this is roughly $O(\frac{N^q - N}{\mu q\ln N})$. 
\end{itemize}
Using this, we can compute the total strength (i.e the sum) of the interaction terms
\begin{align}
    \vert J_{T} \vert \sim \int_{0}^{1/\mu} \mathcal{N}^{1-\alpha\mu}\mathcal{N}^{-1/\mu}_1\mathcal{N}^\alpha \mathrm{d}\alpha = \mathcal{N}^{-1/\mu}_1 \mathcal{N} \frac{1 - \mathcal{N}^{\frac{1}{\mu}-1}}{(\mu - 1)\ln\mathcal{N}} \, .
\end{align}

Given the results above, we can get an estimate of the value of $\mu$ at which the large outliers, although rare, start to become dominant with respect to the weak background, by simply comparing $\vert J_T \vert$ with $\vert J_L \vert$.
By equating the two terms we obtain the cnodition
\begin{align}
    &\mathcal{N}^{-1/\mu}_1 \mathcal{N}^{1/\mu} \sim \mathcal{N}^{-1/\mu}_1 \mathcal{N} \frac{1 - \mathcal{N}^{\frac{1}{\mu}-1}}{(\mu - 1)\ln\mathcal{N}}\notag\\
    \implies &\mathcal{N}^{\frac{1}{\mu}-1} \sim \frac{1 - \mathcal{N}^{\frac{1}{\mu}-1}}{(\mu - 1)\ln\mathcal{N}}\, ,
\end{align}
which is naturally satisfied for $\mu \sim 1^{-}$.
Furthermore, in the limit of large$-N$, the ratio $\vert J_L \vert/\vert J_T \vert$ scales as $\frac{(\mu-1)q\ln N}{N^{\frac{(1-\mu)q}{\mu}}}$ for $\mu \gtrsim 1$ but instead scales as $q \ln N$ for $\mu \lesssim 1$.
Strong ``L\'evy-ness'' in the interaction~\footnote{We may ignore the overall factor of $\gamma(\mu)^{1/\mu}$ since it is common to all terms.} therefore arises for $\mu$ less than $1$. 

However, that itself is unlikely to be enough to cause a deviation from RMT for eigenvalue \emph{correlations}, since an analysis of this kind completely ignores the many-body correlations arising when mapping the $N$ random variables into the many-body Hilbert space.
In the following section, we present a multifractal analysis of the eigenvalue SFD themselves by using a perturbation theory argument.
For this, it is convenient to normalize the spectrum of random variables by the largest one, so that we have
\begin{center}
    $O(\mathcal{N}^{\alpha\mu})$ number of terms having strength $O(\mathcal{N}^{-\alpha})$ for $0 \leq  \alpha \leq \frac{1}{\mu}$. 
\end{center}
This constitutes the fundamental SFD of the L\'evy distribution.
We now turn to the problem of relating the SFD of the L\'evy distribution to the SFD of the eigenvalues of the LSYK model.


\section{The Many-Body Picture}\label{app:AppendixB}

Consider the LSYK Hamiltonian
\begin{align}
    H = \sum_{I}J_{I}\Psi_{I}\;\;\;,\;\Psi_I = \mpsi_{i_1}\mpsi_{i_2}\dots\mpsi_{i_q},\, J_{I} = J_{i_1 i_2 \dots i_q}\, ,
\end{align}
and let us assume to be in the strong L\'evy regime, with one of the $J_I$'s being parametrically larger than the rest.
We rescale the Hamiltonian by this value to ensure that the largest interaction is of order $1$.
Without loss of generality, let us assume that this term corresponds to $I = 1 2 3 \dots q$.
Therefore, the Hamiltonian can be written as 
\begin{align}
    H = \mpsi_{1}\mpsi_2 \mpsi_3 \cdots\mpsi_{q} + \sum_{I\neq 1 2 3\dots q}j_{I}\Psi_{I}
\end{align}
where all $j_{I}$ are smaller than $1$ and distributed according to the L\'evy distribution.
To approximate the eigenvalues perturbatively, it is useful to write the Hamiltonian in the eigenbasis of $\Psi_1 = \mpsi_{1}\mpsi_2 \mpsi_3 \cdots\mpsi_{q}$.
The fermionic states can be represented by first considering the following Majorana$\rightarrow$ Dirac fermion identification
\begin{align}
    \mpsi_{2i} = \frac{c_i + c^\dagger_i}{\sqrt{2}}\,\,\text{and}\,\,\mpsi_{2i-1} = -i\frac{c^\dagger_i - c_i}{\sqrt{2}}\;\; \forall \; i\in \{1,\dots,\frac{N}{2}\}\,.
\end{align}
The action of these operators on a fermionic state $\ket{s}$, with $s = 0,1$ (and corresponding $\bar{s} = 1,0$), is given as
\begin{align}
    \mpsi_{2 i} \ket{s_{i}} = \frac{1}{\sqrt{2}}\ket{\bar{s}_{i}}\;\;\text{and}\;\;\mpsi_{2i-1} \ket{s_{i}} = \frac{i^{2 s_i + 1}}{\sqrt{2}}\ket{\bar{s}_i}\,,\;\; 
\end{align}
which holds $\forall \; i\in \{1,\dots,\frac{N}{2}\}\,$.
The eigenbasis of $\Psi_1$ can then be identified as 
\begin{align}
    \ket{\mathbf{s}}_{1} = \ket{s_1s_2\dots s_{\frac{q}{2}} s_{\frac{q}{2} + 1}\dots s_{\frac{N}{2}}}
\end{align}
with eigenvalues $\epsilon_s = (-1)^{\sum_{i = 1}^{q/2}s_i}$.
This is, however, special to this operator $\Psi_1$.
The remaining terms may have a different structure for their eigenstates, as they may not contain both $\mpsi_{2 i}$ and $\mpsi_{2 i - 1}$ for each $i$.
In these cases, the eigenstate would contain a sum of $2$ such $\ket{\mathbf{s}}_1$ (for different $\mathbf{s}$ values), with appropriate phases.
In this representation, it is also easy to see that the action of an odd number of fermions belonging to $\Psi_1$ on $\ket{\mathbf{s}}_1$ flips the sign of the eigenvalue, and an even number of fermions conserves the sign.

The sub-leading in the Hamiltonian, that comprise $H - \Psi_1$, either commute with $\Psi_1$ (if they have an even number of fermions common with $\Psi_1$) or anti-commute with $\Psi_1$ (if they have an odd number of fermions common with $\Psi_1$).
Let us denote the commuting terms by $\Psi_{A}$ and the anti-commuting terms by $\Psi_{B} $. Then we have
\begin{itemize}
    \item $[\Psi_1,\Psi_A] = 0$. This implies $\Psi_A \ket{\mathbf{s},\pm} = \pm\ket{\mathbf{s'},\pm}$.
    In other words, the commuting terms only contribute to the $+$ and $-$ sectors separately.
    Thus, these only appear in the diagonal blocks ($H_{++}$ and $H_{--}$). 
    \item $\{\Psi_1,\Psi_B\} = 0$.
    This implies $\Psi_B \ket{\mathbf{s},\pm} = \mp \ket{\mathbf{s}',\mp}$.
    In other words, the anti-commuting terms flip the sign from $+$ to $-$ and vice versa.
    Thus, these only appear in the off-diagonal blocks ($H_{+-}$ and $H_{-+}$).
\end{itemize}
Here $\mathbf{s}'$ stands for some configuration that is not necessarily the same as $\mathbf{s}$.
Indeed, $\mathbf{s'}$ can also stand for a linear combination of states with different configurations as long as they correspond to the same eigenvalue of $\Psi_1$. 

Let us now consider the diagonal of the Hamiltonian in this basis.
This corresponds to $\pm 1$ (from $\Psi_1$) and some extra contributions coming from terms that are also diagonalized simultaneously with $\Psi_1$.
A quick numerical check tells us that there are roughly order $N^{r}$ such terms, where $r \gtrsim 1$.
The implication is that the diagonal part can be written as 
\begin{align}
    H_{ii} = \pm 1 + \sum_{l = 1}^{N^r}\alpha^{l}_{i}j_{l}\,.
    \label{eq:e0}
\end{align}
where $\alpha^{l}_{i}$ is some arbitrary phase arising from the orthogonal transformation.
To treat the Hamiltonian eigenvalues in a perturbative manner, we can first consider the diagonal contribution $H_{ii}$ as the $0^\text{th}$ order term (let us denote that by $E^{(0)}_{i}$).
The eigenvectors at $0^\text{th}$ order are simply the eigenvectors of $\Psi_1$.
Let us denote these by $\ket{\psi^{(0)}_i}$. 
Denoting with $H'$ the full LSYK Hamiltonian minus its diagonal part, the first-order contribution to the energy reads $E^{(1)}_{i} = \bra{\psi^{(0)}_i}H'\ket{\psi^{(0)}_i} = 0$.
The first non-vanishing contribution comes therefore at the second order and it reads $E^{(2)}_{i} = \sum_{j\neq i}\frac{\Big\vert\bra{\psi^{(0)}_i}H'\ket{\psi^{(0)}_j}\Big\vert^2}{E^{(0)}_{i} - E^{(0)}_{j}}$. 

The numerator of the $2^\text{nd}$ order term will first have the following contribution 
\begin{align}
    \Big\vert\bra{\psi^{(0),+}_i}H'\ket{\psi^{(0),+}_j}\Big\vert^2 &= \sum_{I}\vert j_I \vert^2 \Big\vert\bra{\psi^{(0),+}_i}\Psi_I \ket{\psi^{(0),+}_j}\Big\vert^2\notag\\
    &= \sum_{I \in A}\vert j_I \vert^2 \Big\vert\bra{\psi^{(0),+}_i}\Psi_I \ket{\psi^{(0),+}_j}\Big\vert^2 \notag\\
    &= \vert j_{I'} \vert^2 \;\;\;,\;\text{where}\;\ \Psi_{I'}\ket{\psi^{(0),+}_j} \propto \ket{\psi^{(0),+}_i}\,.
    \label{eq:ctr1}
\end{align}
That is, only a \textit{single} commuting term contributes.
The other possible contribution comes from the following case
\begin{align}
    \Big\vert\bra{\psi^{(0),+}_i}H'\ket{\psi^{(0),-}_j}\Big\vert^2 &= \sum_{I}\vert j_I \vert^2 \Big\vert\bra{\psi^{(0),+}_i}\Psi_I \ket{\psi^{(0),-}_j}\Big\vert^2\notag\\
    &= \sum_{I \in B}\vert j_I \vert^2 \Big\vert\bra{\psi^{(0),+}_i}\Psi_I \ket{\psi^{(0),-}_j}\Big\vert^2 \notag\\
    &= \vert j_{I'} \vert^2 \;\;\;,\;\text{where}\;\ \Psi_{I'}\ket{\psi^{(0),-}_j} \propto \ket{\psi^{(0),+}_i}\,.
    \label{eq:ctr2}
\end{align}
A term like this contributes across sectors, and thus only participates through anti-commuting terms in the Hamiltonian.
Therefore, in the expression for $E_i^{(2)}$, each value of $j$ will contribute with a single term, either via Eq.~\eqref{eq:ctr1} or via Eq.~\eqref{eq:ctr2}.
The corresponding denominator is then either $+2$ (if $i$ and $j$ belong to opposite sectors) or a term of order $\sum_{l = 1}^{N^r}\alpha^{l}_i j_l$ (if $i$ and $j$ belong to the same sector).
Let us call the denominators arising from terms in the same sectors as $e_{k}$.
The full second-order contribution is then given by
\begin{align}
    E_{i}^{(2)} = \sum_{k = 1}^{\mathcal{N}^{a}} \frac{\vert j_{k} \vert^2}{e_{k}} + \sum_{k' = 1}^{\mathcal{N}^b} \frac{\vert j_{k'} \vert^2}{2}\,.
    \label{eq:e2}
\end{align}
where the sum over $k$ is within the same sector and the sum over $k'$ is across sectors.
At this stage, it is not necessary to pinpoint the exact values of $a$ and $b$.
It is sufficient to know that $0 < a,b < 1$.
The exact values can be estimated at a later moment, if necessary. 

Summarizing, we have obtained a perturbative expression for the many-body energy levels.
Remarkably, the energy levels are \textit{entirely} determined in terms of the L\'evy couplings $j_l$.
The goal now is to use Eq.~\eqref{eq:e0} and Eq.~\eqref{eq:e2} -- together with the distributional properties of the $j_l$ couplings -- to estimate the spacings between energy levels $\Delta E_{ij} = E_i - E_j$.

\subsection{Multifractal spectrum of L\'evy Random Variables}
\label{app:AppendixB1}

As already emphasized when recalling the L\'{e}vy hierarchy theorem, a L\'evy spectrum~\cite{kutlin2024anatomy,monthus2016localization} can be characterized in terms of the strength of the random variable $\vert X \vert$ and the number of random variables of a given strength from $\mathcal{N}$ samples, denoted by $\mathtt{n}(X)$.
More quantitatively, the L\'evy hierarchy theorem translates into the following SFD~\cite{monthus2016localization,kutlin2024anatomy}
\begin{align}
    f_L(\alpha) = \mu\alpha\theta(1-\mu\alpha) + \left(\frac{1}{\mu} + 1 - \alpha\right) \theta(\mu\alpha-1) \, ,
\end{align}
which, graphically, is represented as follows
\begin{center}
\begin{tikzpicture}
    \def\a{5}; 
    \draw[->] (0+\a,0) -- (4+\a,0);
    \draw[->] (0+\a,0) -- (0+\a,3);
    \draw[-,dashed,gray,line width=0.5mm] (0+\a,2) -- (4+\a,2);
    \draw node[rotate=90] at (-0.5+\a,3.0) {$f_L(\alpha)$}; 
    \draw node at (4.0+\a,-0.5) {$\alpha$};
    \draw node at (-0.25+\a,2.0) {$1$};
    \draw node at (2+\a,-0.3) {$\frac{1}{\mu}$};
    \draw[-,dotted,brown,line width=0.5mm] (2+\a,2) -- (2+\a,0);
    \draw[-,blue,line width=0.8mm] (0+\a,0) -- (2+\a,2);
    \draw[-,blue,line width=0.8mm] (2+\a,2) -- (3+\a,0);

    \draw node at (2+\a,3) {$X \sim \vert j_I \vert$};
    
    \draw node[rotate = 45] at (0.8+\a,1.2) {$\mu\alpha$};
    \draw node[rotate = -63.43] at (2.8+\a,1.0) {${1-\alpha + 1/\mu}$};
\end{tikzpicture}
\end{center}
Using the fusion rules of~\cite{kutlin2024anatomy}, we can first compute the distribution of $Z = X+Y$.
The probability distribution for this satisfies
\begin{align}
    p_{X + Y}(\vert Z \vert = N^{-\alpha}) \propto
    \begin{cases}
        &N^{\max\{f_X(\alpha),f_Y(\alpha)\} +\alpha - 1}\;\;\;\;\;\;\;\;\;\;\;\;\;\;\;\;\;\;\;\;\;\;\;\;\;\;\;\;\;\;\,\alpha < \alpha_0 (X), \alpha_0 (Y)\\
        &N^{\max_{\xi \leq \alpha}\{f_X (\xi) + f_Y (\xi) + \xi\}-2} + N^{f_X(\alpha) + \alpha - 1}\;\;\alpha_0 (X) < \alpha_0 < \alpha_0 (Y)\\
        &N^{\max_{\xi \leq \alpha}\{f_X (\xi) + f_Y (\xi) + \xi\}-2}\;\;\;\;\;\;\;\;\;\;\;\;\;\;\;\;\;\;\;\;\;\;\;\;\;\,\alpha > \alpha_0 (X), \alpha_0 (Y)
    \end{cases}
\end{align}
where $f_X$ and $f_Y$ are the SFD of $X$ and $Y$, respectively.
The typical values are denoted by $\alpha_0 (X)$ and $\alpha_0 (Y)$, and without loosing generality we have assumed $\alpha_0 (X) \leq \alpha_0 (Y)$.
When the two random variables come from the same spectrum, the expression simplifies.
Using the expression for $f_L(\alpha)$, we find 
\begin{align}
    p_{X + Y}(\vert Z \vert = N^{-\alpha}) \propto
    \begin{cases}
        &N^{f_L(\alpha) +\alpha - 1}\;\;\;\;\;\;\;\;\;\;\;\;\;\;\;\;\;\;\;\;\;\;\alpha <\frac{1}{\mu}\\
        &2 N^{1/\mu}\;\;\;\;\;\;\;\;\;\;\;\;\;\;\;\;\;\;\;\;\;\;\;\;\;\;\;\;\;\;\;\alpha = 1/\mu\\
        &N^{\max_{\xi \leq \alpha}\{2 f_L (\xi) + \xi\}-2}\;\;\;\;\;\;\;\;\,\alpha > \frac{1}{\mu}
    \end{cases}
\end{align}
It just remains to evaluate the saddle point $\max_{\xi \leq \alpha}\{2 f_L (\xi) + \xi\}$ for $\alpha > \frac{1}{\mu}$.
If we choose $\xi < \frac{1}{\mu}$, then the maximum possible contribution naturally comes from $\xi = \frac{1}{\mu}$, since both $f_L$ and $\xi$ are monotonically increasing functions.
However, the reverse is true for $\xi > \frac{1}{\mu}$, which leads to a competition between $f_L(\xi)$ and $\xi$. 
Therefore we must maximize the expression $2 f_L(\xi) + \xi = -\xi + 2 + \frac{2}{\mu}$.
Naturally, this is maximized for the smallest value of $\xi$ available in this range, which is $\xi = \frac{1}{\mu}$.
Therefore we find that $\max_{\xi \leq \alpha}\{2 f_L (\xi) + \xi\} = 2 f_{L}(1/\mu) + \frac{1}{\mu} = 2 + \frac{1}{\mu}$.
Hence the final result of the combination $Z = X + Y$ is
\begin{align}
    p_{X + Y}(\vert Z \vert = N^{-\alpha}) \propto
    \begin{cases}
        &N^{f_L(\alpha) +\alpha - 1}\;\;\;\;\;\;\;\;\alpha <\frac{1}{\mu}\\
        &2 N^{\frac{1}{\mu}}\;\;\;\;\;\;\;\;\;\;\;\;\;\;\;\;\;\;\;\alpha = 1/\mu\\
        &N^{\frac{1}{\mu}}\;\;\;\;\;\;\;\;\;\;\;\;\;\;\;\;\;\;\;\;\;\alpha > \frac{1}{\mu}
    \end{cases}
\end{align}
Using the relation $p(N^{-\alpha})N^{-\alpha} \equiv N^{f - 1}$, we find that the SFD of $Z = X + Y$ is
\begin{align}
    f_{X + Y}(\alpha) = \mu\alpha \theta(1 - \mu\alpha) + \left(\frac{1}{\mu} + 1 -\alpha\right)\theta(\mu\alpha - 1)
\end{align}
which is exactly the same as $f_L(\alpha)$. 

The next step is to study the square of a random variable with the SFD $f_L(\alpha)$.
This is easily done, since $f_{X^n}(\alpha) = f_X (\alpha/n)$.
Thus the SFD of $\vert j_L \vert^2$ is given by
\begin{align}
    f_{L^2}(\alpha) = \frac{\mu\alpha}{2}\theta(2-\mu\alpha) + \left(\frac{1}{\mu} + 1 - \frac{\alpha}{2}\right)\theta(\mu\alpha - 2)
\end{align}
This can be presented in a similar way as before:
\begin{center}
\begin{tikzpicture}
    \def\a{5}; 
    \draw[->] (0+\a,0) -- (4+\a,0);
    \draw[->] (0+\a,0) -- (0+\a,3);
    \draw[-,dashed,gray,line width=0.5mm] (0+\a,2) -- (4+\a,2);
    \draw node[rotate=90] at (-0.5+\a,3.0) {$f_{L^2}(\alpha)$}; 
    \draw node at (4.0+\a,-0.5) {$\alpha$};
    \draw node at (-0.25+\a,2.0) {$1$};
    \draw node at (2+\a,-0.3) {$\frac{2}{\mu}$};
    \draw[-,dotted,brown,line width=0.5mm] (2+\a,2) -- (2+\a,0);
    \draw[-,red,line width=0.8mm] (0+\a,0) -- (2+\a,2);
    \draw[-,red,line width=0.8mm] (2+\a,2) -- (3+\a,0);

    \draw node at (2+\a,3) {$X \sim \vert j_I \vert^2$};

    \draw node[rotate = 45] at (0.8+\a,1.2) {$\mu\alpha/2$};
    \draw node[rotate = -63.43] at (2.8+\a,1.0) {\scriptsize$1 -\alpha/2 + 1/\mu$};
\end{tikzpicture}
\end{center}
Next, we need to consider the ratio of two L\'evy random numbers.
This is just a minor modification of the rule for the product of random variables with some SFD, and can be written as
\begin{align}
    N^{f_{X/Y}-1} \propto \int\mathrm{d}\xi \mathrm{d}\eta \delta(\alpha - \xi + \eta)N^{f_X (\xi) - 1} N^{f_Y (\eta) - 1} = \int \mathrm{d}\eta N^{f_X(\alpha + \eta) + f_Y (\eta) - 2}
\end{align}
which, by a saddle point approximation, gives us $f_{X/Y}(\alpha) = \max_\eta \{f_X(\alpha + \eta) + f_Y (\eta) - 1\}$.
We particularly focus on the case when $f_X = f_{L^2}$ and $f_{Y} = f_{L}$, which corresponds to $X \sim \vert j_L \vert^2$ and $Y \sim \vert j_L \vert$.
The objective is to evaluate this SFD $f_{X/Y}(\alpha) = \max_{\eta}g(\alpha,\eta)$ where 
\begin{align}
    g(\alpha,\eta) = \frac{\mu(\alpha + \eta)}{2}\theta(2 - \mu\alpha - \mu\eta) + \left(\frac{1}{\mu} + 1 - \frac{\alpha+\eta}{2}\right)\theta(\mu\alpha + \mu\eta - 2) + \mu\eta \theta(1 - \mu\eta) + \left(\frac{1}{\mu} + 1 - \eta\right)\theta(\mu\eta - 1) - 1
\end{align}
In the range of $\eta$, there are effectively $3$ regions that we must investigate.
These are $\mathbf{I}: [0,\min\{\frac{1}{\mu},\frac{2}{\mu}-\alpha\}]$, then $\mathbf{II} : [\min\{\frac{1}{\mu},\frac{2}{\mu}-\alpha\},\max\{\frac{1}{\mu},\frac{2}{\mu}-\alpha\}]$ and $\mathbf{III}: [\max\{\frac{1}{\mu},\frac{2}{\mu}-\alpha\},\infty]$.
Note that $\alpha < \frac{1}{\mu}$ corresponds to $\min\{\frac{1}{\mu},\frac{2}{\mu} - \alpha\} = \frac{1}{\mu}$ and vice-versa.
Let us first being with $\alpha < \frac{1}{\mu}$.
Then we have $g(\alpha,\eta)$ in the three regions as
\begin{align}
    g_{\mathbf{I}}(\alpha,\eta) &= \frac{\mu\alpha}{2} + \frac{3\mu\eta}{2}-1 \implies \eta_{\mathbf{I},\max} = \frac{1}{\mu} \\
    g_{\mathbf{II}}(\alpha,\eta) &= \frac{\mu\alpha}{2} + \frac{1}{\mu} -\eta\left(1 - \frac{\mu}{2}\right) \implies \eta_{\mathbf{II},\max} = \frac{1}{\mu}\\
    g_{\mathbf{III}}(\alpha,\eta) &= \frac{2}{\mu} + 1 -\frac{\alpha}{2} - \frac{3\eta}{2} \implies \eta_{\mathbf{III},\max} = \frac{2}{\mu} - \alpha
\end{align}
Finally, we see that $g_{\mathbf{I},\mathbf{II}}(\alpha,\frac{1}{\mu}) = \frac{\mu\alpha + 1}{2}$ and $g_\mathbf{III}(\alpha,\frac{2}{\mu} - \alpha) = -\frac{1}{\mu} + 1 + \alpha$.
Note that $2 g_{\mathbf{I}}(\alpha,\frac{1}{\mu}) - 2 g_{\mathbf{III}}(\alpha,\frac{2}{\mu} - \alpha) =-1 -(\mu-2)\alpha + \frac{2}{\mu} > 0$ for $\alpha < \frac{1}{\mu}$.
Thus, the maximum value is obtained for $\eta_{\max} = \frac{1}{\mu}$.
Thus $g_{\mathbf{I},\mathbf{II}}(\alpha,\frac{1}{\mu}) = \frac{\mu\alpha + 1}{2}$ is the SFD for $\alpha < \frac{1}{\mu}$.
Now we have to consider the case of $\alpha > \frac{1}{\mu}$.
The three regions are now given by
\begin{align}
    g_{\mathbf{I}}(\alpha,\eta) &= \frac{\mu\alpha}{2} + \frac{3\mu\eta}{2}-1 \implies \eta_{\mathbf{I},\max} = \frac{2}{\mu} -\alpha \\
    g_{\mathbf{II}}(\alpha,\eta) &= \frac{1}{\mu} - \frac{\alpha}{2} + \eta\left(\mu - \frac{1}{2}\right) \implies \eta_{\mathbf{II},\max} = \frac{1}{\mu}\theta_{2\mu > 1} + \left(\frac{2}{\mu} - \alpha\right)\theta_{2\mu < 1}\\
    g_{\mathbf{III}}(\alpha,\eta) &= \frac{2}{\mu} + 1 -\frac{\alpha}{2} - \frac{3\eta}{2} \implies \eta_{\mathbf{III},\max} = \frac{1}{\mu} 
\end{align}
For $\mu < \frac{1}{2}$, we find that $g_{\mathbf{I},\mathbf{II}}(\alpha,\eta_{\max}) = 2 - \mu\alpha$ and $g_{\mathbf{III}}(\alpha,\eta_{\max}) = \frac{1}{2\mu} + 1 - \frac{\alpha}{2}$.
Computing the difference we find that $g_{\mathbf{I},\mathbf{II}}(\alpha,\eta_{\max}) - g_{\mathbf{III}}(\alpha,\eta_{\max}) = 1 - \frac{1}{2\mu} + \alpha\left(\frac{1}{2} - \mu\right)$ which is positive for $\alpha > \frac{1}{\mu}$.
For the opposite case of $\mu > \frac{1}{\mu}$, we find that $g_{\mathbf{I}}(\alpha,\eta_{\max}) = 2 - \mu\alpha$ and $g_{\mathbf{II},\mathbf{III}}(\alpha,\eta_{\max}) = \frac{1}{2\mu} + 1 - \frac{\alpha}{2}$.
Naturally, the same argument holds and we conclude that $g(\alpha,\eta_{\max}) = 2 - \mu\alpha$ is the SFD for $\alpha > \frac{1}{\mu}$. 
Summarizing, the full SFD for the ratio (which we denote by $f_R$) reads
\begin{align}
    f_R(\alpha) = \frac{\mu\alpha + 1}{2}\theta(1-\mu\alpha) + (2-\mu\alpha)\theta(\mu\alpha - 1) \, .
\end{align}
Again, this rule can be represented graphically as follows
\begin{center}
    \begin{tikzpicture}
        \def\a{5}; 
        \def\b{0}; 
        \draw[->] (0+\a,0-\b) -- (4+\a,0-\b);
        \draw[->] (0+\a,0-\b) -- (0+\a,3-\b);
        \draw[-,dashed,gray,line width=0.5mm] (0+\a,2-\b) -- (4+\a,2-\b);
        \draw node[rotate=90] at (-0.5+\a,3.0-\b) {$f_R(\alpha)$}; 
        \draw node at (4.0+\a,-0.5-\b) {$\alpha$};
        \draw node at (-0.25+\a,2.0-\b) {$1$};
        \draw node at (2+\a,-0.3-\b) {$\frac{1}{\mu}$};
        \draw[-,dotted,brown,line width=0.5mm] (2+\a,2-\b) -- (2+\a,0-\b);
        \draw[-,OliveGreen,line width=0.8mm] (0+\a,1-\b) -- (2+\a,2-\b);
        \draw[-,OliveGreen,line width=0.8mm] (2+\a,2.0-\b) -- (3.0+\a,0.0-\b);
        \draw node at (2+\a,3-\b) {$X \sim \vert j_I \vert^2/\vert j_{I'}\vert$};
        \draw node at (-0.25+\a,1) {$\frac{1}{2}$};
        \draw node[rotate = 26.57] at (0.85+\a,1.1-\b) {$\frac{\mu\alpha + 1}{2}$};
        \draw node[rotate = -63.43] at (3.0+\a,1.0-\b) {$2 -\mu\alpha$};
    \end{tikzpicture}
\end{center}
The last ingredient that we need is the SFD of the sum of squares of L\'evy numbers, given by $X \sim \sum_{I = 1}^{N^\beta}\vert j_I \vert^2$.
Computing this quantity is straightforward, as we simply need to shift the SFD of $\vert j_I \vert^2$ by $\beta$, giving us $f_{s} = f_{L^2} + \beta$ and then discarding everything except the first region with $f_{s} < 1$.
This gives us the following SFD, which we denote by $f_{S}$
\begin{align}
    f_{S}(\alpha) = \left(\frac{\mu\alpha}{2} + \beta\right)\tilde{\theta}\left(\frac{2 (1-\beta)}{\mu} - \alpha\right)
\end{align}
where $\tilde{\theta}$ gives $-\infty$ in the region where the argument is negative.
The graphical representation of this SFD is given below
\begin{center}
    \begin{tikzpicture}
        \def\a{5}; 
        \def\b{0.5}; 
        \draw[->] (0+\a,0) -- (4+\a,0);
        \draw[->] (0+\a,0) -- (0+\a,3);
        \draw[-,dashed,gray,line width=0.5mm] (0+\a,2) -- (4+\a,2);
        \draw node[rotate=90] at (-0.5+\a,3.0) {$f_{S}(\alpha)$}; 
        \draw node at (4.0+\a,-0.5) {$\alpha$};
        \draw node at (-0.25+\a,2.0) {$1$};
        \draw node at (1.5+\a,-0.3) {$\frac{2}{\mu}(1-\beta)$};
        \draw[-,dotted,brown,line width=0.5mm] (1.5+\a,2) -- (1.5+\a,0);
        \draw[-,BurntOrange,line width=0.8mm] (0+\a,0+\b) -- (1.5+\a,2);

        \draw node at (2+\a,3) {$X \sim \sum^{N^\beta}_{1}\vert j_I \vert^2$};
        \draw node at (-0.25+\a,0.5) {$\beta$};
        \draw node[rotate = 45] at (0.8+\a,0.9) {$\mu\alpha/2  + \beta$};
    \end{tikzpicture}
\end{center}

Let us recap the main results of this analysis.
Starting from the SFD of a L\'evy variable, we have obtained the SFD for the square of a L\'evy, the ratio between the square of a L\'evy and a L\'evy, and for the extensive sum of squares of L\'evy variables.
They read
\begin{align}
    f_{L}(\alpha) &= \mu\alpha\theta(1-\mu\alpha) + \left(\frac{1}{\mu} + 1 - \alpha\right)\theta(\mu\alpha - 1)\\
    f_{L^2}(\alpha) &= \frac{\mu\alpha}{2}\theta(2-\mu\alpha) + \left(\frac{1}{\mu} + 1 - \frac{\alpha}{2}\right)\theta(\mu\alpha - 2) \\
    f_R(\alpha) &= \frac{\mu\alpha + 1}{2}\theta(1-\mu\alpha) + (2-\mu\alpha)\theta(\mu\alpha - 1)\\
    f_{S}(\alpha) &= \left(\frac{\mu\alpha}{2} + \beta\right)\tilde{\theta}\left(\frac{2 (1-\beta)}{\mu} - \alpha\right)
\end{align}
The graphical representations depend on the value $\beta$ entering in the extensive sum as $N^\beta$: we have the case $\beta > 1/2$ and the case $\beta < 1/2$. 
For $\beta < 1/2$, we have $\frac{1}{\mu} < \frac{2}{\mu}(1-\beta) \frac{2}{\mu}$, while for $\beta > 1/2$ we have $\frac{2}{\mu}(1-\beta) < \frac{1}{\mu} < \frac{2}{\mu}$.
For brevity, we denote $b = 2(1-\beta)$.
In a graphical form, we have the following two cases
\begin{center}
\begin{tikzpicture}
    \def\a{5}; 
    \def\b{0.5}; 

    \draw[->] (0+\a,0) -- (5+\a,0);
    \draw[->] (0+\a,0) -- (0+\a,3);
    \draw[-,dashed,gray,line width=0.5mm] (0+\a,2) -- (5+\a,2);
    \draw node[rotate=90] at (-0.5+\a,3.0) {$f(\alpha)$}; 
    \draw node at (5.0+\a,-0.5) {$\alpha$};
    \draw node at (-0.25+\a,2.0) {$1$};
    \draw node at (-0.25+\a,1.0) {$\frac{1}{2}$};
    \draw node at (-0.25+\a,0.5) {$\beta$};

    \draw node at (2+\a,-0.3) {$\frac{1}{\mu}$};
    \draw[-,dotted,brown,line width=0.5mm] (2+\a,2) -- (2+\a,0);
    \draw node at (3+\a,-0.3) {$\frac{b}{\mu}$};
    \draw[-,dotted,brown,line width=0.5mm] (3+\a,2) -- (3+\a,0);
    \draw node at (4+\a,-0.3) {$\frac{2}{\mu}$};
    \draw[-,dotted,brown,line width=0.5mm] (4+\a,2) -- (4+\a,0);

    \draw[-,blue,line width=0.8mm] (0+\a,0) -- (2+\a,2);
    \draw[-,blue,line width=0.8mm] (2+\a,2) -- (2.8+\a,0);

    \draw[-,BurntOrange, line width=0.8mm] (0+\a,0.5) -- (3.0+\a,2);
    \draw[-,OliveGreen,line width=0.8mm] (0+\a,1.0) -- (2+\a,2.0);
    \draw[-,OliveGreen,line width=0.8mm] (2+\a,2.0) -- (4+\a,0.0);

    \draw[-,red,line width=0.8mm] (0+\a,0.0) -- (4+\a,2.0);
    \draw[-,red,line width=0.8mm] (4+\a,2.0) -- (5+\a,0.4);
    \draw node at (2+\a,3) {$\beta < 1/2$};
    \draw node at (2+\a,2.5) {\scriptsize ${\color{blue} f_L}, {\color{red} f_{L^2}}, {\color{OliveGreen} f_R}, {\color{BurntOrange} f_S}$};

    \def\a{12}; 
    \def\b{0.5}; 

    \draw[->] (0+\a,0) -- (5+\a,0);
    \draw[->] (0+\a,0) -- (0+\a,3);
    \draw[-,dashed,gray,line width=0.5mm] (0+\a,2) -- (5+\a,2);
    \draw node[rotate=90] at (-0.5+\a,3.0) {$f(\alpha)$}; 
    \draw node at (5.0+\a,-0.5) {$\alpha$};
    \draw node at (-0.25+\a,2.0) {$1$};
    \draw node at (-0.25+\a,1.0) {$\frac{1}{2}$};
    \draw node at (-0.25+\a,1.5) {$\beta$};

    \draw node at (2+\a,-0.3) {$\frac{1}{\mu}$};
    \draw[-,dotted,brown,line width=0.5mm] (2+\a,2) -- (2+\a,0);
    \draw node at (1+\a,-0.3) {$\frac{b}{\mu}$};
    \draw[-,dotted,brown,line width=0.5mm] (1+\a,2) -- (1+\a,0);
    \draw node at (4+\a,-0.3) {$\frac{2}{\mu}$};
    \draw[-,dotted,brown,line width=0.5mm] (4+\a,2) -- (4+\a,0);

    \draw[-,blue,line width=0.8mm] (0+\a,0) -- (2+\a,2);
    \draw[-,blue,line width=0.8mm] (2+\a,2) -- (2.8+\a,0);

    \draw[-,BurntOrange, line width=0.8mm] (0+\a,1.5) -- (1.0+\a,2);
    \draw[-,OliveGreen,line width=0.8mm] (0+\a,1.0) -- (2+\a,2.0);
    \draw[-,OliveGreen,line width=0.8mm] (2+\a,2.0) -- (4+\a,0.0);

    \draw[-,red,line width=0.8mm] (0+\a,0.0) -- (4+\a,2.0);
    \draw[-,red,line width=0.8mm] (4+\a,2.0) -- (5+\a,0.4);
    \draw node at (2+\a,3) {$\beta > 1/2$};
    \draw node at (2+\a,2.5) {\scriptsize ${\color{blue} f_L}, {\color{red} f_{L^2}}, {\color{OliveGreen} f_R}, {\color{BurntOrange} f_S}$};
\end{tikzpicture}
\end{center}
This can now be used to estimate the scaling of the eigenvalue difference.

\subsection{Level Spacing SFD}\label{app:AppendixB2}

In this section, we will provide an estimate for the scaling of level spacings $E_i - E_j$.
Since the DOS splits into bubbles, corresponding to the eigenvalues of the largest interaction term, we can focus on a single bubble and investigate the level spacing there. 

Let us recall the expression for the eigenvalues
\begin{align}
    E_{i} = 1 + \sum_{k}^{N^r}w_{i k}j_{k} + \sum_{m}^{N^a}\frac{\vert j_{m} \vert^2}{e_{m}} + \sum_{n}^{N^b}\frac{\vert j_n \vert^2}{2}
\end{align}
where $w_{i k}$ are $O(1)$ phases, and $e_{m} = \sum_{k'}^{N^r}w_{m_1 k'}j_{k'} - \sum_{k}^{N^r}w_{m_2 k''}j_{k''}$ for some pair of indices $m_1, m_2$.
Since the eigenvalues are not necessarily sorted in this basis, aside from the $0^\text{th}$ order contribution of $\pm 1$, the signs of $e_m$ are not fixed.
This leads to each term in the second sum having the SFD $f_R$ but with arbitrary signs.
Since we are interested in the eigenvalue differences, we can ignore the leading term $+1$, and evaluate the SFD of the remaining combination.
It is important to note that we have $\mathcal{N} = \binom{N}{q} \sim N^q$ independent $j_k$ couplings, but the eigenvalues in each bubble are roughly $\mathcal{D}/2$ in number.
This significant discrepancy is the underlying reason behind the presence of a crossover rather than a transition in the eigenvalue correlation statistics. 

First, we note that sums of L\'evy random variables with arbitrary signs give the same SFD as that of the original variables which we add up.
Therefore, we end up with the SFD $f_L(\alpha)$ for the sum $\sum_{k}^{N^r}w_{i k}j_{k}$.
The next step is evaluating the sum $\sum_{m}^{N^a}\frac{\vert j_{m} \vert^2}{e_{m}}$.
Since this is a sum of numbers with arbitary phases and individual SFD given by $f_R$, the overall sum will also have an SFD given by $f_R$.
The last term is the extensive sum of L\'evy random variables, whose SFD is given by $f_S (\alpha)$ with the $\beta$ corresponding to $b$ in this case.
We now have to sum up the second and third term, $\sum_{k}^{N^r}w_{i k}j_{k} + \sum_{m}^{N^a}\frac{\vert j_{m} \vert^2}{e_{m}}$, which is equivalent to performing the sum $Z = X + Y$ with respective SFDs $f_L$ and $f_R$.
As the maxima for both these SFDs is obtained for $\alpha = 1/\mu$, we have to use the following rule
\begin{align}
    p_{X + Y}(\vert Z \vert = N^{-\alpha}) \propto
    \begin{cases}
        &N^{\max\{f_L(\alpha),f_R(\alpha)\} +\alpha - 1}\;\;\;\;\;\;\;\;\;\;\;\;\;\;\;\;\;\;\;\;\;\;\;\;\;\;\;\;\;\;\,\alpha < \frac{1}{\mu}\\
        &N^{\max_{\xi \leq \alpha}\{f_L (\xi) + f_R (\xi) + \xi\}-2} + N^{f_L(\alpha) + \alpha - 1}\;\;\,\alpha = \frac{1}{\mu}\\
        &N^{\max_{\xi \leq \alpha}\{f_L (\xi) + f_R (\xi) + \xi\}-2}\;\;\;\;\;\;\;\;\;\;\;\;\;\;\;\;\;\;\;\;\;\;\;\;\;\,\alpha >\frac{1}{\mu}
    \end{cases}
\end{align}
Let us denote this combined SFD as $f_{LR}$.
From the previous subsection, we know that $f_R > f_L\,\forall \alpha$.
Therefore, for $\alpha < \frac{1}{\mu}$, $f_{LR} = f_{R} = \frac{\mu\alpha + 1}{2}$.
The next step is evaluate $\max_{\xi \leq \alpha}\{f_L (\xi) + f_R (\xi) + \xi\}$ for $\alpha > \frac{1}{\mu}$.
If we choose $\xi < \frac{1}{\mu}$ then $f_L(\xi) + f_R(\xi) + \xi = \frac{3\mu\xi + 1}{2} + \xi$.
This is naturally maximized for $\xi = \frac{1}{\mu}$ in this range. In the other range $\xi > \frac{1}{\mu}$, we obtain $f_L(\xi) + f_R(\xi) + \xi = 3+\frac{1}{\mu}-\mu\xi$ which is again maximized for $\xi = \frac{1}{\mu}$. Therefore, the SFD $f_{LR}$ can be written as
\begin{align}
    f_{LR}(\alpha) = \frac{\mu\alpha + 1}{2}\theta(1-\mu\alpha) + \left(\frac{1}{\mu} + 1 - \alpha\right)\theta(\mu\alpha - 1)
\end{align}
This is represented graphically as follows.
\begin{center}
\begin{tikzpicture}
    \def\a{5}; 
    \def\b{0.5}; 
    \draw[->] (0+\a,0) -- (4+\a,0);
    \draw[->] (0+\a,0) -- (0+\a,3);
    \draw[-,dashed,gray,line width=0.5mm] (0+\a,2) -- (4+\a,2);
    \draw node[rotate=90] at (-0.5+\a,3.0) {$f_{LR}(\alpha)$}; 
    \draw node at (4.0+\a,-0.5) {$\alpha$};
    \draw node at (-0.25+\a,2.0) {$1$};
    \draw node at (-0.25+\a,1.0) {$\frac{1}{2}$};
    \draw node at (2+\a,-0.3) {$\frac{1}{\mu}$};
    \draw[-,dotted,brown,line width=0.5mm] (2+\a,2) -- (2+\a,0);
    \draw[-,Periwinkle,line width=0.8mm] (2+\a,2) -- (2.8+\a,0);
    \draw[-,Periwinkle,line width=0.8mm] (0+\a,1.0) -- (2+\a,2.0);
    \draw node[rotate = 26.57] at (0.85+\a,1.1) {$\frac{\mu\alpha + 1}{2}$};
    \draw node[rotate = -68.2] at (2.8+\a,1.0) {${1-\alpha + 1/\mu}$};
\end{tikzpicture}
\end{center}
The final step is to combine this with the sum $\sum_{n}^{N^b}\frac{\vert j_n \vert^2}{2}$.
There are two cases, corresponding to $b > 1/2$ and $b < 1/2$. For $b > 1/2$, we have $\frac{2}{\mu}(1-b) < \frac{1}{\mu}$.
Therefore, the combination rule becomes
\begin{align}
    p_{X + Y}(\vert Z \vert = N^{-\alpha}) \propto
    \begin{cases}
        &N^{\max\{f_S(\alpha),f_{LR}(\alpha)\} +\alpha - 1}\;\;\;\;\;\;\;\;\;\;\;\;\;\;\;\;\;\;\;\;\;\;\;\;\;\;\;\;\;\;\,\alpha < \frac{2}{\mu}(1-b)\\
        &N^{\max_{\xi \leq \alpha}\{f_S (\xi) + f_{LR} (\xi) + \xi\}-2} + N^{f_S(\alpha) + \alpha - 1}\;\;\,\frac{2}{\mu}(1-b) < \alpha < \frac{1}{\mu}\\
        &N^{\max_{\xi \leq \alpha}\{f_S (\xi) + f_{LR} (\xi) + \xi\}-2}\;\;\;\;\;\;\;\;\;\;\;\;\;\;\;\;\;\;\;\;\;\;\;\;\;\,\alpha >\frac{1}{\mu}
    \end{cases}
\end{align}
Let us denote the SFD of the full combination as $f_E$.
In the region $\alpha < \frac{2}{\mu}(1-b)$, the SFD is dominated by $f_S$.
So we have $f_E = f_S (\alpha) \theta\left(\frac{2}{\mu}(1-b) - \alpha\right)$.
Note that since $f_S (\alpha > \frac{2}{\mu}(1-b)) = -\infty$ we must have $\max_{\xi \leq \alpha}\{f_S (\xi) + f_{LR} (\xi) + \xi\} = f_S (\frac{2}{\mu}(1-b)) + f_{LR} (\frac{2}{\mu}(1-b)) + \frac{2}{\mu}(1-b) = \frac{1}{2} -b + \frac{2(1-b)}{\mu}$.
This gives us the SFD
\begin{align}
    f_{E}(\alpha)\vert_{b > 1/2} = \left(\frac{\mu\alpha}{2} + b\right) \theta\left(\frac{2(1-b)}{\mu} - \alpha\right) + \left(\frac{3}{2} - b + \frac{2(1-b)}{\mu} - \alpha\right) \theta\left(\alpha - \frac{2(1-b)}{\mu}\right)
\end{align}
Now, for $b < \frac{1}{2}$, we have $\frac{1}{\mu} < \frac{2}{\mu}(1-b)$ and correspondingly must switch $f_S$ and $f_{LR}$ in the analysis above.
This gives us the following result (noting that $\{f_S (\xi) + f_{LR} (\xi) + \xi\}$ is maximized for $\xi = \frac{2}{\mu}(1-b)$.
This give us
\begin{align}
    f_{E}(\alpha)\vert_{b<1/2} = \frac{\mu\alpha + 1}{2}\theta(1-\mu\alpha) + \left(1 + \frac{1}{\mu} - \alpha\right)\theta(\mu\alpha - 1)
\end{align}
Thus the two cases can be represented graphically as follows
\begin{center}
\begin{tikzpicture}
    \def\a{5}; 
    \def\b{0.5}; 
    \draw[->] (0+\a,0) -- (4+\a,0);
    \draw[->] (0+\a,0) -- (0+\a,3);
    \draw[-,dashed,gray,line width=0.5mm] (0+\a,2) -- (4+\a,2);
    \draw node[rotate=90] at (-0.5+\a,3.0) {$f_{E}(\alpha)$}; 
    \draw node at (4.0+\a,-0.5) {$\alpha$};
    \draw node at (-0.25+\a,2.0) {$1$};
    \draw node at (-0.25+\a,1.0) {$\frac{1}{2}$};
    \draw node at (2+\a,-0.3) {$\frac{1}{\mu}$};
    \draw[-,dotted,brown,line width=0.5mm] (2+\a,2) -- (2+\a,0);
    \draw[-,Periwinkle,line width=0.8mm] (2+\a,2) -- (2.8+\a,0);
    \draw[-,Periwinkle,line width=0.8mm] (0+\a,1.0) -- (2+\a,2.0);
    \draw node[rotate = 26.57] at (0.85+\a,1.1) {$\frac{\mu\alpha + 1}{2}$};
    \draw node[rotate = -68.2] at (2.8+\a,1.0) {${1-\alpha + 1/\mu}$};
    \draw node at (2.0+\a,3.0) {$b < 1/2$};

    \def\a{10}; 
    \def\b{0.5}; 
    \draw[->] (0+\a,0) -- (4+\a,0);
    \draw[->] (0+\a,0) -- (0+\a,3);
    \draw[-,dashed,gray,line width=0.5mm] (0+\a,2) -- (4+\a,2);
    \draw node[rotate=90] at (-0.5+\a,3.0) {$f_{E}(\alpha)$}; 
    \draw node at (4.0+\a,-0.5) {$\alpha$};
    \draw node at (-0.25+\a,2.0) {$1$};
    \draw node at (-0.25+\a,1.5) {$b$};
    \draw node at (1.2+\a,-0.3) {$\frac{2}{\mu}(1-b)$};
    \draw[-,dotted,brown,line width=0.5mm] (1.2+\a,2) -- (1.2+\a,0);
    \draw[-,BurntOrange,line width=0.8mm] (1.2+\a,1.2) -- (1.68+\a,0);
    \draw[-,BurntOrange, line width=0.8mm] (0+\a,1.5) -- (1.2+\a,2);
    \draw node[rotate = 22.62] at (0.54+\a,1.4) {$\frac{\mu\alpha}{2} + b$};
    \draw node[rotate = -68.2] at (1.6+\a,0.8) {${\mathbf{b} - \alpha}$};
    \draw node at (2.0+\a,3.0) {$b > 1/2$};
\end{tikzpicture}
\end{center}
where, in the figure above, we have introduced  $\mathbf{b} = \frac{3}{2} -b + \frac{2(1-b)}{\mu}$.
This is the SFD of the (perturbative) eigenvalues.
From this, we can compute the eigenvalue difference (between any two eigenvalues).
The SFD of $\Delta E = E_i - E_j$, which we denote by $f_\Delta (\alpha)$.
For the case of $b < \frac{1}{2}$, it follows from the previous analysis that $f_\Delta = f_E$.
In the opposite case of $b > 1/2$, we similarly have $f_\Delta = \left(\frac{\mu\alpha}{2} + b\right)\theta\left(\frac{2(1-b)}{\mu} - \alpha\right)$ (for $\alpha < \frac{2(1-b)}{\mu}$).
For the $\alpha > \frac{2(1-b)}{\mu}$, we have find $\max_{\xi \leq \alpha}\{\mu\xi + 2 b + \xi\}$, which is satisfied for $\xi = \frac{2(1-b)}{\mu}$, giving us the following SFD:
\begin{align}
    f_\Delta (\alpha) = \left(\frac{\mu\alpha}{2} + \max\left\{b,\frac{1}{2}\right\}\right)\theta(1-\alpha_0) + \left(1 + \alpha_0 - \alpha\right)\theta(\alpha_0 - 1)
\end{align}
where $\alpha_0 = \min\{\frac{1}{\mu},\frac{2(1-b)}{\mu}\}$.
A graphical representation of the same would be given by:
\begin{center}
\begin{tikzpicture}
    \def\a{5}; 
    \def\b{0.5}; 
    \draw[->] (0+\a,0) -- (4+\a,0);
    \draw[->] (0+\a,0) -- (0+\a,3);
    \draw[-,dashed,gray,line width=0.5mm] (0+\a,2) -- (4+\a,2);
    \draw node[rotate=90] at (-0.5+\a,3.0) {$f_{\Delta}(\alpha)$}; 
    \draw node at (4.0+\a,-0.5) {$\alpha$};
    \draw node at (-0.25+\a,2.0) {$1$};
    \draw node at (-0.8+\a,1.0) {$\max\{b,\frac{1}{2}\}$};
    \draw node at (2+\a,-0.3) {\scriptsize $\alpha_0 = \min\{\frac{1}{\mu},\frac{2(1-b)}{\mu}\}$};
    \draw[-,dotted,brown,line width=0.5mm] (2+\a,2) -- (2+\a,0);
    \draw[-,RubineRed,line width=0.8mm] (2+\a,2) -- (2.8+\a,0);
    \draw[-,RubineRed,line width=0.8mm] (0+\a,1.0) -- (2+\a,2.0);
    \draw node[rotate = 26.57] at (0.95+\a,1.2) {\scriptsize $\frac{\mu\alpha}{2} + \max\{b,\frac{1}{2}\}$};
    \draw node[rotate = -68.2] at (2.8+\a,1.0) {${1-\alpha + \alpha_0}$};
\end{tikzpicture}
\end{center}
Not surprisingly, we find that SFD of the eigenvalue difference is dominated by extensive sums of L\'evy random numbers (with or without arbitrary phases).

\subsection{Crossover from Chaotic to Integrable Eigenspectrum}\label{app:AppendixB3}

Having studied the SFD of the eigenvalue spacings, we are finally ready to address the emerging integrability observed numerically in the eigenvalue correlations of the L\'evy SYK.
Let us begin by considering the action of a single term in the Hamiltonian on an eigenstate 
\begin{align}
    j_{i}\Psi_{i}\ket{\psi_n} = j_{i}\sum_{l = 1}^{M}\zeta_{l n}\ket{\psi_{m}}\,.
    \label{eq:act-eigst}
\end{align}
where $\zeta_{l n}$ are $O(1)$ weights. On a related note, let us recall that the first-order correction to the \emph{eigenstates} is given by 
\begin{align}
    \ket{\psi^{(1)}_{k}} &= \ket{\psi^{(0)}_k} + \sum_{n \neq k}\frac{\bra{\psi^{(0)}_{n}}H'\ket{\psi^{(0)}_{k}}}{E^{(0)}_n - E^{(0)}_k}\ket{\psi^{(0)}_{n}}\notag\\
    &= \ket{\psi^{(0)}_k} + \sum_{m = 1}^{N^a}\frac{j_m}{e_{m}}\ket{\psi^{(0)}_{m}} + \sum_{n = 1}^{N^b}\frac{j_n}{2}\ket{\psi^{(0)}_n}
\end{align}
where the first sum in the second line is coming from the eigenstates in the same sector (corresponding to the eigenvalue of $\Psi_1$) and the second sum consists of eigenstates from the complementary sector.
The coefficient $\frac{j_m}{e_{m}}$ has the SFD $f_R$ while the coefficient $\frac{j_n}{2}$ has the SFD $f_L$. 
Since $f_R > f_L$ always, the correction term in the eigenstate will be dominated by the states in the same sector.
This reinforces the ``bubble-splitting'' picture, as also observed in the density of states. 

Returning to~\eqref{eq:act-eigst}, we observe that the probability of a state jumping into another is proportional to $j_{i}$, \textit{i.e.} to the interaction strength.
As a result, we can argue that a given eigenstate can be hybridised into another eigenstate under the action of the Hamiltonian if the term $j_i$ that causes this jump is of the same order as the eigenvalue difference between the two states.

Therefore, we propose the following criterion for chaotic eigenspectrum:
\emph{the mean level spacing within a sector, $\Delta_{\pm}$, is comparable to the typical interaction strength $j_{\text{typ}}$}. 

Mathematically, the condition for integrable level correlation statistics can be recast as follows
\begin{align}
    j_{\text{typ}} \ll \Delta_{\pm} = \frac{\sum_{k} e_{k}}{\mathcal{D}^2} 
\end{align}
where $e_{k}$ are the eigenvalue differences that we consider. 

First, recall that we have $\mathcal{N}$ number of random interactions.
This means that the eigenvalues have to be sums of this fixed number of terms.
Thus, the typical sum of $e_{k}$ can be chosen to be a sum of $\mathcal{N}$ numbers which follow the SFD given by $f_\Delta$.
The peak of $f_\Delta$ tells us the typical value of each $e_k$, which is $\mathcal{N}^{-\alpha_0}$.
To estimate the sum, we can replace each $e_k$ by $\mathcal{N}^{-\alpha_0} = \mathcal{N}^{-\min\{\frac{1}{\mu},\frac{2(1-b)}{\mu}\}}$.
The sum of $\mathcal{N}$ terms is of the order $\mathcal{N}^{\frac{1}{\mu} - \min\{\frac{1}{\mu},\frac{2(1-b)}{\mu}\}}$, which again follows from the extensivity of L\'evy sums.
From the multifractal analysis of $j_{i}$, we know that the (normalized) typical values are of the order $\mathcal{N}^{-\frac{1}{\mu}}$.
This allows us to perform the comparison 
\begin{align}
    \mathcal{N}^{-\frac{1}{\mu}} \sim N^{-\frac{q}{\mu}} \lesssim \frac{N^{\frac{qc}{\mu}}}{N^{N \frac{\ln 2}{\ln N}}}    
\end{align}
where we have used the notation $c = 1 - \min\{1,2(1-b)\} < 1$, and $\mathcal{D} \sim N^{\frac{N \ln 2}{2 \ln N}}$.
This finally leads us to the condition
\begin{align}
    N^{\frac{qc + q}{\mu}} \gtrsim N^{\frac{N \ln 2}{\ln N}} \implies \mu \lesssim \frac{\ln N}{(q c + q) N\ln 2} \sim \frac{\ln N}{N}\,.
    \label{eq:mu-bound}
\end{align}
where we have used the fact that $q \geq 4$ in most of the cases that we study.
This tells us that for emergent integrability, we must scale $\mu$ with the number of fermions $N$ at least as $\frac{\ln N}{N}$.
Our numerical investigations of the eigenvalues tell us that the critical $\mu$ (denoted by $\mu_c$, at which we find deviation from chaoticity) scales as $\frac{1}{N}$, which is consistent with the bound \eqref{eq:mu-bound} derived in this section.
A simple consequence of this approach is to consider the case where the Hilbert space dimension is $\mathcal{N} \sim N^q$, which leads us to the condition
\begin{align}
    &N^{-\frac{q}{\mu}} \lesssim \frac{N^{\frac{q c}{\mu}}}{N^{2q}}\notag\\
    \implies &N^{\frac{q + qc}{\mu}} \gtrsim N^{2 q}
\end{align}
This gives the condition $\mu \lesssim \frac{q + q c}{2 q} = \frac{1 + c}{2}$, which is an $N-$independent constant.
This result is in agreement with the idea that the model is now effectively behaving as a structureless L\'evy random matrix.

\subsection{Zeroth order contribution}\label{app:AppendixB4}

An interesting aspect of this construction is to consider what happens if we simply consider the $0^\text{th}$ order contribution of the eigenvalues.
Recall that at this order, the eigenvalues can be written as
\begin{align}
    E_{i}^{\pm} = \pm1 + \sum_{k}^{N^r}w_{i k}^{\pm}j^{\pm}_{k}
\end{align}
Since we are concerned with the eigenvalue differences within a bubble, we can ignore the leading $\pm 1$ and focus on the sum over $N^r$ terms.
As demonstrated in the previous section, the SFD of this term is $f_{L}(\alpha)$.
Naturally, the difference of any two such eigenvalues will also have the same SFD by the same argument. 
\begin{center}
\begin{tikzpicture}
    \def\a{5}; 
    \draw[->] (0+\a,0) -- (4+\a,0);
    \draw[->] (0+\a,0) -- (0+\a,3);
    \draw[-,dashed,gray,line width=0.5mm] (0+\a,2) -- (4+\a,2);
    \draw node[rotate=90] at (-0.5+\a,3.0) {$f_L(\alpha)$}; 
    \draw node at (4.0+\a,-0.5) {$\alpha$};
    \draw node at (-0.25+\a,2.0) {$1$};
    \draw node at (2+\a,-0.3) {$\frac{1}{\mu}$};
    \draw[-,dotted,brown,line width=0.5mm] (2+\a,2) -- (2+\a,0);
    \draw[-,blue,line width=0.8mm] (0+\a,0) -- (2+\a,2);
    \draw[-,blue,line width=0.8mm] (2+\a,2) -- (3+\a,0);

    \draw node at (2+\a,3) {$X \sim \vert j_I \vert$};
    
    \draw node[rotate = 45] at (0.8+\a,1.2) {$\mu\alpha$};
    \draw node[rotate = -63.43] at (2.8+\a,1.0) {${1-\alpha + 1/\mu}$};
\end{tikzpicture}
\end{center}
Let us note that there are $\sim O(N)$ number of such interaction terms $j_k$ that contribute to $e_{i j} = E_i - E_j$.
The sum over such eigenvalue differences $\sum_{i < j}e_{i j}$ will therefore have a large number of terms that are $0$.
If we estimate that there are $\sim O(N^2)$ such terms, then we can see that the extensive sum $\sum_{i < j}e_{i j}$ is of the order $N^{-2/\mu}N^{2/\mu} \sim N^{0}$.
Thus, at the first order, the sum over level spacings does not scale with $N$.
Thus at this order, there is \emph{always} an $O(1)$ number of interaction terms that can cause hopping between the eigenstates or combinations thereof.
This is a reflection of the fact that we are simply considering the $0^\text{th}$ order contribution of the eigenvalue spacings and as such are forced with a large number of degeneracies.
Once we include higher order corrections, the sums of spacings are expected to scale with $N$.
Note that this is \emph{not} a non-perturbative effect since individual eigenvalues $E_{i}$ pick up parametrically smaller or equivalent contributions (as evidenced by the fact that the peak of the SFD either remains at $1/\mu$ or shifts to the right).
The argument for integrability then rests on the condition that the mean level spacing (at a high-enough order where scaling with $N$ is reflected) is greater than the typical interaction $j_I$.

\section{Details of Numerical Computations}\label{app:AppendixC}

In this section, we describe the numerical computations presented in the main text, in further detail. We shall discuss the long- and short-range eigenvalue probes that we study (i.e. mean level spacing ratio and Spectral Form Factor, respectively).
We then discuss the methods used for processing the disordered data (filtering, averaging etc.) and issues regarding exact diagonalization of L\'evy SYK Hamiltonian.
Finally, we discuss the method that we employ to quantify the deviation from RMT predictions for the spectral statistics.

\subsection{Probes of Short-range and Long-range correlations}\label{app:AppendixC1}

The short-range (i.e. nearest-neighbour, in our case) correlations are captured by computing the mean of the ratio of nearest-neighbour level spacings.
A simple and useful metric, when doing numerical studies, is the mean \(\langle r \rangle\) of the nearest neighbour spacing \emph{ratio} distribution, which is defined as
\begin{align}
    r_{n} = \mathrm{min}\left(\overline{r}_{n},\frac{1}{\overline{r}_n}\right)\;\;,\;\;
    \overline{r}_{n} = \frac{s_{n}}{s_{n - 1}}\,.
    \label{levelspcratio}
\end{align}
Their computation does not require unfolding of the spectrum, and therefore, it turns out to be more practical than the spacing distribution.
The ratios $\overline{r}_{n}$ are independent of spectral unfolding procedure.
The values $\langle r\rangle$ for the Gaussian ensembles are known to be $\langle r \rangle = 0.5307\,(\text{GOE}),\,\,0.5996\,(\text{GUE})$ and $0.6744\,(\text{GSE})$.
For the case of uncorrelated spectra, the value $\langle r \rangle$ are usually much lower, with $\langle r\rangle \approx 0.39$ for Poissonian spectra.
For a given Hamiltonian \(\hat{H}\), we compute its spectrum \(\{\epsilon_{n}\}\) via exact diagonalization.
From the eigenvalues, we compute the mean $\langle r \rangle$ and study it as a function of the scale parameter $\mu$ (and averaging over many realizations of the disorder). 

To capture long-range correlations, we use the \emph{spectral form factor} (SFF).
The SFF $K(t)$ is defined as the (averaged) norm-square of the partition function $Z(\beta, t)$ at inverse temperature $\beta$. 
\begin{align}
    K(\beta, t) = \left\langle \frac{\vert Z(\beta,t) \vert^2}{\vert Z(\beta,0) \vert^2} \right\rangle,\;\; Z(\beta, t) = \text{Tr}\left(e^{-\beta H + i t H}\right)\,.
    \label{eqn:sff-defn}
\end{align}
It is also the Fourier transform of the spectral $2$-point function $\rho(\lambda_1,\lambda_2)$.
Different types of eigenvalue correlations contribute to the behavior of the SFF $K(\beta, t)$~\cite{mehta2004random,Cotler2017chaos}.
For Gaussian RMT, the SFF is characterized by the slope-dip-ramp-plateau behavior (often referred to as the \emph{correlation hole}).
The behavior of the SFF has been explored in the Gaussian SYK model~\cite{cotler2016black}, and it is found to agree strongly with RMT predictions.
To eliminate the effects of local density of states, we unfold the spectrum $\{\epsilon_{1} \leq \epsilon_{2} \leq \cdots \leq \epsilon_{L}\}$ i.e. scale the eigenvalues such that $\rho(\epsilon) = \frac{1}{L}$, where $L$ is the Hilbert space dimension.
We also scale the time axis to $\tau = t/\tmh$, where $\tmh = 2\pi$ is the Heisenberg time (i.e. the inverse of the mean level spacing) for Gaussian RMT.
Due to unfolding, the mean level spacing is set to \(1\), and correspondingly $\tmh = 2\pi$.
We are interested in capturing the deviation of the LSYK SFF from the Gaussian SYK case~\cite{cotler2016black} upon introducing L\'evy disorder.
We set \(\beta = 0\) to remove temperature effects and we scale the SFF in Eq.~\eqref{eqn:sff-defn} by $\vert Z(0,0) \vert^2 = L^2$ so that the long time saturation value is $K(\tau \rightarrow \infty) = 1$. 

\begin{figure}[ht]
    \centering
    \includegraphics[width=0.45\linewidth]{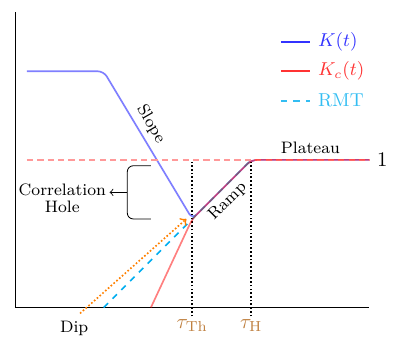}
    \caption{
        Schematic representation of the connected and full spectral form factors (on the log-log scale), with the Heisenberg time $\tau_{\mathrm{H}}$ and the Thouless time $\tmt$ marked.
        The general expected structure for chaotic systems -- slope-dip-ramp-plateau -- is highlighted and marked.
    }
    \label{fig:schemsff}
\end{figure}

The early time behaviour of the SFF~\eqref{eqn:sff-defn} is dominated by it's \emph{disconnected part}, which is defined as
\begin{align}
    K_{dc}(\tau) = \frac{\left\vert \sum_{k} \langle e^{- 2\pi i \epsilon_{k} \tau} \rangle \right\vert^2}{\vert Z(0,0) \vert^2}\,.
    \label{eqn:sffdc-defn}
\end{align}
Subtracting \eqref{eqn:sffdc-defn} from \eqref{eqn:sff-defn} gives the \emph{connected} SFF
\begin{align}
    K_{c}(\tau) \equiv K(\tau) - K_{dc}(\tau)\,,
    \label{eqn:sffc-defn}
\end{align}
which solely contains eigenvalue spacing contributions (of the form $\epsilon_{i} - \epsilon_{j}$). The schematic behavior of the SFF for the Gaussian SYK~\cite{Cotler2017chaos} is shown in Fig.~\ref{fig:schemsff}.

\subsection{Diagonalization, Filtering and Averaging}\label{app:AppendixC2}
\label{app:1p1}

In the numerical calculations discussed in the main text, we have worked with eigenvalues of LSYK up to system size $N = 32$.
The Hilbert space dimension for a given $N$ (in the parity-resolved sector) is $2^{\frac{N}{2} - 1}$.
To diagonalize such large, dense matrices it is convenient to focus on parts of the eigenspectrum and use Arnoldi methods.
However, such approaches fail for LSYK (and more advanced cases such as Chebyshev Filtering as well~\cite{andreanov2025from,pieper2016high,chebf}) since depending on the particular disorder realisation, the density of state tends to be peaked around the edges (as discussed in Section IV of the main text).
Therefore, we are forced to use full exact diagonalization to extract the eigenvalues. 

While computing the spectral form factor, there are strong noisy effects due to both the disordered nature of the Hamiltonian, as well as very different contributions from long- and short- range correlations.
Therefore, it is necessary to suppress these in a meaningful way, in order to extract relevant physics.
This is done via two methods: averaging and filtering.
Following~\cite{suntas2020quantum}, we employ a Gaussian filter function $F_\eta(E_n) = e^{- \frac{(E_n - \overline{E})}{2 (\eta \Gamma)^2}}$, where $\overline{E}$ and $\Gamma$ are the mean and variance of the eigenvalue spectrum.
The strength of the filter is controlled by $\eta$.
This function ensures that the eigenvalues near the middle of the spectrum are given the highest weight when computing the SFF, thus minimizing anomalous effects which may arise from the edge of the spectrum (for a given realisation).
We find similar results for $0.1 \leq \eta \leq 0.5$.
Naturally, we also average over disorder realisations.
It is worth noting that both these processes can be tied to the introduction of non-unitary channels~\cite{apollo2023unitarity}, despite which the SFF still remains a useful probe for random matrix behavior.
To further smoothen the results, we perform a moving average of $100$ data points for the SFF (where each SFF curve consists of $> 10^6$ data points), following the approach of~\cite{suntas2020quantum}.

\subsection{Crossover: Deviation from RMT}\label{app:AppendixC3}

\begin{figure}[htbp]
    \centering
    \subfigure[]{
        \includegraphics[width=0.45\linewidth]{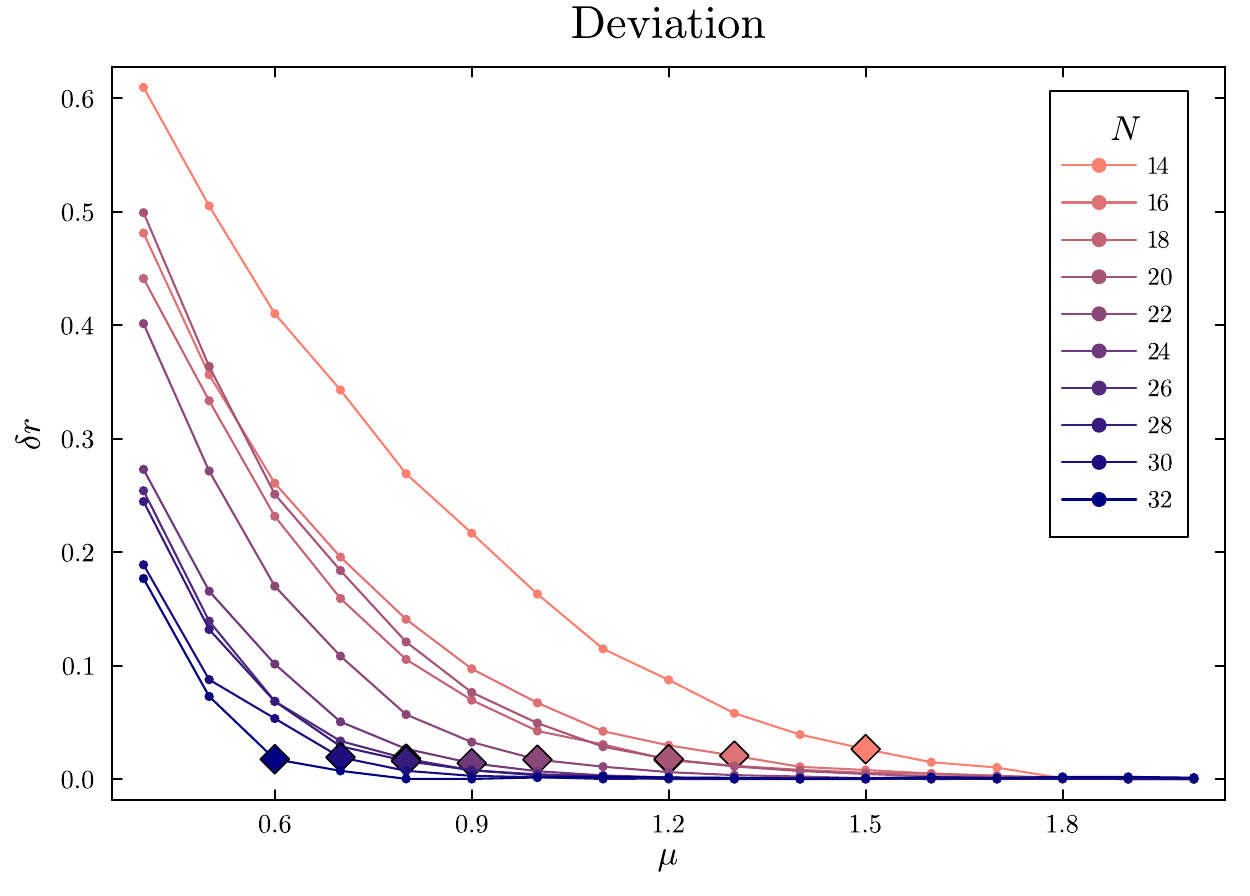}
        \label{fig:lvlspc-dev}
    }
    \subfigure[]{
        \includegraphics[width=0.45\linewidth]{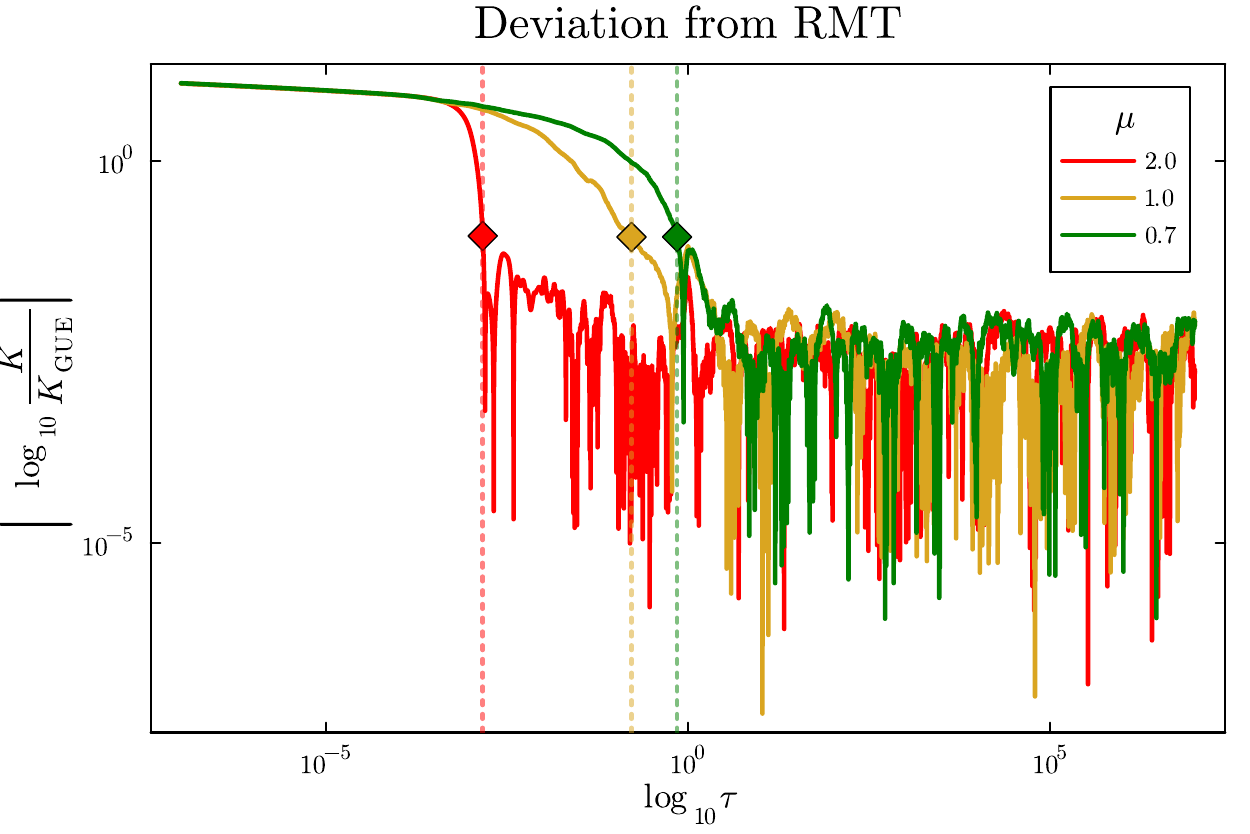}
        \label{fig:deviation-sff}
    }
    \caption{
        (a) Deviation of the $r$-ratio from the RMT value for each system size $N$, with the crossover values mark separately, and
        (b) Deviation of the SFF ratio from the RMT curve for $N = 30$ and $\mu = 0.7, 1.0$ and $2.0$, with the crossover values (i.e. Thouless time $\tau_{\mathrm{Th}}$) marked separately.
        For $N = 30$, the RMT class is GUE.
    }
    \label{fig:rmt-deviation}
\end{figure}

To compute the deviation from random matrix theory for the short-range correlations, we first compute the level spacing ratio \eqref{levelspcratio}.
From this, the deviation is calculated
\begin{align}
    \delta r = \frac{r - r_{\text{RMT}}}{r_{\text{RMT}}}
\end{align}
where $r_{\text{RMT}}$ is the random matrix value for the corresponding universality class.
We identify the crossover value $\mu_c$ by determining the closest value of $\mu$ that corresponds to a deviation $> 0.01$ (i.e. a $1\%$ deviation from the RMT value).
The result is provided in Fig.~\ref{fig:lvlspc-dev}.
Considering other threshold values (up to $10\%$) produces similar scaling of $\mu_c$ with $N$ as presented in the main text. 

To determine the Thouless times, we compute the difference between the numerically evaluated SFF and the analytical RMT curve, following the approach of~\cite{suntas2020quantum}.
To evaluate the difference between the actual SFF $K$ and the RMT curve $K_{\mathrm{GOE/GUE/GSE}}$ we compute $\Delta K = \Big\vert\log_{10}\frac{K}{K_{\mathrm{GOE/GUE/GSE}}} \Big\vert$ and observe it as a function of rescaled time $\tau = t/t_{H}$ where $t_{H}$ is the Heisenberg time.
Since we work with unfolded spectra, the Heisenberg time is by definition $t_{H} = 2\pi$.
To capture the Thouless time $\tau_{\text{Th}}$, we identify the $\tau$ value where $\Delta K \leq 0.1$ (i.e. $ < 10\%$ deviation from RMT curve).
An example is provided in Fig.~\ref{fig:deviation-sff}. 

To determine the crossover value $\mu_{c,2}$, we simply identify the value of $\mu$ for which the Thouless time becomes close to $1$.
Due to the limited parameter values for which we have numerical results, we sometimes find cases where $\tau_{\mathrm{Th}}(\mu_1) < 1.0$ and $\tau_{\mathrm{Th}}(\mu_2 = \mu_1 + 0.1) \gtrsim 1.0$.
In such cases, we have the freedom of choosing $\mu_{c,2} = \mu_1$ or $\mu_{c,2} = \mu_2$ or $\mu_{c,2} = \frac{\mu_{1} + \mu_{2}}{2}$.
We find the choices do not qualitatively alter our conclusions, with the crossover parameter scaling as $\mu_{c,2} \sim N^{-\eta_2}$ where $1.1 \leq \eta_2 \leq 1.7$.

\section{Deformations and Other Numerical Results}\label{app:AppendixD}

In this section, we present some additional numerical results.
These support the results of the main manuscript but are not essential to the conclusions presented there.
We shall present the density of states, the level spacing distribution, behavior of the ground state energy distribution and some results regarding the spectral form factor.

\subsection{Eigenvalue Distributions}\label{app:AppendixD1}

\begin{figure}[htbp]
    \centering
    \subfigure[]{
    \includegraphics[width=0.45\linewidth]{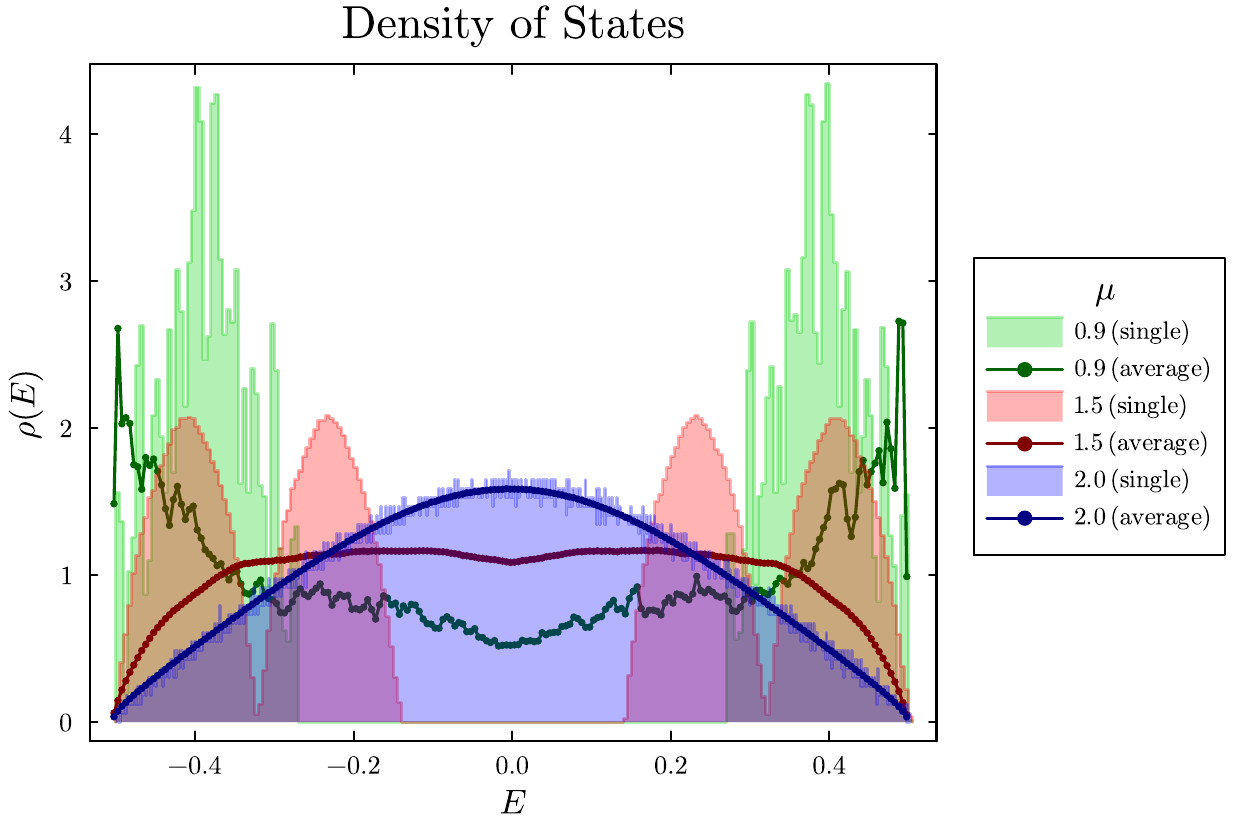}\label{fig:full-rho-supp}}
    \subfigure[]{
    \includegraphics[width=0.45\linewidth]{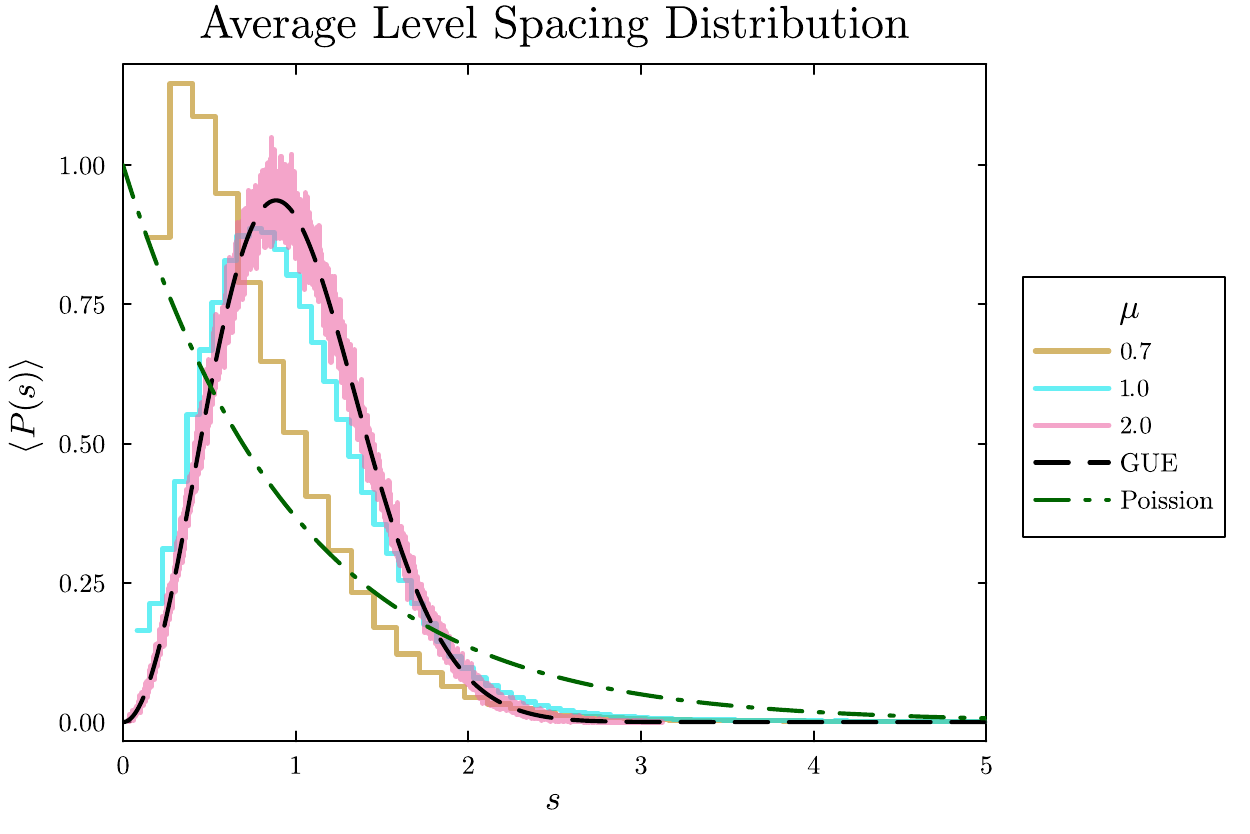}\label{fig:avg-lvlsp-dist-supp}}
    \caption{
        (a) The average density of states \(\langle \rho(E) \rangle\) and single representative realization $\rho(E)$, for several values of \(\mu\) (for \(N=30\) fermions).
        For each disorder realization the eigenvalues were rescaled into the range \([-0.5,0.5]\).
        (b) The average level spacing distribution $\langle P(s) \rangle$ for different values of $\mu$ and $N = 30$.
    }
    \label{fig:eigen-stat}
\end{figure}

Here we discuss the behavior of the density of states and the nearest-neighbor spacing distribution.
As discussed in the main manuscript, the eigenvalues are dominated by the largest interaction terms, thus leading to a split in the density of states.
The DOS thus splits into ``bubbles'' centred at $\pm J_{\text{largest}}$, with small fluctuations generated by contributions from parametrically smaller terms which commute with the Majorana string corresponding to $J_{\text{largest}}$.
As discussed in the main text, it follows from the Ger\v{s}gorin Theorem that the remaining eigenvalues will lie within these bubbles.
We observe this phenomenon for $N = 30$ in Fig.~\ref{fig:full-rho-supp}.
This effect is explained by the dominance of $J_{\text{largest}}$ over the remaining interaction terms.
This is almost surely true for $\mu < 1$ (though it can be observed for $\mu > 1$ as well in some realizations), which can be determined by comparing the strength of all the interactions aside from the largest one with the magnitude of the largest interaction.
This is discussed in the following section.
This is an effect that follows from the L\'evy distribution, instead of being a genuine many-body effect.

The nearest-neighbor spacing distribution, on the other hand, captures a genuine many-body effect.
In the main text we argue that this will lead to a scaling in the value of $\mu$ at which chaotic behavior (see~\cite{atas2013distribution}) is lost.
It is reflected in the spacing distribution as well, as observed in Fig.~\ref{fig:avg-lvlsp-dist-supp}.
Note that even at $\mu = 1$, there is only a slight shift from the curve at $\mu = 2$.
This is a result of the scaling with $N$, since this result is for $N = 30$ and so $\mu_c$ is expected to be small.

\subsection{Ground state energy and SFF}\label{app:AppendixD2}

\begin{figure}[htbp]
    \centering
    \subfigure[]{
        \includegraphics[width=0.45\linewidth]{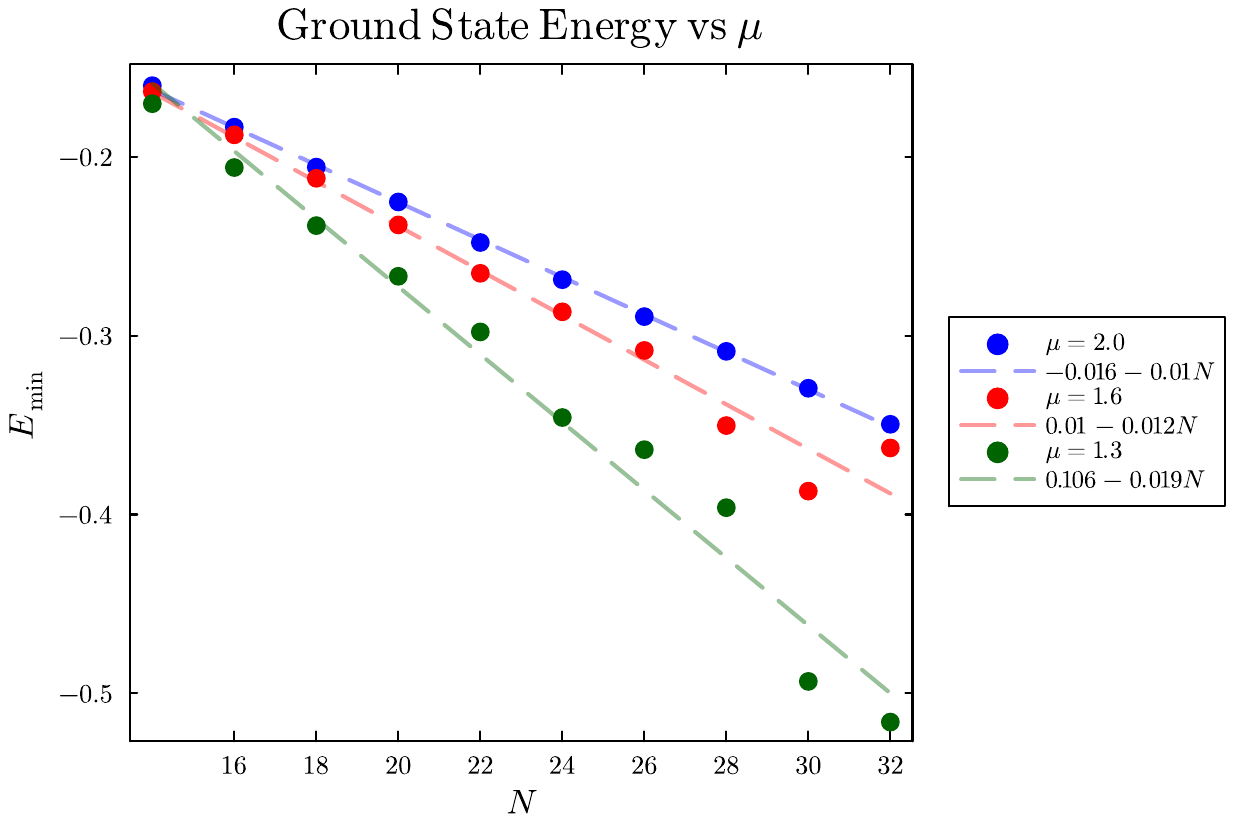}
        \label{fig:emin-scaling-supp}
    }
    \subfigure[]{
        \includegraphics[width=0.45\linewidth]{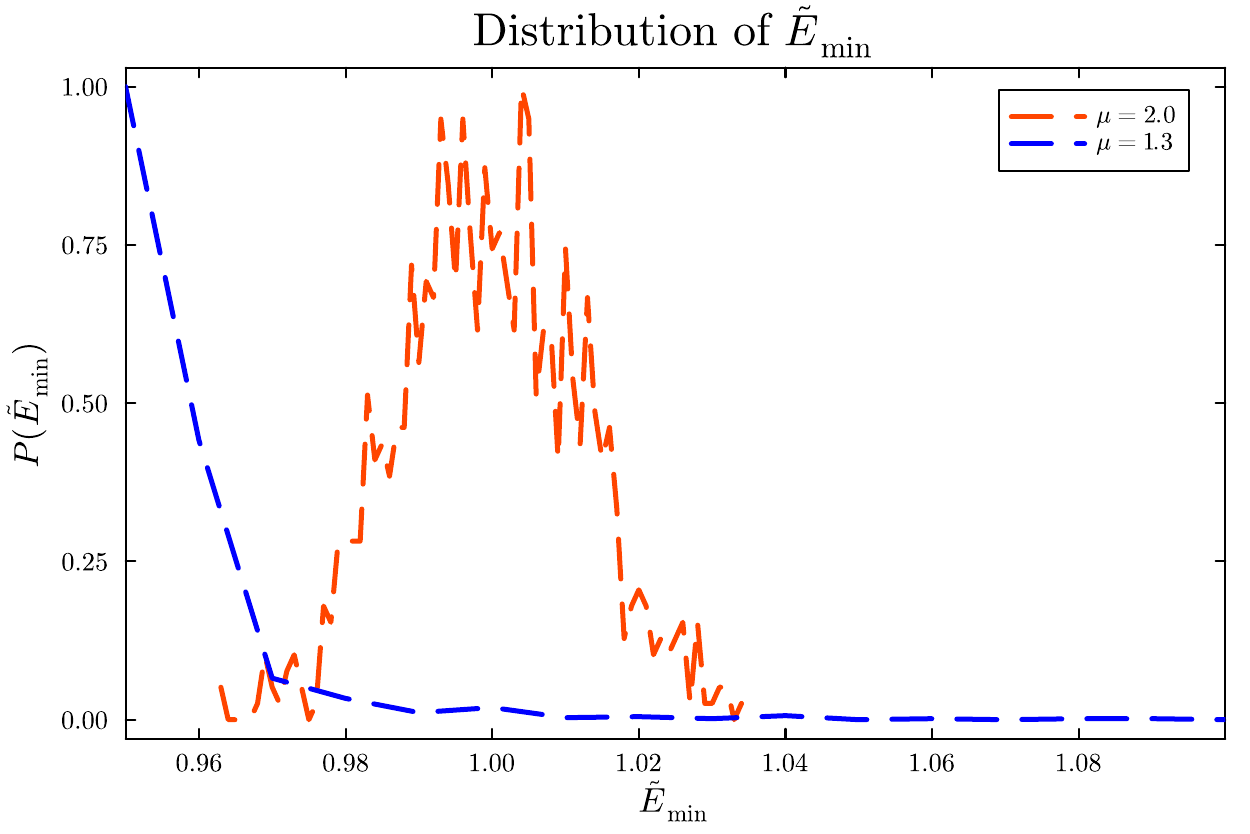}
        \label{fig:e0-dist-supp}
    }
    \caption{
        (a)  Scaling of the ground state energy with \(N\) for different values of \(\mu\).
        The linear fits are the corresponding dashed lines.
        (b) Distribution of the minimum eigenvalue $E_{\mathrm{min}}$.
        The energies (\(x\) axis) are scaled for comparison.
        The system size considered is $N = 30$.
    }
    \label{fig:eigen-stat}
\end{figure}

As discussed in the main text, the ground state energy of an SYK model is expected to be close to a configuration where all fermions are in the lowest available single-particle states, following Pauli exclusion principle.
The average $\langle E_{\min} \rangle$ is then expected to scale with $N$.
The same should broadly be true for LSYK, but now in competition with the parametrically large interaction.
The L\'evy disorder will introduce strong fluctuations in $\langle E_{\min} \rangle$.
This is observed in Fig.~\ref{fig:emin-scaling-supp}, where the mean ground state energy does not scale smoothly with \(N\): a linear fit becomes progressively poorer as \(\mu\) decreases.
This is a consequence of the strong L\'evy fluctuations.
We also compute the distribution of the $E_{\mathrm{min}}$ for different $\mu$, and present the result in Fig.~\ref{fig:e0-dist-supp}.
The distribution shifts from a Tracy-Widom-like~\cite{Tracy1994,nadal2011a} to a Poisson-like shape, as expected from corresponding results for random matrices~\cite{auffinger2016extreme}.

\begin{figure}[ht]
    \centering
    \includegraphics[width=0.45\linewidth]{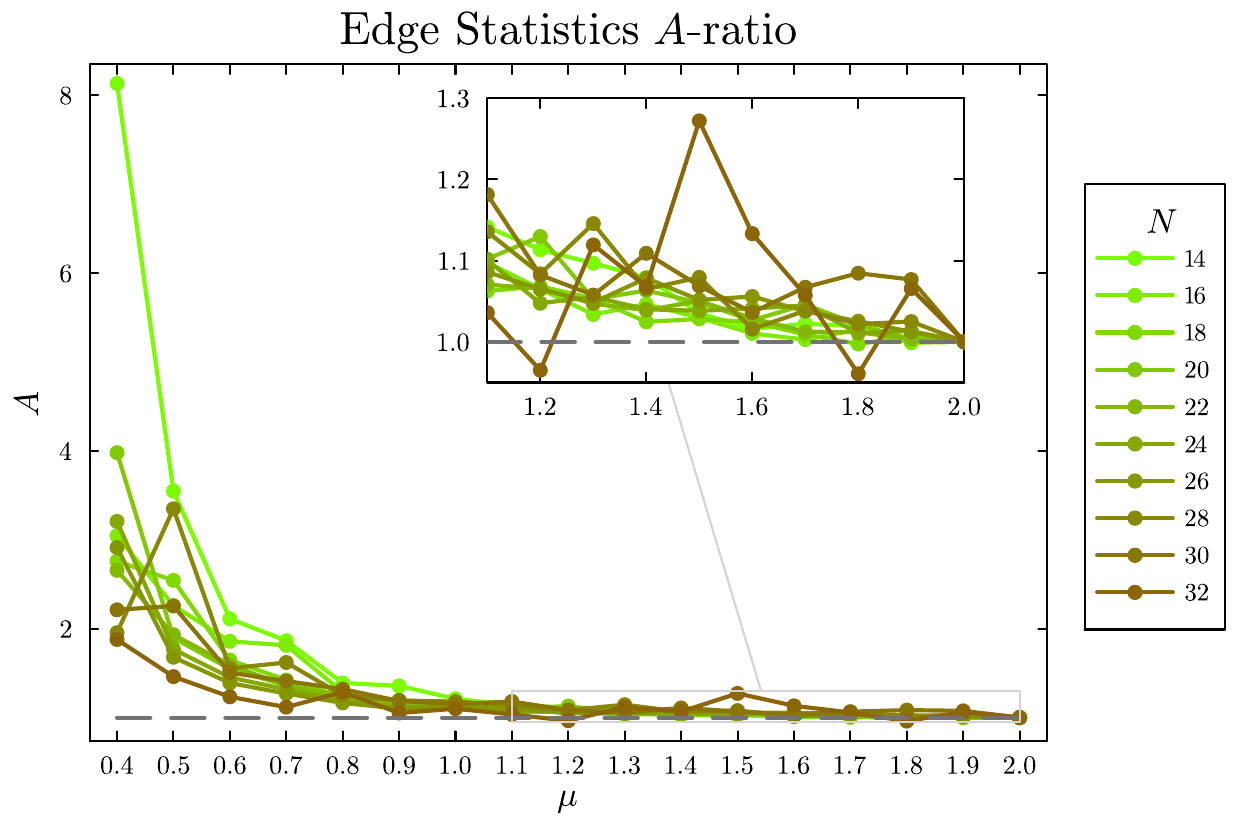}
    \caption{
        Increase of the ratio $A$ as a function of $\mu$, for different system sizes $N$.
        The ratio is scaled with the corresponding RMT value, which sets the value at $\mu = 2.0$ to $A = 1.0$.
        For $\mu < 2.0$, an increase in $A$ is observed.
        The dashed line corresponds to $1.0$ (i.e. the RMT value).
        (Inset) A close-up of the scaling near $\mu = 2.0$.
    }
    \label{fig:A-ratio}
\end{figure}

Another probe of edge statistics was proposed in~\cite{sun2020periodic}, capturing the fluctuations in the gap from the ground state.
\begin{align}
    A = \frac{\langle (E_2 - E_1)^2 \rangle}{\langle E_2 - E_1 \rangle^2}\,
    \label{eqn:Aratio}
\end{align}
where $E_1$ and $E_2$ are the lowest and second-lowest eigenvalues.
Extreme statistics are sensitive to the $10$ Altland-Zirnbauer classes of RMT (as well as the \(38\) non-Hermitian classes)~\cite{cipollini2021edge,xiao2024universal}, with exact values of \(A\) known for each of the classes~\cite{sun2020periodic}.
It is also known that $A$ decreases upon breaking integrability, going down to the RMT values when the system becomes fully chaotic.
It is important to note that there are various other quantities which can be used to explore the properties of the gap, such as $r-$ ratio of first and second level spacings, or other moments of $E_2 - E_1$.
In the absence of further motivation to study a particular quantity, we choose to study the gap through $A$, defined in \eqref{eqn:Aratio}. 
It is instructive to study how $A$ changes with $\mu$ in LSYK.
We study the behavior of \(A\) for different system sizes $N$ and determine the values of $\mu$ at which the edge statistics quantified by \(A\) deviates significantly from the RMT values.
The results are presented in Fig.~\ref{fig:A-ratio}.
We observe that $A$ increases as $\mu$ decreases, indicating emergent integrability, in agreement with \(r\)-ratios.
Interestingly, the system size dependence is remains weak down to \(\mu\approx 1.0\), and especially near \(\mu = 2.0\).
The curves for different system sizes scale similarly for \(\mu\) close to $2.0$, indicating weak system size dependence.
Recall that we observe a clear system-size dependence for \(r\)-ratios, as well as in the long-range statistics discussed below.

We can now discuss some of the supporting numerical results on the spectral form factor.
In the main text, we present the disconnected SFF, since it demonstrates the L\'evy peak. However, the full SFF is also equally sensitive to the deviation from RMT (despite not showing the peak):
this is expected, since the deviation arises from eigenvalue correlations $E_i - E_j$, which are present in both SFFs. 
\begin{figure}[htbp]
    \centering
    \subfigure[]{
        \includegraphics[width=0.45\linewidth]{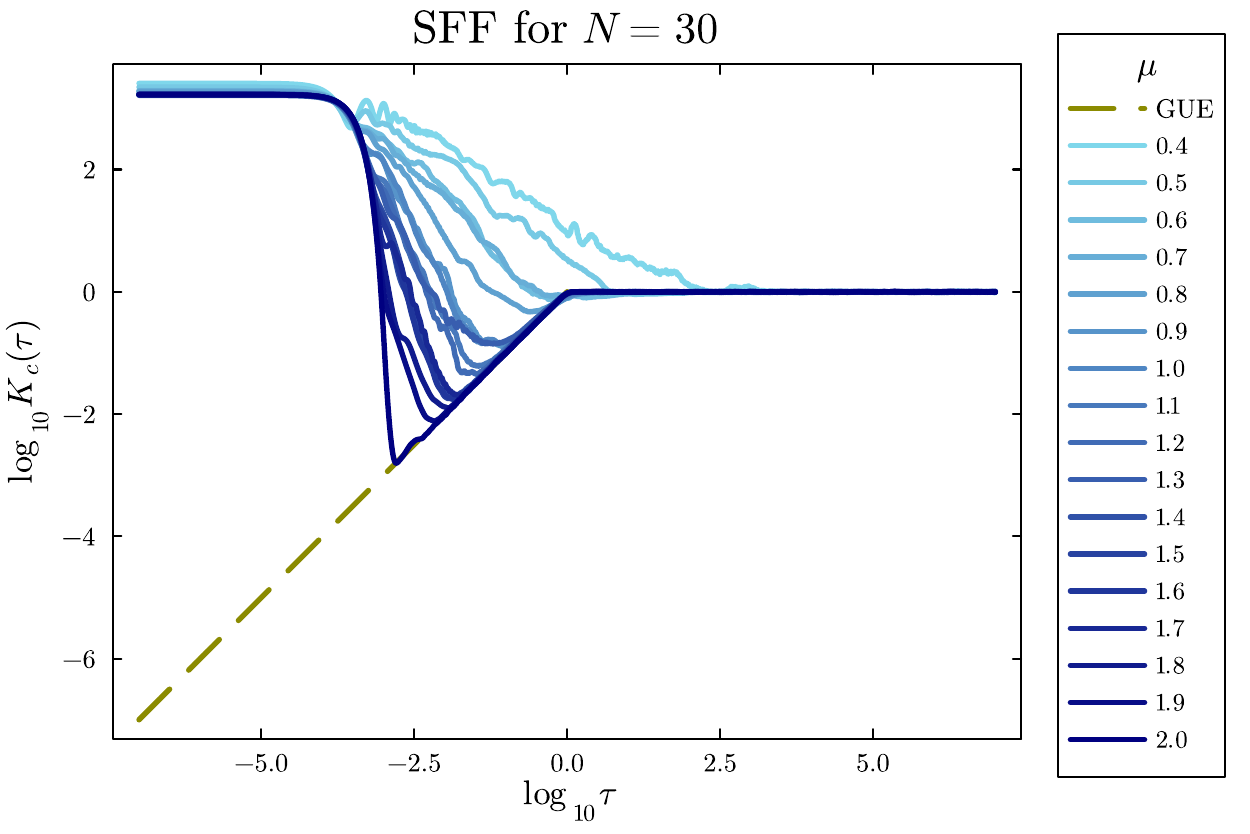}
        \label{fig:unf-sff-30-supp}
    }
    \subfigure[]{
        \includegraphics[width=0.45\linewidth]{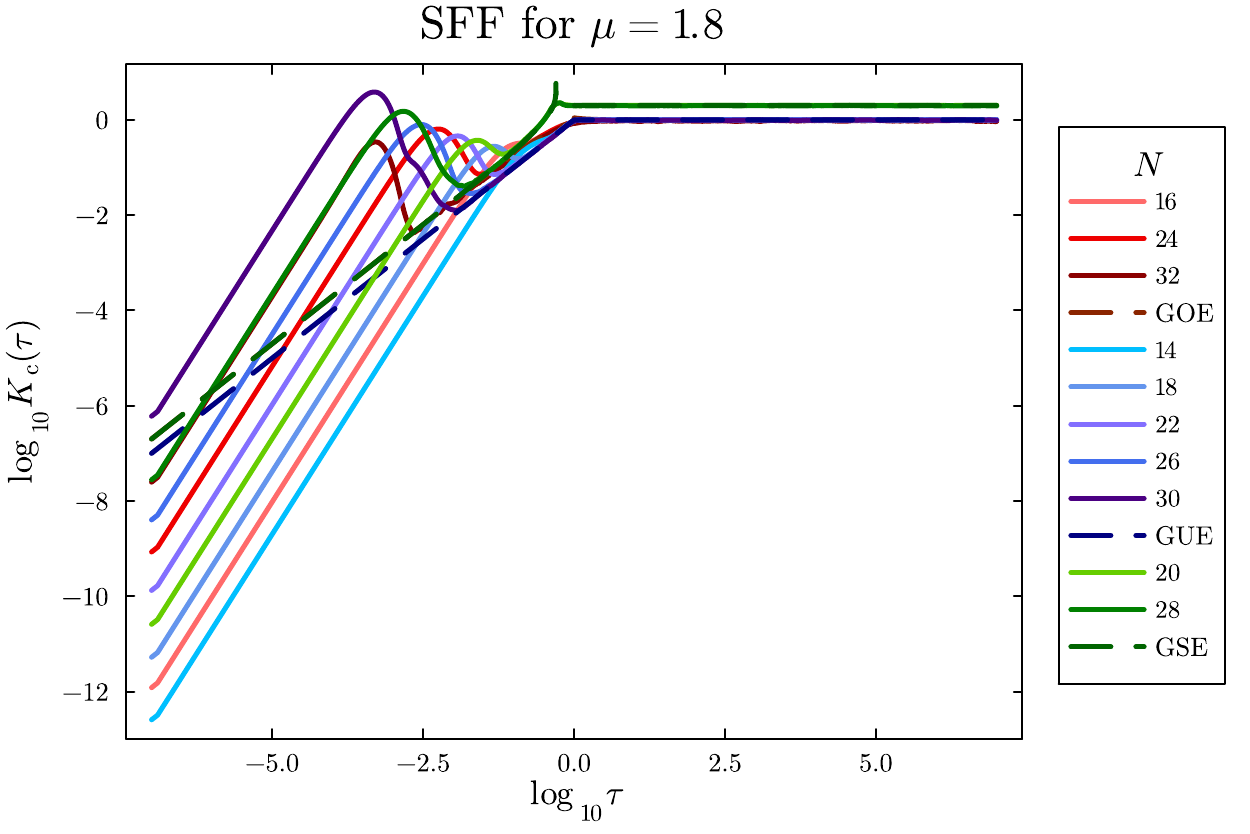}
        \label{fig:unf-ALLN-sff-conn-15-supp}
    }
    \caption{
        (a)  SFF computed using unfolded spectrum for \(N = 30\) fermions and a range of Levy parameters \(\mu\).
        The dashed line represents the corresponding Gaussian RMT behavior of the unfolded SFF (GUE for \(N=30\) fermions).
        (b) Connected SFF for $\mu = 1.8$ and system sizes \(N = 14\dots 32\).
        As \(N\) is increased, the peak time $\tau^{*}$ decreases, while the peak amplitude $K_{c}(\tau^{*})$ increases.
        The dashed lines indicate the RMT curves corresponding to GOE, GUE and GSE classes.
    }
    \label{fig:unf-sff-supp}
\end{figure}
As an example, we present the full SFF for $N = 30$ and various values of $\mu$ in Fig.~\ref{fig:unf-sff-30-supp}. 
The deviation from chaotic behaviour is apparent as $\mu$ decreases.
However, the L\'evy peak is absent. This is instead seen in the disconnected SFF in Fig.~\ref{fig:unf-ALLN-sff-conn-15-supp}, where we find that the location of the peak $\tau^{*}$ decreases with system size $N$.
In the main text, this scaling law is presented, which goes as $\log_{10} \tau^{*} \sim -0.15 N$. 
\begin{figure}[htbp]
    \centering
    \subfigure[]{
        \includegraphics[width=0.45\linewidth]{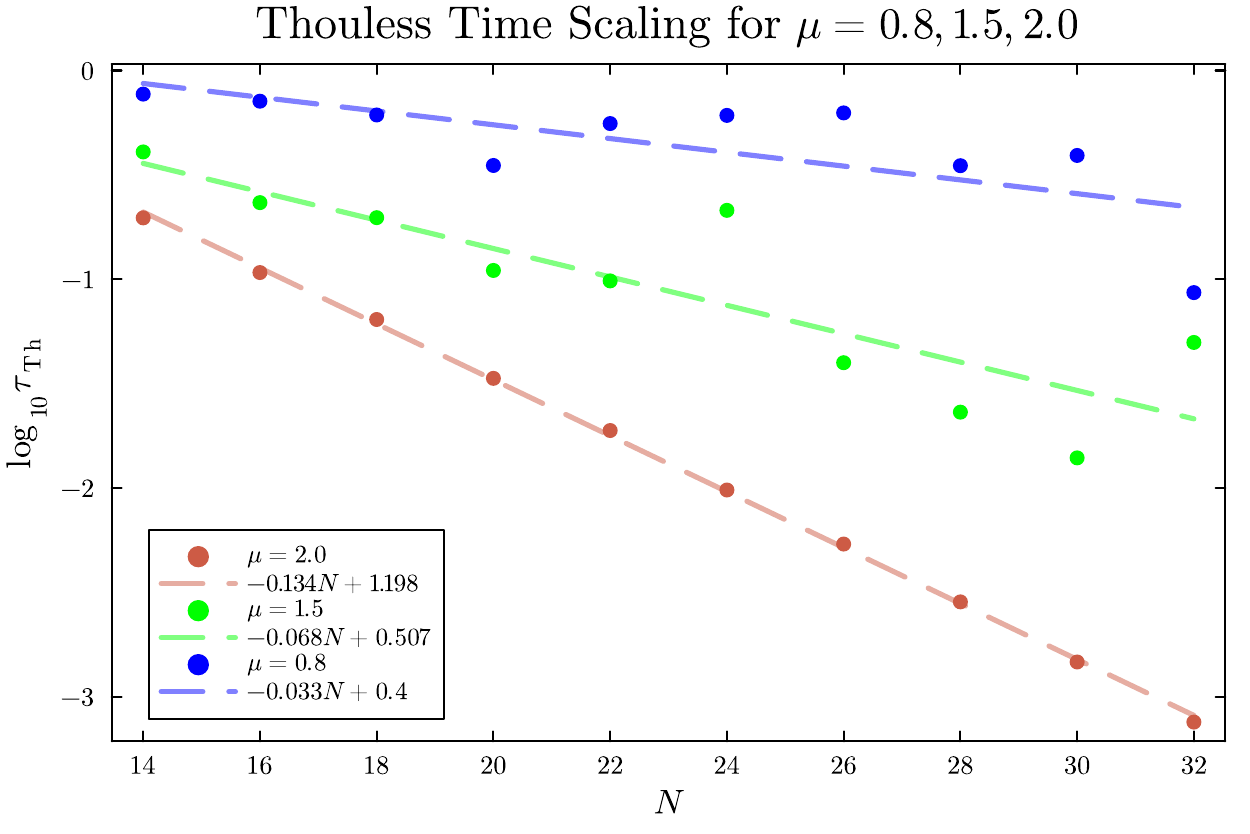}
        \label{fig:thouless-threesyk-supp}
    }
    \subfigure[]{
        \includegraphics[width=0.45\linewidth]{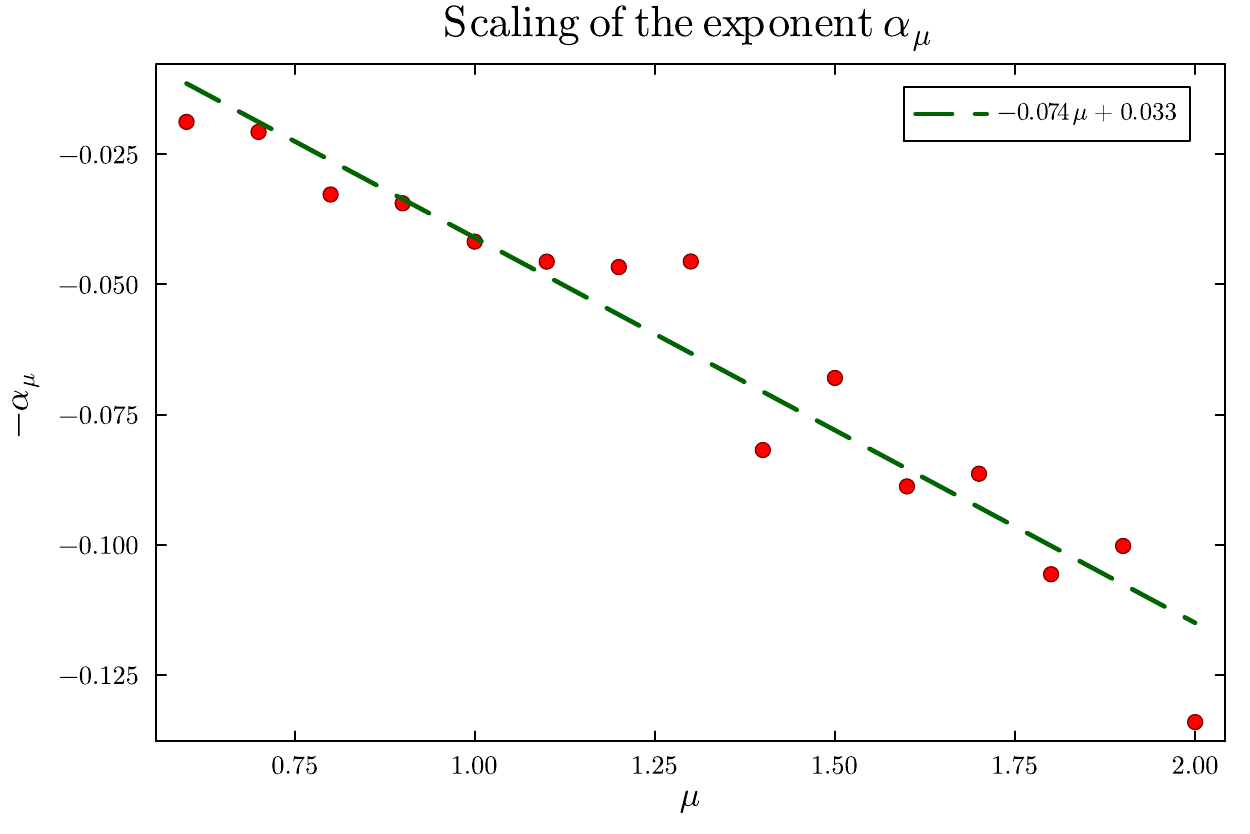}
        \label{fig:scaling-alpha-supp}
    }
    \caption{
        (a) Scaling of the Thouless time $\tmt$ with system size $N$ for $\mu = 0.8, 1.5, 2.0$.
        (b) Scaling of the exponent $\alpha_{\mu}$ as a function of the L\'evy parameter $\mu$.
        The scaling behavior is roughly linear, especially close to $\mu = 2.0$.
    }
    \label{fig:tmt-sff-supp}
\end{figure}
Similarly, we also investigate the scaling of the Thouless time with $N$ for various values of $\mu$.
For the usual SYK (i.e \(\mu = 2\)), we find that $\log_{10}\tmt$ decreases linearly with $N$.
This is presented for a few $\mu$ values in Fig.~\ref{fig:thouless-threesyk-supp}.
Note that the scaling of the $\tmt$ is weaker than the scaling of $\tau^{*}$.
This is natural, since $\tau^{*} < \tmt$. For arbitrary $\mu$, we can determine the scaling coefficient $\alpha_\mu$ which controls the scaling of $\tmt$: $\log_{10}\tmt \sim -\alpha_\mu N$.
We note in Fig.~\ref{fig:scaling-alpha-supp} that $\alpha_\mu$ scales roughly linearly with $\mu$, especially near $\mu = 2$.

\subsection{Deformations}\label{app:AppendixD3}
\begin{figure}[htbp]
    \centering
    \subfigure[]{
    \includegraphics[width=0.45\linewidth]{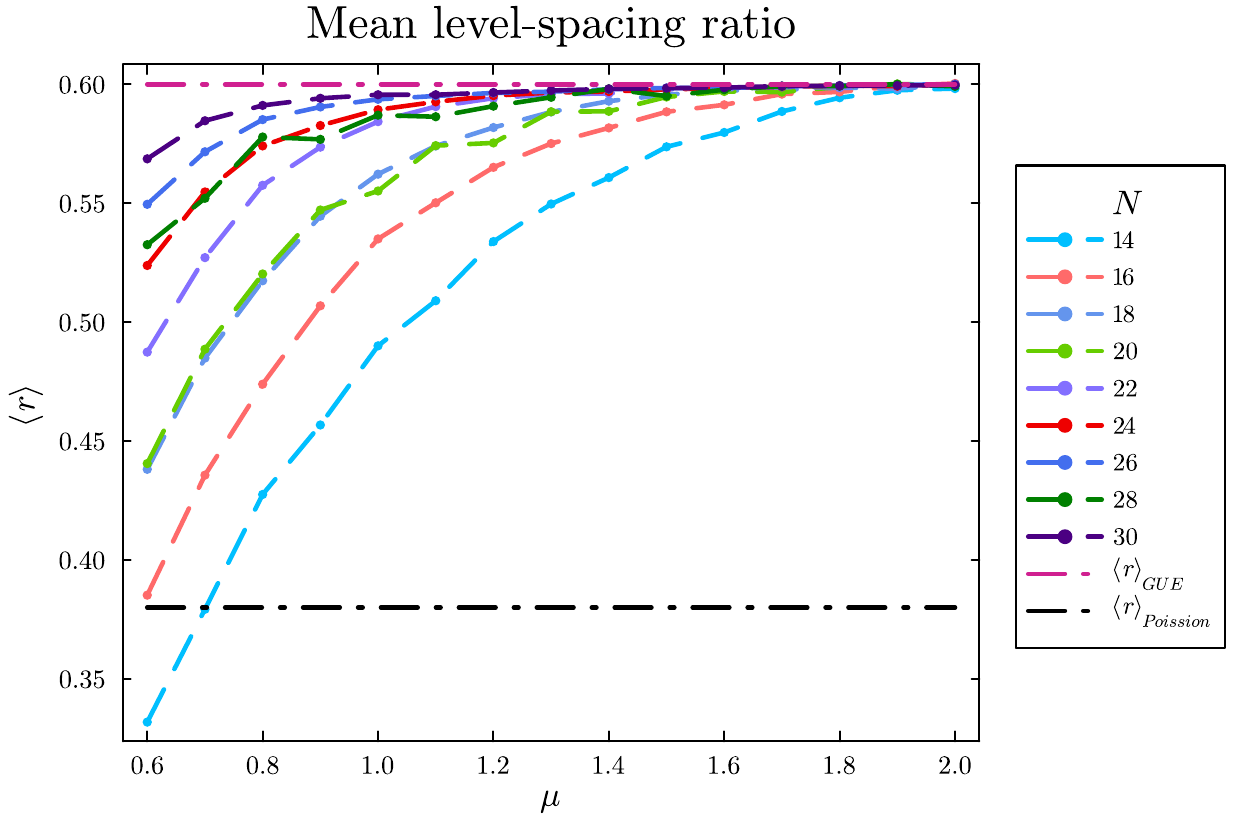}}
    \subfigure[]{
    \includegraphics[width=0.45\linewidth]{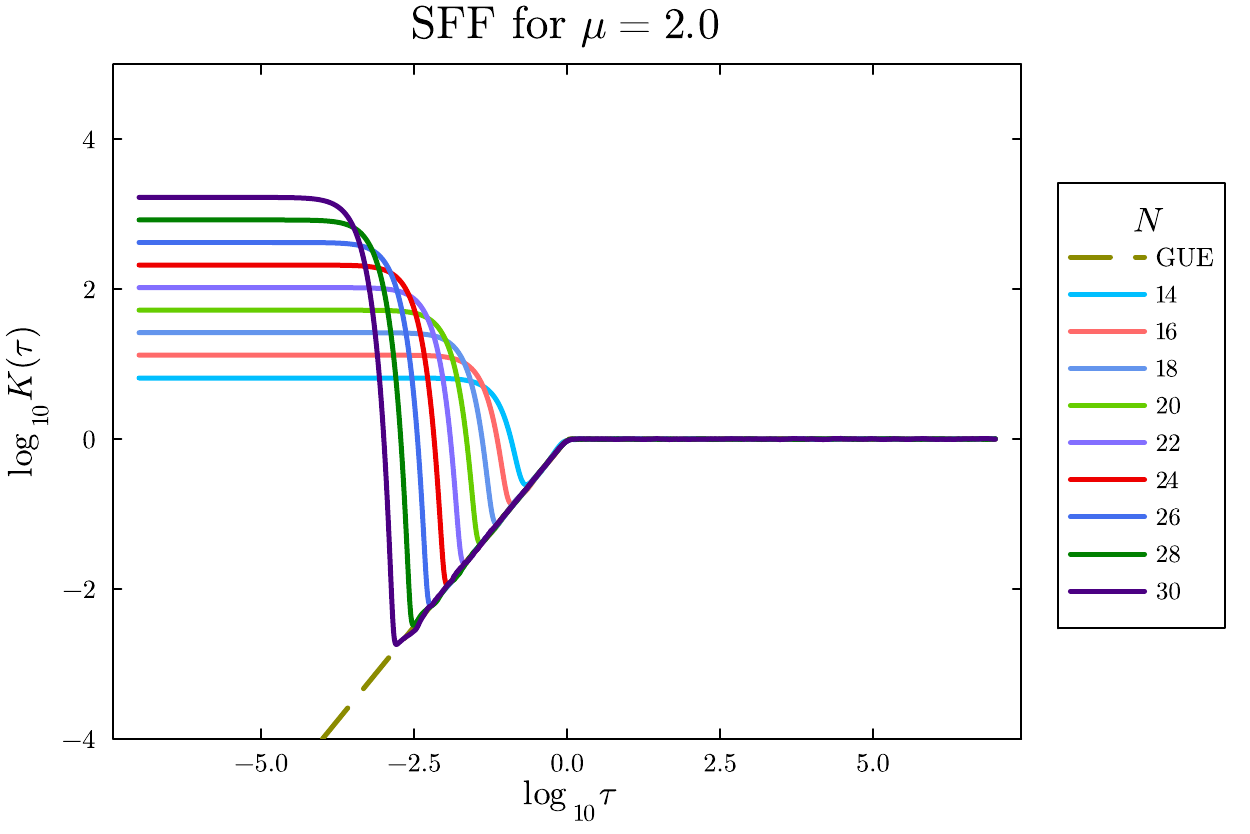}}
    \caption{
        Short- and long-range correlations for the defect-deformed L\'evy SYK ($H^{(1)}_{SYK}$) captured via (a) the $r$-ratio and (b) spectral form factor.
        The main effect is to collapse all universality classes on to GUE.
    }
    \label{fig:deformation-stat}
\end{figure} 

\begin{figure}[ht]
    \centering
    \subfigure[]{
    \includegraphics[width=0.45\linewidth]{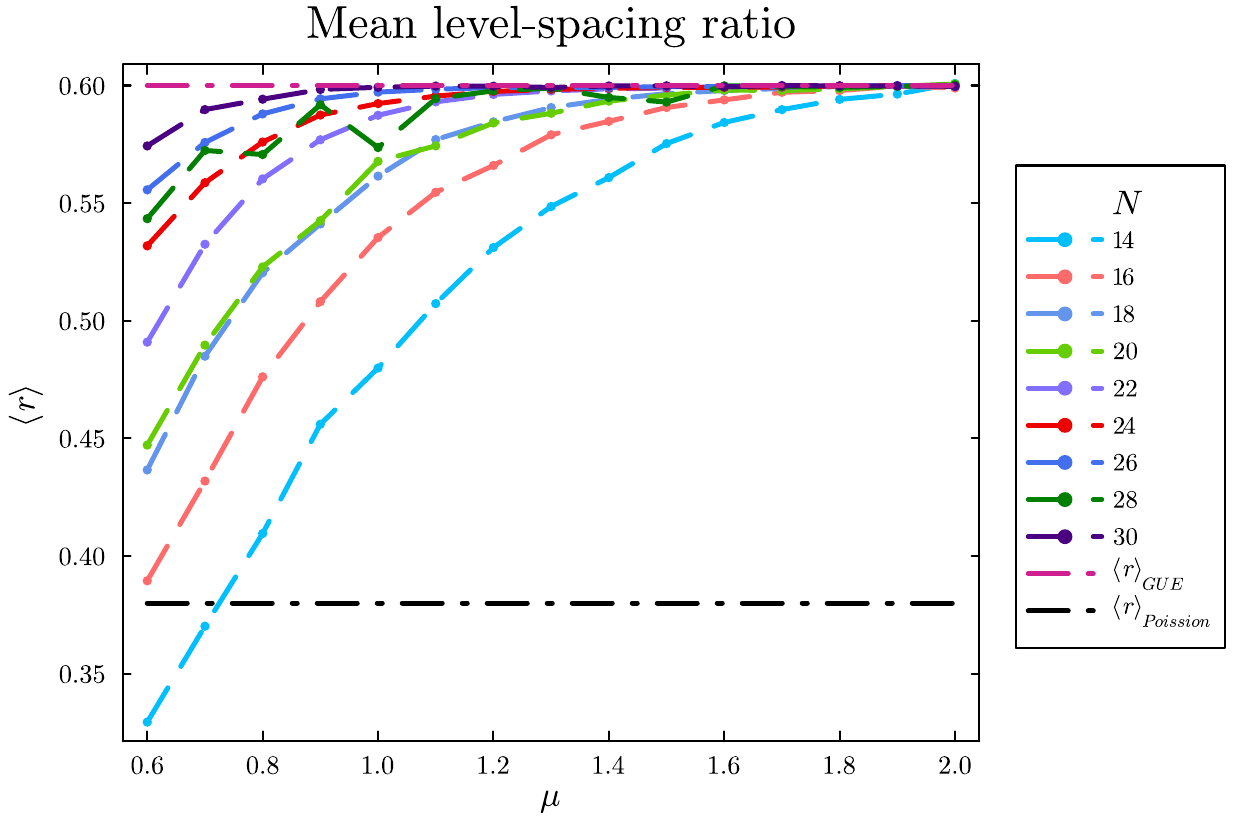}}
    \subfigure[]{
    \includegraphics[width=0.45\linewidth]{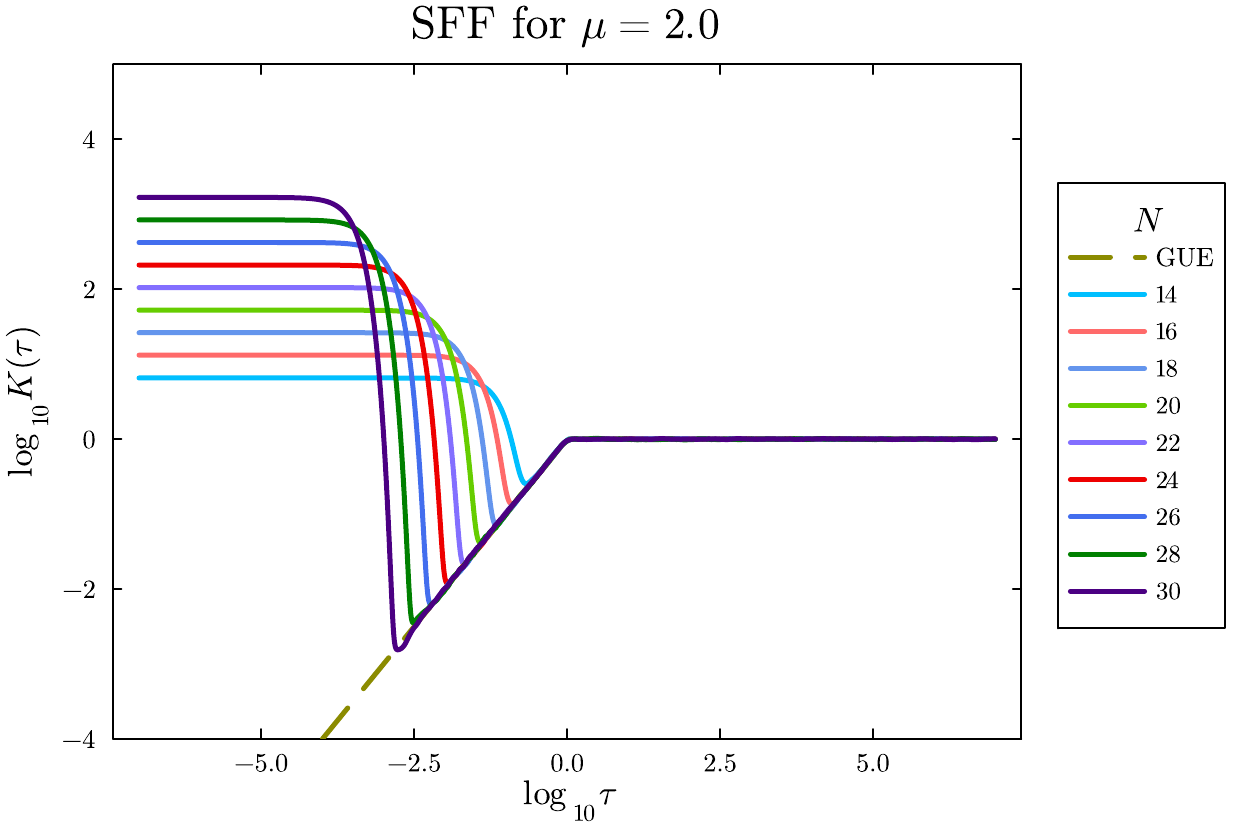}}
    \caption{
        Short- and long-range correlations for the mass-deformed L\'evy SYK ($H^{(1)}_{SYK}$) captured via (a) the $r$- ratio and (b) the spectral form factor.
        The main effect is to collapse all universality classes on to GUE.
    }
    \label{fig:massdef-stat}
\end{figure}

It is known that due to the Majorana algebra, the SYK model (irrespective of the disorder distribution) has certain (anti-) unitary symmetries which causes different system sizes to fall into different Gaussian universality classes Ref.~\onlinecite {garcia2016spectral}.
The symmetry classes repeat as $N\,\text{mod}\;8$.
The system sizes $N = 14,18,22,26\,\&\,30$ correspond to GUE, $N = 16, 24\,\&\,32$ correspond to GOE and $N = 20\,\&\,28$ correspond to GSE.
For numerical computations, it is sometimes convenient to break these symmetries by introducing small deformations.
There are two deformations that we consider: defect deformation and mass deformation.
\begin{align}
    H^{(1)}_{SYK} &= H_{SYK} + \frac{1}{N}\gamma_i \;\;\;\;\;\;\;\;\;\;\;\;\, \text{Defect deformation} \notag \\
    H^{(2)}_{SYK} &= H_{SYK} + \frac{1}{N}\sum_{i < j}\gamma_{i}\gamma_{j} \;\;\; \text{Mass deformation} \notag
\end{align}
Both deformations break these symmetries, forcing the GOE and GSE classes to revert to GUE.
The eigenvalues can then be evaluated, and the level statistics (short- and long-range correlations) can be studied.
The conclusions remain the same as in the non-deformed case (given that the deformation strength is small), with the major change being that the RMT results correspond to GUE for all $N$. 

The $r$-ratio and spectral form factor can be evaluated for these systems in an identical fashion as before.
These results are presented in Fig.~\ref{fig:deformation-stat} for the defect deformation.

Similarly, we study the spectral correlations for the mass-deformed case, with the main results presented in Fig.~\ref{fig:massdef-stat}.

\twocolumngrid

\bibliography{ergodicity,general,glass,mbl,software,local}

@article{sun2020periodic,
  title = {Periodic Table of the Ordinary and Supersymmetric Sachdev-Ye-Kitaev Models},
  author = {Sun, Fadi and Ye, Jinwu},
  journal = {Phys. Rev. Lett.},
  volume = {124},
  issue = {24},
  pages = {244101},
  numpages = {6},
  year = {2020},
  month = {Jun},
  publisher = {American Physical Society},
  doi = {10.1103/PhysRevLett.124.244101},
  url = {https://link.aps.org/doi/10.1103/PhysRevLett.124.244101}
}

@article{atas2013distribution,
 author = {Atas, Y.Y. and Bogomolny, E. and Giraud, O. and Roux, G.},
 title = {Distribution of the Ratio of Consecutive Level Spacings in Random Matrix Ensembles},
 journal = {Phys. Rev. Lett.},
 volume = {110},
 issue = {8},
 pages = {084101},
 numpages = {5},
 year = {2013},
 month = feb,
 publisher = {American Physical Society},
 doi = {10.1103/PhysRevLett.110.084101},
 url = {https://link.aps.org/doi/10.1103/PhysRevLett.110.084101},
}

@article{garcia2021sparse,
  title = {Sparse Sachdev-Ye-Kitaev model, quantum chaos, and gravity duals},
  author = {Garc\'{\i}a-Garc\'{\i}a, Antonio M. and Jia, Yiyang and Rosa, Dario and Verbaarschot, Jacobus J. M.},
  journal = {Phys. Rev. D},
  volume = {103},
  issue = {10},
  pages = {106002},
  numpages = {28},
  year = {2021},
  month = {May},
  publisher = {American Physical Society},
  doi = {10.1103/PhysRevD.103.106002},
  url = {https://link.aps.org/doi/10.1103/PhysRevD.103.106002}
}

@article{tezuka2023binary,
  title = {Binary-coupling sparse Sachdev-Ye-Kitaev model: An improved model of quantum chaos and holography},
  author = {Tezuka, Masaki and Oktay, Onur and Rinaldi, Enrico and Hanada, Masanori and Nori, Franco},
  journal = {Phys. Rev. B},
  volume = {107},
  issue = {8},
  pages = {L081103},
  numpages = {7},
  year = {2023},
  month = {Feb},
  publisher = {American Physical Society},
  doi = {10.1103/PhysRevB.107.L081103},
  url = {https://link.aps.org/doi/10.1103/PhysRevB.107.L081103}
}

@article{garcia2016spectral,
  title = {Spectral and thermodynamic properties of the Sachdev-Ye-Kitaev model},
  author = {Garc\'{\i}a-Garc\'{\i}a, Antonio M. and Verbaarschot, Jacobus J. M.},
  journal = {Phys. Rev. D},
  volume = {94},
  issue = {12},
  pages = {126010},
  numpages = {13},
  year = {2016},
  month = {Dec},
  publisher = {American Physical Society},
  doi = {10.1103/PhysRevD.94.126010},
  url = {https://link.aps.org/doi/10.1103/PhysRevD.94.126010}
}

@book{nolan2020univariate,
    author = "John P. Nolan",
    title = "Univariate Stable Distributions",
    publisher = "Springer Cham",
    doi = "https://doi.org/10.1007/978-3-030-52915-4",
    year = "2020"
}

@book{kolmogorov1954limit,
    author = "{ B. V. Gnedenko and A. N. Kolmogorov}",
    title = "{Limit distributions for sums of independent random variables}" ,
    publisher = "{Addison-Wesley.}",
    year = "1954"
}

@book{mehta2004random,
  title = {Random Matrices},
  address = {Amsterdam},
  author = {Mehta, M. L.},
  address = {Amsterdam},
  booktitle = {Random matrices},
  edition = {3rd ed.},
  isbn = {0120884097},
  publisher = {Elsevier/Academic Press},
  series = {Pure and applied mathematics: v. 142},
  title = {Random matrices / Madan Lal Mehta.},
  year = {2004},
}

@article{burda2002free,
  title = {Free random L\'evy matrices},
  author = {Burda, Zdzis\l{}aw and Janik, Romuald A. and Jurkiewicz, Jerzy and Nowak, Maciej A. and Papp, Gabor and Zahed, Ismail},
  journal = {Phys. Rev. E},
  volume = {65},
  issue = {2},
  pages = {021106},
  numpages = {5},
  year = {2002},
  month = {Jan},
  publisher = {American Physical Society},
  doi = {10.1103/PhysRevE.65.021106},
  url = {https://link.aps.org/doi/10.1103/PhysRevE.65.021106}
}

@article{burda2007free,
  title = {Free random L\'evy and Wigner-L\'evy matrices},
  author = {Burda, Zdzis\l{}aw and Jurkiewicz, Jerzy and Nowak, Maciej A. and Papp, Gabor and Zahed, Ismail},
  journal = {Phys. Rev. E},
  volume = {75},
  issue = {5},
  pages = {051126},
  numpages = {11},
  year = {2007},
  month = {May},
  publisher = {American Physical Society},
  doi = {10.1103/PhysRevE.75.051126},
  url = {https://link.aps.org/doi/10.1103/PhysRevE.75.051126}
}

@article{maldacena2016remarks,
  title = {Remarks on the Sachdev-Ye-Kitaev model},
  author = {Maldacena, Juan and Stanford, Douglas},
  journal = {Phys. Rev. D},
  volume = {94},
  issue = {10},
  pages = {106002},
  numpages = {43},
  year = {2016},
  month = {Nov},
  publisher = {American Physical Society},
  doi = {10.1103/PhysRevD.94.106002},
  url = {https://link.aps.org/doi/10.1103/PhysRevD.94.106002}
}

@incollection{haake1991quantum,
	author = {Haake, Fritz},
	booktitle = {Quantum Coherence in Mesoscopic Systems},
	publisher = {Springer},
	title = {Quantum signatures of chaos},
	year = {1991}
}

@article{janzen2010thermodynamics,
  title = {Thermodynamics of the L\'evy spin glass},
  author = {Janzen, K. and Engel, A. and M\'ezard, M.},
  journal = {Phys. Rev. E},
  volume = {82},
  issue = {2},
  pages = {021127},
  numpages = {14},
  year = {2010},
  month = {Aug},
  publisher = {American Physical Society},
  doi = {10.1103/PhysRevE.82.021127},
  url = {https://link.aps.org/doi/10.1103/PhysRevE.82.021127}
}

@article{sachdev1993spin,
  title = {Gapless spin-fluid ground state in a random quantum Heisenberg magnet},
  author = {Sachdev, Subir and Ye, Jinwu},
  journal = {Phys. Rev. Lett.},
  volume = {70},
  issue = {21},
  pages = {3339--3342},
  numpages = {0},
  year = {1993},
  month = {May},
  publisher = {American Physical Society},
  doi = {10.1103/PhysRevLett.70.3339},
  url = {https://link.aps.org/doi/10.1103/PhysRevLett.70.3339}
}

@misc{kitaevLectures,
author = "Alexei Kitaev",
title = "{A simple model of quantum holography}",
journal = "{\em KITP strings seminar and Entanglement 2015 program} (Feb. 12, April 7, and May 27, 2015)", 
url = "{http://online.kitp.ucsb.edu/online/entangled15/}"
}

@article{Orman:2024mpw,
    author = "Orman, Patrick and Gharibyan, Hrant and Preskill, John",
    title = "{Quantum chaos in the sparse SYK model}",
    eprint = "2403.13884",
    journal="",
    archivePrefix = "arXiv",
    primaryClass = "hep-th",
    month = "3",
    year = "2024"
}

@article{Cotler2017chaos,
    author = "Cotler, Jordan and Hunter-Jones, Nicholas and Liu, Junyu and Yoshida, Beni",
    title = "{Chaos, Complexity, and Random Matrices}",
    eprint = "1706.05400",
    archivePrefix = "arXiv",
    primaryClass = "hep-th",
    doi = "10.1007/JHEP11(2017)048",
    journal = "JHEP",
    volume = "11",
    pages = "048",
    year = "2017"
}

@article{cotler2016black,
    author = "Cotler, Jordan S. and Gur-Ari, Guy and Hanada, Masanori and Polchinski, Joseph and Saad, Phil and Shenker, Stephen H. and Stanford, Douglas and Streicher, Alexandre and Tezuka, Masaki",
    title = "{Black Holes and Random Matrices}",
    eprint = "1611.04650",
    archivePrefix = "arXiv",
    primaryClass = "hep-th",
    reportNumber = "SU-ITP-16-19, SU-ITP-16/19, YITP-16-124",
    doi = "10.1007/JHEP05(2017)118",
    journal = "JHEP",
    volume = "05",
    pages = "118",
    year = "2017",
    note = "[Erratum: JHEP 09, 002 (2018)]"
}

@article{majumdar2010extreme,
  title={Extreme eigenvalues of Wishart matrices: application to entangled bipartite system},
  author={Majumdar, Satya N},
  journal={arXiv preprint arXiv:1005.4515},
  year={2010}
}

@article{auffinger2016extreme,
title = {Extreme eigenvalues of sparse, heavy tailed random matrices},
journal = {Stochastic Processes and their Applications},
volume = {126},
number = {11},
pages = {3310-3330},
year = {2016},
issn = {0304-4149},
doi = {https://doi.org/10.1016/j.spa.2016.04.029},
url = {https://www.sciencedirect.com/science/article/pii/S0304414915300089},
author = {Antonio Auffinger and Si Tang}
}

@article{cipollini2021edge,
author={Cipolloni, Giorgio
and Erd{\H{o}}s, L{\'a}szl{\'o}
and Schr{\"o}der, Dominik},
title={Edge universality for non-Hermitian random matrices},
journal={Probability Theory and Related Fields},
year={2021},
month={Feb},
day={01},
volume={179},
number={1},
pages={1-28},
issn={1432-2064},
doi={10.1007/s00440-020-01003-7},
url={https://doi.org/10.1007/s00440-020-01003-7}
}

@article{xiao2024universal,
  title = {Universal hard-edge statistics of non-Hermitian random matrices},
  author = {Xiao, Zhenyu and Shindou, Ryuichi and Kawabata, Kohei},
  journal = {Phys. Rev. Res.},
  volume = {6},
  issue = {2},
  pages = {023303},
  numpages = {33},
  year = {2024},
  month = {Jun},
  publisher = {American Physical Society},
  doi = {10.1103/PhysRevResearch.6.023303},
  url = {https://link.aps.org/doi/10.1103/PhysRevResearch.6.023303}
}

@article{winn2023extreme,
  title={Extreme eigenvalues of random matrices from Jacobi ensembles},
  author={Winn, B},
  journal={arXiv preprint arXiv:2302.12082},
  year={2023}
}

@article{dean2006large,
  title = {Large Deviations of Extreme Eigenvalues of Random Matrices},
  author = {Dean, David S. and Majumdar, Satya N.},
  journal = {Phys. Rev. Lett.},
  volume = {97},
  issue = {16},
  pages = {160201},
  numpages = {4},
  year = {2006},
  month = {Oct},
  publisher = {American Physical Society},
  doi = {10.1103/PhysRevLett.97.160201},
  url = {https://link.aps.org/doi/10.1103/PhysRevLett.97.160201}
}

@article{nadal2011a,
doi = {10.1088/1742-5468/2011/04/P04001},
url = {https://dx.doi.org/10.1088/1742-5468/2011/04/P04001},
year = {2011},
month = {apr},
publisher = {},
volume = {2011},
number = {04},
pages = {P04001},
author = {Nadal, Celine and Majumdar, Satya N},
title = {A simple derivation of the Tracy–Widom distribution of the maximal eigenvalue of a
Gaussian unitary random matrix},
journal = {Journal of Statistical Mechanics: Theory and Experiment}
}

@article{Tracy1994,
author={Tracy, Craig A.
and Widom, Harold},
title={Level-spacing distributions and the Airy kernel},
journal={Communications in Mathematical Physics},
year={1994},
month={Jan},
day={01},
volume={159},
number={1},
pages={151-174},
issn={1432-0916},
doi={10.1007/BF02100489},
url={https://doi.org/10.1007/BF02100489}
}

@article{a1995spectral,
  title = {Spectral statistics in the quantized cardioid billiard},
  author = {B\"acker, A. and Steiner, F. and Stifter, P.},
  journal = {Phys. Rev. E},
  volume = {52},
  issue = {3},
  pages = {2463--2472},
  numpages = {0},
  year = {1995},
  month = {Sep},
  publisher = {American Physical Society},
  doi = {10.1103/PhysRevE.52.2463},
  url = {https://link.aps.org/doi/10.1103/PhysRevE.52.2463}
}

@article{bujisman2020sensitivity,
  title = {Sensitivity of the spectral form factor to short-range level statistics},
  author = {Buijsman, Wouter and Cheianov, Vadim and Gritsev, Vladimir},
  journal = {Phys. Rev. E},
  volume = {102},
  issue = {4},
  pages = {042216},
  numpages = {6},
  year = {2020},
  month = {Oct},
  publisher = {American Physical Society},
  doi = {10.1103/PhysRevE.102.042216},
  url = {https://link.aps.org/doi/10.1103/PhysRevE.102.042216}
}

@article{A2002quantum,
  title = {Quantum Chaos and $1/f$ Noise},
  author = {Rela\~no, A. and G\'omez, J. M. G. and Molina, R. A. and Retamosa, J. and Faleiro, E.},
  journal = {Phys. Rev. Lett.},
  volume = {89},
  issue = {24},
  pages = {244102},
  numpages = {4},
  year = {2002},
  month = {Nov},
  publisher = {American Physical Society},
  doi = {10.1103/PhysRevLett.89.244102},
  url = {https://link.aps.org/doi/10.1103/PhysRevLett.89.244102}
}

@article{andreanov2025from,
  title = {From Dyson models to many-body quantum chaos},
  author = {Andreanov, Alexei and Carrega, Matteo and Murugan, Jeff and Olle, Jan and Rosa, Dario and Shir, Ruth},
  journal = {Phys. Rev. B},
  volume = {111},
  issue = {3},
  pages = {035147},
  numpages = {9},
  year = {2025},
  month = {Jan},
  publisher = {American Physical Society},
  doi = {10.1103/PhysRevB.111.035147},
  url = {https://link.aps.org/doi/10.1103/PhysRevB.111.035147}
}

@article{apollo2023unitarity,
  title = {Unitarity breaking in self-averaging spectral form factors},
  author = {Matsoukas-Roubeas, Apollonas S. and Beau, Mathieu and Santos, Lea F. and del Campo, Adolfo},
  journal = {Phys. Rev. A},
  volume = {108},
  issue = {6},
  pages = {062201},
  numpages = {11},
  year = {2023},
  month = {Dec},
  publisher = {American Physical Society},
  doi = {10.1103/PhysRevA.108.062201},
  url = {https://link.aps.org/doi/10.1103/PhysRevA.108.062201}
}

@article{fu2017susy,
  title = {Supersymmetric Sachdev-Ye-Kitaev models},
  author = {Fu, Wenbo and Gaiotto, Davide and Maldacena, Juan and Sachdev, Subir},
  journal = {Phys. Rev. D},
  volume = {95},
  issue = {2},
  pages = {026009},
  numpages = {20},
  year = {2017},
  month = {Jan},
  publisher = {American Physical Society},
  doi = {10.1103/PhysRevD.95.026009},
  url = {https://link.aps.org/doi/10.1103/PhysRevD.95.026009}
}

@article{hanada2024a,
    author = "Hanada, Masanori and Jevicki, Antal and Liu, Xianlong and Rinaldi, Enrico and Tezuka, Masaki",
    title = "{A model of randomly-coupled Pauli spins}",
    eprint = "2309.15349",
    archivePrefix = "arXiv",
    primaryClass = "hep-th",
    reportNumber = "RIKEN-iTHEMS-Report-23",
    doi = "10.1007/JHEP05(2024)280",
    journal = "JHEP",
    volume = "05",
    pages = "280",
    year = "2024"
}

@article{sunderhauf2019quantum,
author={S{\"u}nderhauf, Christoph
and Piroli, Lorenzo
and Qi, Xiao-Liang
and Schuch, Norbert
and Cirac, J. Ignacio},
title={Quantum chaos in the Brownian SYK model with large finite N : OTOCs and tripartite information},
journal={Journal of High Energy Physics},
year={2019},
month={Nov},
day={07},
volume={2019},
number={11},
pages={38},
doi={10.1007/JHEP11(2019)038},
url={https://doi.org/10.1007/JHEP11(2019)038}
}

@article{krajewski2019non,
  title = {Non-Gaussian disorder average in the Sachdev-Ye-Kitaev model},
  author = {Krajewski, T. and Laudonio, M. and Pascalie, R. and Tanasa, A.},
  journal = {Phys. Rev. D},
  volume = {99},
  issue = {12},
  pages = {126014},
  numpages = {17},
  year = {2019},
  month = {Jun},
  publisher = {American Physical Society},
  doi = {10.1103/PhysRevD.99.126014},
  url = {https://link.aps.org/doi/10.1103/PhysRevD.99.126014}
}

@article{witten2019an,
doi = {10.1088/1751-8121/ab3752},
url = {https://dx.doi.org/10.1088/1751-8121/ab3752},
year = {2019},
month = {oct},
publisher = {IOP Publishing},
volume = {52},
number = {47},
pages = {474002},
author = {Witten, Edward},
title = {An SYK-like model without disorder},
journal = {Journal of Physics A: Mathematical and Theoretical}
}

@article{monthus2016localization,
doi = {10.1088/1742-5468/2016/09/093304},
url = {https://dx.doi.org/10.1088/1742-5468/2016/09/093304},
year = {2016},
month = {sep},
publisher = {IOP Publishing and SISSA},
volume = {2016},
number = {9},
pages = {093304},
author = {Monthus, Cécile},
title = {Localization transition in random Lévy matrices: multifractality of eigenvectors in the localized phase and at criticality},
journal = {Journal of Statistical Mechanics: Theory and Experiment}
}

@article{jevicki2016bilocal,
    author = "Jevicki, Antal and Suzuki, Kenta and Yoon, Junggi",
    title = "{Bi-Local Holography in the SYK Model}",
    eprint = "1603.06246",
    archivePrefix = "arXiv",
    primaryClass = "hep-th",
    reportNumber = "BROWN-HET-1673",
    doi = "10.1007/JHEP07(2016)007",
    journal = "JHEP",
    volume = "07",
    pages = "007",
    year = "2016"
}

@article{pieper2016high,
title = {High-performance implementation of Chebyshev filter diagonalization for interior eigenvalue computations},
journal = {Journal of Computational Physics},
volume = {325},
pages = {226-243},
year = {2016},
issn = {0021-9991},
doi = {https://doi.org/10.1016/j.jcp.2016.08.027},
url = {https://www.sciencedirect.com/science/article/pii/S0021999116303837},
author = {Andreas Pieper and Moritz Kreutzer and Andreas Alvermann and Martin Galgon and Holger Fehske and Georg Hager and Bruno Lang and Gerhard Wellein}
}

@misc{chebf,
  author = "Dario Rosa",
  title = "{ChebyshevFiltering.jl}",
  year = "2023",
  publisher = "GitHub",
  journal = "GitHub repository",
  url = "https://github.com/Dario-Rosa85/ChebyshevFiltering.jl"
}

@book{varga2011geršgorin,
  title={Ger{\v{s}}gorin and His Circles},
  author={Varga, R.S.},
  isbn={9783540211006},
  lccn={2004104814},
  series={Springer Series in Computational Mathematics},
  url={https://books.google.co.kr/books?id=q4g0sMe3Ss8C},
  year={2011},
  publisher={Springer Berlin Heidelberg}
}

@article{winer2022hydrodynamic,
  title = {Hydrodynamic Theory of the Connected Spectral form Factor},
  author = {Winer, Michael and Swingle, Brian},
  journal = {Phys. Rev. X},
  volume = {12},
  issue = {2},
  pages = {021009},
  numpages = {24},
  year = {2022},
  month = {Apr},
  publisher = {American Physical Society},
  doi = {10.1103/PhysRevX.12.021009},
  url = {https://link.aps.org/doi/10.1103/PhysRevX.12.021009}
}

@article{rossini2020quantum,
  title = {Quantum Advantage in the Charging Process of Sachdev-Ye-Kitaev Batteries},
  author = {Rossini, Davide and Andolina, Gian Marcello and Rosa, Dario and Carrega, Matteo and Polini, Marco},
  journal = {Phys. Rev. Lett.},
  volume = {125},
  issue = {23},
  pages = {236402},
  numpages = {6},
  year = {2020},
  month = {Dec},
  publisher = {American Physical Society},
  doi = {10.1103/PhysRevLett.125.236402},
  url = {https://link.aps.org/doi/10.1103/PhysRevLett.125.236402}
}

@article{caceras2022spectral,
author={C{\'a}ceres, Elena and Misobuchi, Anderson and Raz, Amir},
title={Spectral form factor in sparse SYK models},
journal={Journal of High Energy Physics},
year={2022},
month={Aug},
day={23},
volume={2022},
number={8},
pages={236},
issn={1029-8479},
doi={10.1007/JHEP08(2022)236},
url={https://doi.org/10.1007/JHEP08(2022)236}
}

@misc{xu2020a,
    author = {Xu, Shenglong and Susskind, Leonard and Su, Yuan and Swingle, Brian},
    title = {A Sparse Model of Quantum Holography},
    eprint = {2008.02303},
    archivePrefix = {arXiv},
    primaryClass = {cond-mat.str-el},
    month = {8},
    year = {2020}
}

@article{cizeau1994theory,
  title = {Theory of L\'evy matrices},
  author = {Cizeau, P. and Bouchaud, J. P.},
  journal = {Phys. Rev. E},
  volume = {50},
  issue = {3},
  pages = {1810--1822},
  numpages = {0},
  year = {1994},
  month = {Sep},
  publisher = {American Physical Society},
  doi = {10.1103/PhysRevE.50.1810},
  url = {https://link.aps.org/doi/10.1103/PhysRevE.50.1810}
}

@article{kutlin2024anatomy,
  title={{Anatomy of the eigenstates distribution: A quest for a genuine multifractality}},
  author={Anton Kutlin and Ivan M. Khaymovich},
  journal={SciPost Phys.},
  volume={16},
  pages={008},
  year={2024},
  publisher={SciPost},
  doi={10.21468/SciPostPhys.16.1.008},
  url={https://scipost.org/10.21468/SciPostPhys.16.1.008},
  eprint={2309.06468},
  archivePrefix={arXiv},
  primaryClass={cond-mat.stat-mech}
}

@article{tarquini2016level,
  title = {Level Statistics and Localization Transitions of L\'evy Matrices},
  author = {Tarquini, E. and Biroli, G. and Tarzia, M.},
  journal = {Phys. Rev. Lett.},
  volume = {116},
  issue = {1},
  pages = {010601},
  numpages = {5},
  year = {2016},
  month = {Jan},
  publisher = {American Physical Society},
  doi = {10.1103/PhysRevLett.116.010601},
  url = {https://link.aps.org/doi/10.1103/PhysRevLett.116.010601}
}

@article{suntas2020quantum,
  title = {Quantum chaos challenges many-body localization},
  author = {\ifmmode \check{S}\else \v{S}\fi{}untajs, Jan and Bon\ifmmode \check{c}\else \v{c}\fi{}a, Janez and Prosen, Toma\v{z} and Vidmar, Lev},
  journal = {Phys. Rev. E},
  volume = {102},
  issue = {6},
  pages = {062144},
  numpages = {12},
  year = {2020},
  month = {Dec},
  publisher = {American Physical Society},
  doi = {10.1103/PhysRevE.102.062144},
  url = {https://link.aps.org/doi/10.1103/PhysRevE.102.062144}
}

@article{biroli2021levy,
  title = {L\'evy-Rosenzweig-Porter random matrix ensemble},
  author = {Biroli, G. and Tarzia, M.},
  journal = {Phys. Rev. B},
  volume = {103},
  issue = {10},
  pages = {104205},
  numpages = {20},
  year = {2021},
  month = {Mar},
  publisher = {American Physical Society},
  doi = {10.1103/PhysRevB.103.104205},
  url = {https://link.aps.org/doi/10.1103/PhysRevB.103.104205}
}

@article{oganesyan2007localization,
 author = {Oganesyan, Vadim and Huse, David A.},
 title = {Localization of interacting fermions at high temperature},
 journal = {Phys. Rev. B},
 volume = {75},
 issue = {15},
 pages = {155111},
 numpages = {5},
 year = {2007},
 month = apr,
 publisher = {American Physical Society},
 doi = {10.1103/PhysRevB.75.155111},
 url = {https://link.aps.org/doi/10.1103/PhysRevB.75.155111},
}




\end{document}